%% file: main.tex
\newif\iftikz
\tikzfalse
\newif\ifarxiv
\arxivtrue % uncomment both liens to compile arxiv version

\newif\ifanonymize
\input{preamble}

\begin{document}
\title{Misspecifications in structural equation modeling: The choice of latent variables, causal-formative constructs or composites}
\ifarxiv
\maketitle
\onehalfspacing
\abstract[Abstract]{Empirical research in many social disciplines involves constructs that are not directly observable, such as behaviors and intentions. 
To model these constructs, they must be operationalized using their relations with indicators.
Structural equation modeling (SEM) is the primary approach for this purpose.
In SEM, three types of constructs are distinguished: latent variables, causal-formative constructs, and composites.
To estimate the parameters of the different models, various estimators have been developed, such as a \gls{ml} estimator and partial least squares path modeling. 
To examine the estimation performances of different estimators for the different construct types, many Monte Carlo studies have been conducted.
One aspect evaluated in such studies is the consequences of construct misspecification, i.e.\ the true construct type differs from the modeling choice, on the parameter estimates and the model fit. 
% Shortcoming of papers that blame estimator A
For example, the parameter bias of models that misspecify latent variables as composites is frequently attributed to the chosen estimator, although the parameters of the models are based on different estimators, making it impossible to examine the factors individually.
This is an example of the entanglement of construct misspecification and parameter estimation in the literature.
%\par
% What are we doing?
This article aims to disentangle the two issues by a comprehensive Monte Carlo study of all combinations between true and assumed construct types. 
In order to focus on construct misspecification, we used the same estimator for all models, namely the \gls{ml} estimator.
To generalize our findings beyond the usage of \gls{ml}, we then replicate the simulation using estimators based on partial least squares.
Consequently, we aim to identify construct misspecification and not the choice for an estimator as the driving factor for biased path coefficients. 
In particular, we show that the misspecification of constructs leads to biased path coefficient estimates.
%Although the operationalization of constructs as causal-formative turns out asymptotically and Fisher consistent for all construct types, estimates are particularly biased for small samples.
%Assuming the construct to be a composite yields the smallest estimation variance, but a bias that increases with the sample size.
%The misspecification of a construct as latent variable produces a much larger bias when the true construct is a composite than with causal-formative constructs.
Further, we evaluate whether fit measures can tell apart models with correct construct operationalization from those with misspecified constructs.
We find that none of the criteria considered in this article is suited for this purpose.
These findings underscore the importance of thoughtful construct specification in SEM analyses and the necessity for further research on the detection of misspecification.
}

\ifarxiv
\else
{
\singlespacing
\maketitle
}

\bmsection*{Conflict of interest}
The authors declare no potential conflict of interests.
\fi 
\section{Introduction} \label{sec:intro}
% Below: Old description
%Many disciplines face research questions involving constructs that cannot be observed directly, such as behavior research \citep[e.g.,][]{Asif2023}, innovation and technology management \citep[e.g.,][]{Waqar2023} and social sciences \citep[e.g.,][]{Rasheed2024TooQuality}. 
% Why talking about constructs at all and SEM in particular?
Unobserved constructs play an important role in many disciplines such as social sciences \citep[e.g.,][]{Tarka2018AnSciences, Carvalho2014ApplicationsResearch}, economics \citep[e.g.,][]{Bollen2011StructuralBehavior}, and psychology \citep[e.g.,][]{Martens2005TheResearch}.
Such \textit{constructs} can, for example, be intentions \citep[e.g., buying intention in][]{Asif2023} or behaviors \citep[e.g., leadership behavior in][]{Rasheed2024TooQuality}.
\gls{sem} is one approach to model constructs as well as relations between them \citep{Kline2023}. 
\gls{sem} is widely popular because it extends regression analysis and factor analysis methods to incorporate different types of variables as well as errors in explanatory and explained variables \citep{Bollen1989}.
Nowadays, \gls{sem} is applicable in scenarios with multiple groups \citep[e.g.,][]{Cheah2020}, latent classes \citep[e.g.,][]{Bollen2005LatentPerspective, Tuma2013}, ordinal indicators \citep[e.g.,][]{Millsap2004AssessingMeasures}, higher-order constructs \citep[e.g.,][]{Edwards2001MultidimensionalFramework}, non-linear relationships \citep[e.g.,][]{Joreskog1996NonlinearEffects}, prior knowledge \citep[e.g.,][]{Kaplan2013BayesianMethods}, or outliers \citep[e.g.,][]{Schamberger2020}.
Moreover, powerful and user-friendly software such as the \texttt{R} packages \texttt{lavaan} \citep{Rosseel2012} and \texttt{cSEM} \citep{Rademaker2020} make \gls{sem} easily accessible by applied scientists. 
\par
% How can constructs be operationalized in SEM?
As constructs cannot be observed directly, they must be \textit{operationalized}, i.e.\ modeled using their relations with variables that are observed, so-called indicators.
There are three ways to do this in \gls{sem}.
First, a construct can be modeled as a \textit{latent variable}.
This means that it is the natural and underlying cause of its indicators \citep{Joreskog1970a}.
Since these indicators are measures of the latent variables, they are subject to measurement errors.
Second, a construct can be modeled as \textit{causal-formative} construct.
Here, the direction of causality is reversed such that the construct is caused by the indicators \citep{Diamantopoulos2008AdvancingModels}.
Such a construct represents the conceptual unity that the indicators share \citep{Bollen2017}.
All variation beyond the influence by their indicators is captured by the construct's disturbance term.
Third, a construct can be modeled as \textit{composite}.
That is, an artificial composition of the indicators, which is forged by human design \citep{Henseler2021} or a collection of heterogeneous causes \citep{Grace2008RepresentingVariables}. 
Since this type of construct is designed to serve a purpose, it assumes no causal relationship to the indicators and has thus no error term \citep{Bollen2017}.
As all three construct types are characterized by their relationships with the indicators, we henceforth refer to the operationalization of constructs as \gls{icm}.
\par
The \gls{icm} determines the meaning and interpretation of constructs \citep{Diamantopoulos2011IncorporatingModels}.
% Beispiel?
%Misspecifying, for instance a composite model as a causal-formative model can have significant implications \citep{Hardin2011FormativeTheory}.
Consequently, misspecification of one construct type by another affects the interpretations of the model and compromises the conclusions drawn from it. 
Hereafter, we will refer to this problem as \gls{icm} misspecification.
Numerous research studies have been conducted concerning the consequences of different types of \gls{icm} misspecification.
%
% Literature review:
The misspecification of latent variables as composites has been found to produce biased parameter estimates \citep[e.g.,][]{McDonald1996PathVariables, Ronkko2013AModeling, Aguirre-Urreta2014,Hwang2010AModeling, Reinartz2009AnSEM}.
% The other way around!
The same is true, but with a bias in the opposite direction, when misspecifying composites as latent variables  \citep[e.g.,][]{Sarstedt2016, Rhemtulla2020, Cho2020AnModels, Hwang2021AnAnalysis,Cho2022ARepresentations}.
Furthermore, the misspecification of causal-formative constructs as latent variables can also yield biased path coefficients \citep[e.g.,][]{Jarvis2003AResearch, Petter2007SpecifyingResearch, MacKenzie2005TheSolutions, Aguirre-Urreta2012RevisitingForm, Diamantopoulos2008AdvancingModels, Aguirre-Urreta2024}.
% Goodness of fit:
The apparent parameter bias has encouraged research into the detectability of such \gls{icm} misspecification.
Fit measures could neither detect the misspecification of causal-formative constructs as latent variables \citep[e.g.,][]{MacKenzie2005TheSolutions, Diamantopoulos2008AdvancingModels}, nor the misspecification of composites as latent variables \citep[e.g.,][]{Rhemtulla2020} accurately.
A more recent study, on the other side, found that misspecified models often fail model quality checks \citep{Aguirre-Urreta2024}.
\par 
Monte Carlo studies are frequently used to evaluate statistical models, %a statistician's Petri dish experiment
since they allow to isolate and then examine a particular issue under controlled conditions.
The aforementioned studies, however, almost exclusively considered multiple issues simultaneously.
If, for instance, different estimators are compared, often each choice is consistent on exactly one \gls{icm}, but inconsistent for the others \citep[see, e.g.,][]{Cho2022ARepresentations}.
Hence, the extent to which such findings are due to a particular misspecification, specific to the applied estimator or another confounding factor remains unknown.
Furthermore, many studies were interested in the suitability \textit{of} some estimator on particular \glspl{icm}, but at the same time misspecified some \glspl{icm} \textit{by} the choice for an estimator.
In general, the two issues of parameter estimation and \gls{icm} misspecification have been strongly entangled in the \gls{sem} literature. 
\par
% What do we bring to the table?
This article considers \gls{icm} (mis-)specifications independently of the choice for an estimator. 
In particular, we investigate the impact of \gls{icm} misspecification on the estimation performance, i.e.\ bias and variance of path coefficients, and various fit measures through a comprehensive Monte Carlo study. 
We investigate nine combinations of \gls{icm} variants for the \gls{dgp} and for the assumed model. 
All models are estimated once using \glsfirst{ml} \citep[see, e.g.,][]{Joreskog1969AAnalysis} and a second time using \gls{pls}\footnote{In the literature, different names are used for particular estimators. E.g., our notion of \glsfirst{plspm} is sometimes called PLS-SEM or simply PLS. In this article, we denote as \gls{pls} the whole of \gls{sem} estimators that are based on the PLS algorithm.} techniques \citep[see, e.g.,][]{Wold1975, Dijkstra2015ConsistentModeling}.
This allows us to disentangle the issues of parameter estimation and \gls{icm} misspecification and examine the extent to which previous findings also apply to other \glspl{dgp} and \gls{icm} variants.
\par
% How is this work structured?
The remainder of this article is structured as follows: 
All three \glspl{icm} are introduced in Section \ref{sec:ICMs} including examples and parameter estimation. 
Subsequently, we present the design of our Monte Carlo simulation in %Section \ref{sec:design}. 
Section \ref{sec:sim}, followed by an illustration of the results of our study with respect to Fisher consistency, admissibility of estimates, estimation performance and detection of misspecification.
Lastly, Section \ref{sec:discuss} discusses our findings and limitations of our study. 

\section{Indicator-construct models in SEM} \label{sec:ICMs}
The \gls{icm} describes hypotheses about the nature of each construct by its relations to indicators.
This section introduces the three types of \glspl{icm} which we denote as \textit{latent variable model}, \textit{causal-formative model} and \textit{composite model}. 
We present examples, and discuss estimation techniques for every model type. 
If multiple construct types are involved in a model, our nomenclature refers to the construct in question.
In addition, we assume hereafter that each variable has an expectation of zero.
\subsection{Latent variable models} \label{sec:LVModel}
Latent variable models are based on the principle of common cause \citep[see, e.g.,][]{Reichenbach1956TheTime.}.
Thus, the variance-covariance structure of $K$ indicators $x_{1}, \dots, x_{K}$ is explained by a latent variable $\eta$ \citep[see, e.g.,][]{Joreskog1970a, Bollen1989, Hitchcock2021ReichenbachsPrinciple}.
Figure~\ref{fig:refl_ex} illustrates a path diagram of a latent variable model.
\begin{figure}[ht]
\centering        \includegraphics[width=0.4\textwidth]{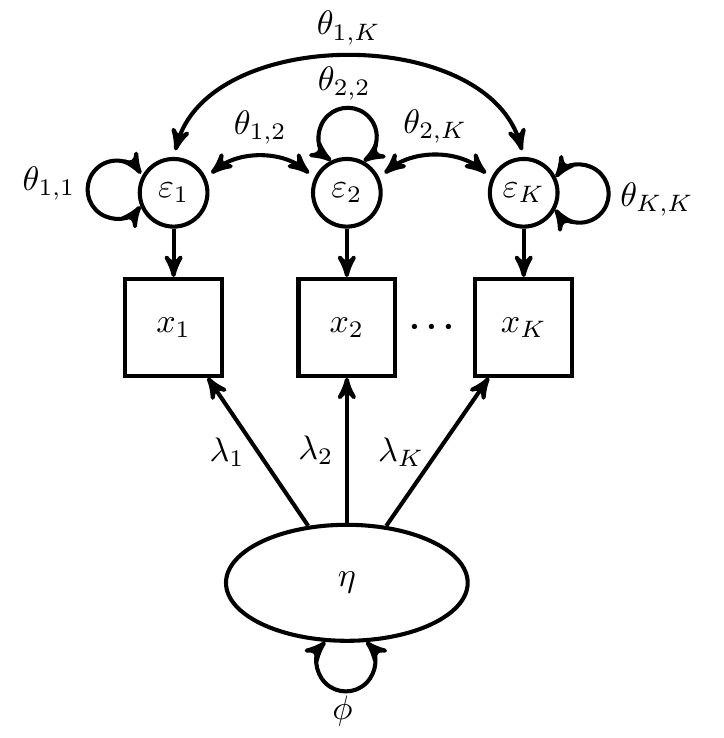}
%\begin{minipage}{0.4\textwidth}%\input{graphics/example_reflK.tikz}
%\end{minipage}
    \caption{Illustration of a latent variable~$\eta$ with variance $\phi$ which causes indicators~$x_j, j=1,\dots,K$. Factor loadings are denoted by $\lambda_j, j=1,\dots, K$ and measurement errors by $\varepsilon_j,j,= 1\dots, K$. These measurement errors have \newline(co-)variances $\theta_{jk}, j,k=1,\dots,K$.}
    \label{fig:refl_ex}
\end{figure}
Latent variables are visualized by ovals and indicators by rectangles.
Given that all indicator-construct relations are linear, the \gls{icm} of a latent variable can be written as follows:
\begin{equation}
\underbrace{\begin{pmatrix}
    x_{1} \\ \vdots \\ x_{K} 
\end{pmatrix}}_{=\bm{x} }=\underbrace{ \begin{pmatrix}
    \lambda_{1} \\ \vdots \\ \lambda_{K} 
\end{pmatrix}}_{=\bm{\lambda}} \cdot \eta + \underbrace{ \begin{pmatrix}
    \varepsilon_{1} \\ \vdots \\ \varepsilon_{K} 
\end{pmatrix}}_{=\bm{\varepsilon}}. \label{eq:ref}
\end{equation}
The elements of the vector $\bm{\lambda}$ are called \textit{factor loadings} and quantify, on average, the effect that a one-unit change in the latent variable has, ceteris paribus, on the associated indicators \citep{Bollen1989}.
In addition, each indicator $x_i$ is also contaminated by a measurement error $\varepsilon_{i}$ which captures the part of the indicator that is not explained by the latent variable.
The covariances between indicators are completely conveyed by these measurement errors and the common construct $\eta$ \citep{Joreskog1970a}.
In addition, all measurement errors are assumed to be uncorrelated with the latent variable%, i.e. $\text{cov}(\eta, \bm\varepsilon)=\bm 0$
.
%The variance-covariance matrix of the measurement errors, $\bm \Theta$, is often and hereafter assumed to be a diagonal matrix, i.e. all measurement errors are assumed to be uncorrelated. 
\par
Instead of starting from the statistical model and describing its assumptions, it is also common to derive latent variable models by defining the latent variable first and then constructing the model with sets of assumptions.
One way of doing so is the latent state-trait theory, which is revised in \citet{Steyer2015ARevised}.
\par
%
% Identification 
%
To ensure identification of the latent variable model, further restrictions need to be imposed.
A necessary condition for the identification of latent variable models is that the latent variables need to be scaled \citep{Bollen2009CausalTesting}. 
This can be achieved, for example, by fixing either one loading, e.g., $\lambda_1=1$, or the variance of the latent variable~$\eta$, e.g., $\phi=1$. 
However, to ensure identification further restrictions need to be imposed.
For more details on identification of latent variable models, we refer to \citet{Kline2023}.
\par
%
% ML Estimation
%
Several techniques have been developed to estimate the parameters of latent variable models, which can be divided into covariance-based estimators and variance-based estimators.
Covariance-based estimators such as \gls{ml} estimate the model parameters by minimizing the discrepancy between the indicators' empirical and model-implied variance-covariance matrix \citep{Joreskog1969AAnalysis}.
%Due to the structure of latent variable models, this optimization can be solved numerically \citep[see, e.g.,][]{Joreskog1970a}.
%As all model parameters are optimized simultaneously \gls{ml} is also called a full-information method \citep{Kline2023}.
\gls{sem} with \gls{ml} yields consistent as well as asymptotically efficient and normally distributed parameter estimates under the assumptions of multivariate normality of the indicators and a correctly specified model \citep[see, e.g.,][]{Bollen1989, Kline2023}.
Many software applications for \gls{ml} in \gls{sem} are available, e.g., AMOS \citep{Arbuckle2011IBMGuide} in SPSS, LISREL \citep{Joreskog2018LISRELWindows} and \texttt{lavaan} in R \citep{RCoreTeam2024}.
% PLSc estimation
An alternative estimator for latent variable models is \gls{plsc} \citep{Dijkstra2015ConsistentModeling}.
\gls{pls}-based estimators first build proxies for the theoretical constructs and then estimate the model parameters based on these proxies in a second step \citep[see e.g.,][]{Joreskog1975EstimationVariable, Ronkko2013AModeling, Kroonenberg1990LatentSquares}.
\gls{plsc} is an adaption of \gls{plspm} \citep[see, e.g.,][]{Wold1982, Wold1975, Tenenhaus2005PLSModeling, Chin1998TheModeling} and has been developed especially for latent variables because \gls{plspm} overestimates their factor loadings and underestimates correlations between them due to an attenuation bias \citep{Dijkstra2014ConsistentModels, Rigdon2016}.
Hence, \gls{plsc} applies a correction for disattenuation to the parameter estimates obtained by \gls{plspm} to ensure consistent estimates for latent variable models.
%This additional step of PLSc ``exploits the lack of correlation between some of the measurement errors within blocks" \citep{Dijkstra2017}.
\gls{plsc} requires less information from the data than \gls{ml}.
For this reason, it is advocated as distribution-free,  more robust against model misspecifications and does not rely on asymptotics \citep[see, e.g.,][]{Vilares2010}.
\par
%
%
%Example
%
An example for a latent variable is the 'attitude' of potential customers.
Consider, for instance, the question why consumers do or do not buy a specific product.
Behavioral research identified a person's 'attitude' towards the product as a fundamental, but latent determinant for this kind of decisions \citep{Sutton1998}.
Finding measurable indicators that reflect such 'attitudes' would thus be invaluable for companies.
However, it is likely that these indicators are imperfectly explained by 'attitude', but are affected by measurement errors \citep{Bentler1981AttitudesAnalysis.}.
Among many others, \gls{sem} with latent variables is frequently used in the research areas of consumer behavior \citep{Bitrian2024}, corporate social responsibility \citep{Stojanovic2020CorporatePerformance}, criminology \citep[see, e.g.,][]{Kleck2006WhatJustice}, ecology \citep{Abate2020ValuationValuation}, entrepreneurship \citep{Bachmann2024WhatIntention}, epidemiology \citep{Barillari2021AStudy}, innovation and technology \citep{ Gao2023TheTransformation}, psychology \citep{Arslan2020StudentAnalysis}, tourism research \citep{Nunkoo2013UseFuture} and the social sciences \citep[see, e.g.,][]{Bollen2002LatentSciences}.

\subsection{Causal-formative models} \label{sec:CFModel}
A causal-formative construct~$\eta$ represents the one conceptual dimension which is spanned jointly by a set of $K$ associated indicators $x_{1},\dots, x_{K}$ \citep{Bollen2017}.
Figure~\hyperref[fig:caus_ex1]{2 (a)} depicts an example of such a causal-formative model.
In strict contrast to the latent variable model, the cause-effect relationship is the other way around, i.e.\ causal-formative indicators cause the construct.
Hence, the directions of the arrows connecting the indicators and the construct in Figure~\hyperref[fig:caus_ex1]{2 (a)} have been reversed as well and now point to the construct~$\eta$.
\begin{figure}[ht]
    \centering
    \begin{subfigure}[t]{0.48\textwidth}
        \includegraphics[width=2\textwidth]{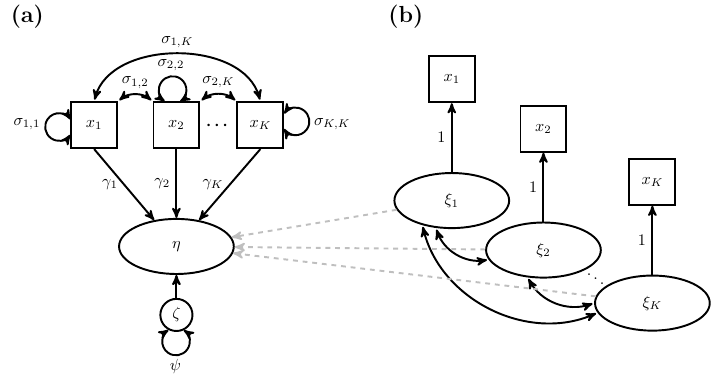}
        \caption{Illustration of a causal-formative construct~$\eta$ which is caused by indicators $x_j$,~$j=1,\dots,K$ and afflicted with a disturbance term~$\zeta$. 
   Weights are denoted by  $\gamma_1,\dots, \gamma_K$ and the elements of the variance-covariance matrix of the indicators by $\sigma_{1,1},\sigma_{1,2},\dots, \sigma_{K,K}$.}
   \label{fig:caus_ex1}
    \end{subfigure}
    \hfill
    \begin{subfigure}[t]{0.48\textwidth}
        \caption{Representation of the causal-formative construct $\eta$ by perfectly measured latent variables~$\xi_1,\dots,\xi_K$.}
        \label{fig:caus_ex2}
    \end{subfigure}
    \hspace*{0.05\linewidth}

\end{figure}
The strength of the effect that the $i$th indicator has on $\eta$ is represented by the corresponding \textit{weight}~$\gamma_{i}$ of the following multivariate regression:
\begin{equation}
\eta = \gamma_{1} x_{1} +\dots+ \gamma_{K} x_{K} + \zeta = \bm\gamma^T \bm x + \zeta.  \label{eq:caus}    
\end{equation}
The indicators~$x_1,\ldots,x_K$ are allowed to (co-)vary freely, which is captured by parameters for their variances~$\sigma_{1,1}, \dots, \sigma_{K,K}$ and covariances~$\sigma_{1,2}, \dots, \sigma_{K-1,K}$.
The consequences of omitting an indicator from Equation \ref{eq:caus} can be severe and depend on its unique contribution to $\eta$.
If the information provided by that indicator is redundant, the remaining indicators can compensate for its omission.
However, if the indicator represents a distinct facet of $\eta$, omitting it also removes that part of the construct \citep{Bollen1991ConventionalPerspective}.
The parts of $\eta$ that cannot be explained by its indicators are summarized in a disturbance term $\zeta$ with a variance parameter denoted by $\phi$ \citep{Bollen2011}.
This disturbance is assumed to be independent of all indicators, i.e.\ $\text{Cov}(\zeta, x_{i})=0$ for all $i$. 
\par
%
% Identification
%
The identification issue of causal-formative models can be more complex than that of latent variable models \citep{Bollen2017}.
However, one can represent each indicator~$x_j$ of a causal-formative construct by a perfectly measured and correlated single-indicator latent variable~$\xi_j$, and then verify model identification exactly as for latent variable models \citep[see, e.g.,][]{Bollen2009CausalTesting}. 
Figure~\hyperref[fig:caus_ex2]{2 (b)} depicts this procedure for the causal-formative construct from \hyperref[fig:caus_ex1]{2 (a)}.
The initial construct $\eta$ can then be modeled using these augmented latent variables.
Another necessary but not sufficient identification condition for causal-formative constructs is the '2+ Emitted Paths rule' \citep[see, e.g.,][]{Bollen2009CausalTesting}.
It states that each construct with an unrestricted (disturbance) variance must appear as explanatory variable in at least two other equations. 
Graphically, this means that two directed paths have to point from this construct to two other variables with unrestricted error variances \citep{Bollen2009CausalTesting}.
Thus, the causal-formative model from Figure~\hyperref[fig:caus_ex1]{2 (a)} would not be identified.
\par
%
% Estimation with ML
%
Parameter estimation for identified causal-formative models can be carried out with \gls{ml} similarly to latent variable models.
The identifiability constraints, however, can even contradict theoretical considerations \citep{Hair2012AnResearch}.
%
% PLS-PM
%
\gls{plspm} is an alternative to \gls{ml} and particularly useful when model prediction is the goal, or when \gls{ml} assumptions are likely to be violated \citep{Kline2023}.
Since the proxies that \gls{plspm} builds have no disturbance term, it tends to inflate causal-formative weights \citep{Cenfetelli2009InterpretationResearch}.% and lacks an assessment of how well the indicators explain the construct \citep{Diamantopoulos2011IncorporatingModels}.
\par
%
% Example:
% perceived company size.
The socioeconomic status is an example for a causal-formative construct.
As this construct is unobservable, it must be approximated by observable determinants like education and income \citep[e.g.,][]{Bollen2000AIndicators.}. 
Even though a person's education and income are probably correlated, it seems unlikely that a change in the socioeconomic status is responsible for changes in these variables, but the other way round.
Moreover, the socioeconomic status of two individuals with the same indicator values can still be different. 
This stochasticity is contained in its disturbance term.
% 'Perceived company size' serves as an example for a causal-formative construct.
% Consumers often lack direct knowledge about the true company size and thus approximate this construct by observable determinants like the number of employees, revenue and market share \citep{Flaswinkel2022ConsumersMouth}. 
% Even though revenue and market share are probably correlated, one would not expect that a change in the 'perceived company size' is responsible for changes in these variables, but the other way round.
% Moreover, the three indicators are unlikely to explain the 'perceived company size' perfectly.
% The influence of other determinants on the 'perceived company size' are contained in its disturbance term.
Applications with causal-formative constructs can be found in the research areas of 'consumer–brand relationship' \citep{Rahman2021BrandModel}, 
'digital transformation' \citep{Battistoni2023AdoptionSME}, 
'employee well-being' \citep{Khatri2019DevelopmentModel}, 
'quality of life' \citep{Testa2021ResponseIndicators}, 'psychology' \citep{Willoughby2016MeasuringMeasurement.}, 'supply chain management' \citep{Chand2022DirectTheory}, 
'sustainable employability' \citep{Fleuren2018HandlingEmployability}, and 'tourism' \citep[see, e.g.,][]{Berbekova2025UnderstandingMisspecification}.
\subsection{Composite models} \label{sec:CoModel}
Composites are proxies that are formed by a set of indicators.
Instead of representing a common cause or conceptual unity, composites acquire meaning by purposefulness \citep{Henseler2021}. % Source for no conceptual unity was Bollen 2017
Figure~\hyperref[fig:comp_ex]{3 (a)} gives an example of a composite $\eta$ that is forged from of a block of indicators $x_{1},\dots, x_{K}$.
Analogue to causal-formative models, the relations between indicators and the composite~$\eta$ are represented by weights~$w_{1}, \dots, w_{L}$.
In addition, the indicators that form composites are also allowed to freely vary and covary, which is expressed by the parameters $\sigma_{1,1}, \dots, \sigma_{K,K}$.
%The elements of the variance-covariance matrix~$\bm\Sigma$ of the indicators must also be estimated (i.e.\ $\sigma_{ij}$ in Figure~\ref{fig:comp_ex}).
Composites are usually visualized as hexagons to distinguish them from the other variants in a path diagram.

\begin{figure*}[ht]
    \centering
    \begin{subfigure}[t]{0.48\textwidth}
        \centering
        \includegraphics[width=1\textwidth]{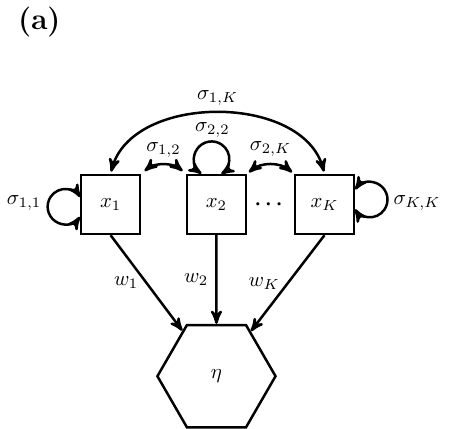}
        \caption{Illustration of a composite~$\eta$ which is forged out of indicators $x_j, j=1,\dots,K$.
    Weights are denoted by $w_1,\dots, w_K$ and the elements of the variance-covariance matrix of the indicators by $\sigma_{1,1},\sigma_{1,2},\dots, \sigma_{K,K}$.}
    \label{fig:comp_ex}
    \end{subfigure}%
    \hfill
    \begin{subfigure}[t]{0.48\textwidth}
        \centering
        \includegraphics[width=0.92\textwidth]{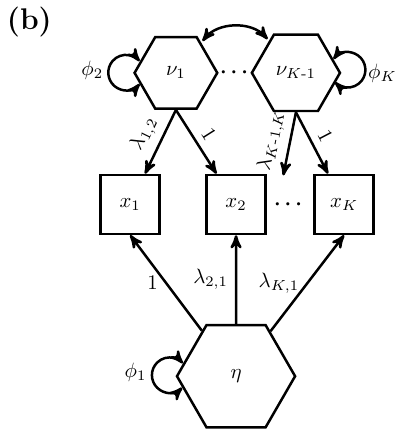}
        \caption{Representation of $\eta$ by the refined \gls{hospec} \citep{Yu2023} using excrescent variables~$\nu_1,\dots, \nu_{K-1}$ and composite loadings~$\lambda_{1,1}, \dots, \lambda_{K,K}$.}
    \label{fig:compHO}
    \end{subfigure}
\end{figure*}
In contrast to causal-formative constructs, composites are assumed to be fully determined by their related indicators and thus are not affected by a disturbance term \citep{Bollen2017}.
Instead, composites are the following one-dimensional projection from the space spanned by the indicators:
\begin{equation}
\eta = w_1x_1+ \dots + w_Kx_K= \bm{w}^T \bm{x}.   \label{eq:comp} 
\end{equation}
%However, the relations of indicators with other constructs could also be modeled directly \citep{Bollen2011}.
%The goal of summarizing the indicators by a composite is to make interpretations more convenient. 
%
\par
%
% Identification
%
The composite weights $\bm w = (w_{1}, \dots, w_{K})$ must either be pre-specificied or estimated empirically \citep{Bollen2017}.
Pre-specified weights are used, for instance, to build sum scores \citep[see, e.g.,][]{Rose2019Model-BasedModels, Schuberth2025TheModeling}. Analogue to latent variable models, one can define composites with pre-specified weights in a first step and then derive the model by sets of additional assumptions \citep[see, e.g.,][]{Mayer2012AComponents}.
Consequently, composite models with pre-specified weights are easily identified since the variances and covariances of a set of indicators can be estimated by their empirical counterparts. 

Even with empirically estimated weights, composite models have less restrictive identifiability constraints than causal-formative models.
Nevertheless, composites must also be assigned a scale, e.g., by fixing their variance to be one.
Moreover, composites must have at least one path connecting it to another variable beyond its indicators (unlike the isolated composite in Figure~\hyperref[fig:comp_ex]{3 (a)}) to be identified because the model-implied variance-covariance matrix of the indicators is unconstrained \citep{Henseler2021}.
In the following, we assume that the composite weights are estimated.
\par
% Estimation with ML / H--O
In principle, model parameters of composite models can also be estimated with \gls{ml}.
However, the iterative computations of the likelihood and its derivatives become mathematically and computationally more challenging.
For this reason, the \textit{\gls{hospec}} \citep{Henseler2021, Schuberth2021a, Yu2023} has been developed.
The \gls{hospec} uses augmentation and reparameterization of composite models to enable seamless integration into standard \gls{ml} software tools \citep{Yu2023}.
This is achieved by augmenting the model by  \textit{excrescent variables}~$\nu_1,\dots, \nu_{K-1}$ so that it contains as many constructs as the initial composite $\eta$ has indicators.
The projection from the augmented space of the constructs $(\eta, \bm \nu)$ to that of the indicators is then given by the following invertible rotation:
\begin{align}
    \binom{\eta}{\bm{\nu}} = \bm{W}' \bm{x} \iff \bm{\Lambda} \binom{\eta}{\bm{\nu}} = \bm{x}
\end{align}
Thus, the relations between composites and their indicators can be expressed with a matrix~$\bm{\Lambda}$ of composite loadings, instead of weights \citep{Schuberth2021a}.
% The \gls{hospec} expresses relations between composites and their indicators with so-called composite loadings and \textit{excrescent variables}~$\nu_1,\dots, \nu_{K-1}$, instead of weights \citep{Schuberth2021a}.
% The excrescent variables extend the composite model, such that it has as many constructs as the initial composite $\eta$ had indicators.
% Doing so, the \gls{hospec} enables seamless integration of composite models into standard \gls{ml} software tools \citep{Yu2023}.
% The set of excrescent variables~$\bm{\nu}$ ensures that the transformation 
% \begin{align}
%     \binom{\eta}{\bm{\nu}} = \bm{W}' \bm{x} \iff \bm{\Lambda} \binom{\eta}{\bm{\nu}} = \bm{x}
% \end{align}
% is an invertible rotation, from which a composite loading matrix $\bm{\Lambda}$ can be obtained.
%Second, \gls{hospec}s are not a cure for unidentifiability, but only have a unique \gls{ml} solution if the original composite model has been identified.
Both, the weight matrix~$\bm W$ and the factor loading matrix~$\bm\Lambda$ are $K$-dimensional square matrices.
$\bm W$ projects from the space spanned by the indicators to the augmented space of the composites~$(\eta,\nu_1,\dots,\nu_{K-1})$, whereas $\bm \Lambda$ describes the inverse projection.
The $i$th%\textsuperscript{th}
row of $\bm W$ contains the weights to form the $i$th composite.
Figure~\hyperref[fig:comp_ex]{3 (b)} shows the corresponding \gls{hospec} for the composite on the left.
The entries of $\bm \Lambda$ are visualized along the paths.
The $K-1$ excrescent variables are are allowed to covary with each other, but not with $\eta$.
More details about the \gls{hospec}, such as the necessary parameter restrictions to ensure identification of the model, can be found in \citet{Yu2023}.
% PLS-PM
Besides the \gls{ml} estimator, \gls{plspm} can also be used to derive consistent parameter estimates of composite models \citep[see, e.g.,][]{Becker2013PredictiveRelevance, Sarstedt2016,Hair2017}.
This is due to the fact that \gls{plspm} builds proxies in each iteration of the alternating algorithm, which essentially are composites.
\par
%
% Example:
'Desire' is an example for a composite that consists of a person's attitudes, norms and emotions \citep{Nascimento2024}.
Analyzing, for instance, the impact of the 'desire' on a consumer's 'buying intention' can be more convenient than interpreting their attitude, norms and emotions in a fixed proportion.
The challenge is to develop meaningful composites and evaluate them \citep{Henseler2021}. 
Just as, ceteris paribus, changing the attitude towards environmental problems does not necessarily increase the desire for a product, the meaning and usefulness of a composite depends entirely on the weights.
Composites should preserve all necessary information from their indicators, so that every correlation between indicators from different composites is conveyed solely through those composites \citep{Dijkstra2017}. 
%
% Applications
%
Although composite models are the younger and less widespread approach, they gradually catch up with respect to the number of published applications \citep{Hair2017}.
For instance, composites have been used in research areas such as 'innovation and technology management' \citep{Yoshikuni2024StrategicPerformance}, 'consumer behavior' \citep{Schleiden2020DoesBehaviour}, 'sociology' \citep{Mustillo2021EvaluatingApproach}, 'tourism' \citep{Liu2022ModelingAnalysis}, and 'well-being' \citep{Tian2025AImplementation}. 

All three introduced indicator-construct types differ conceptually and parametrically.
Consequently, aspects of the estimation procedure such as the likelihood function or construct proxies differ as well.
For these reasons, the specification of an \gls{icm} affects the choice of the estimator and is crucial for meaningful statistical inference \citep{Kline2023}.
The next section demonstrates the consequences of \gls{icm} misspecification on the estimation performance and the model fit by means of a Monte Carlo simulation.

\section{Simulation Study} \label{sec:sim}
The previous section described the three possible operationalizations of constructs in the \gls{icm}. 
In empirical research, one needs to choose (i) one of these three options for each construct and (ii) an appropriate method to estimate the model parameters.
Both choices will have implications on the results. 
In the following, we investigate the implications of the choice of an \gls{icm} on the estimation results by means of a simulation study.
For this purpose, we consider a scenario in which the \gls{dgp}, i.e.\ the true model, corresponds to one of the three \glspl{icm}.
When estimating the parameters, the assumed \gls{icm} is either specified correctly, or it is misspecified. 
To ensure that our findings can be generalized beyond the use of a specific estimator, we investigate the consequences of misspecification both for \gls{ml} and using \gls{pls}-based estimation. 
These are measured with respect to (i) estimation performance and (ii) the model fit using common fit measures.
Figure~\ref{fig:sim_flow} provides an overview of the structure of the simulation study and the combinations of (mis-)specifications and estimation methods that will be examined.
\begin{figure}[ht]
    \centering
    \includegraphics[width=0.75\textwidth]{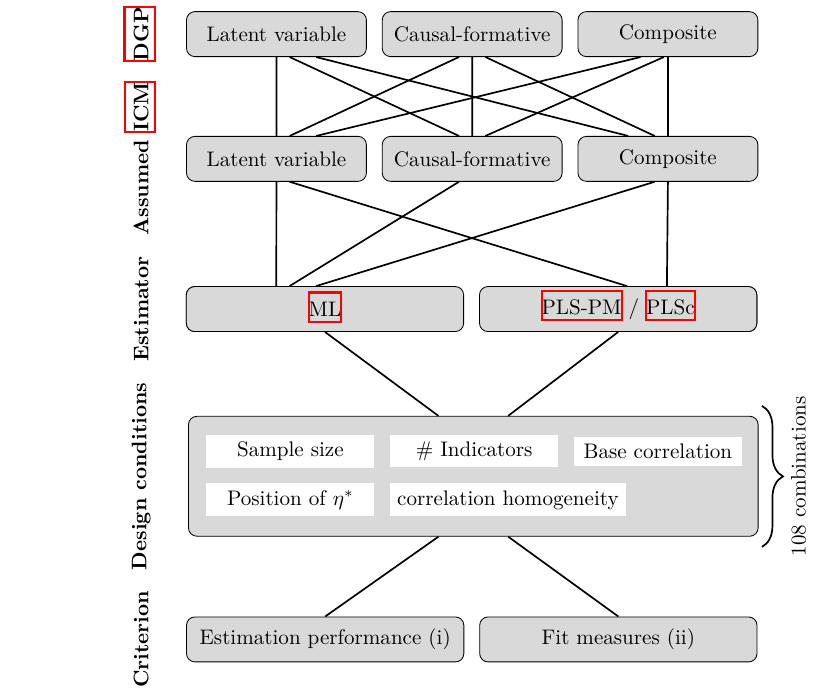}
    % \begin{minipage}{0.75\linewidth}
    % \input{graphics/sim_flow.tikz}
    % \end{minipage}
    \caption{Flowchart of the different combinations in our Monte Carlo study regarding true and assumed construct type, estimator, additional design conditions and evaluation criteria. From the $1944=3 \times 3 \times 2 \times 108$ possible combinations, some are excluded from our simulation study. In particular, causal-formative models are generally not estimated with \gls{pls}-based estimators and with \gls{ml} only if $\eta^*$ is exogenous.
    We evaluate the samples from all possible design conditions using two types of criteria.}
\label{fig:sim_flow}
\end{figure}
%Figure~\ref{fig:sim_flow} depicts the scope of \gls{icm} (mis-)specifications in our study.
%Assuming a composite \gls{icm}, for instance, reflects the composite \gls{dgp} correctly, but would be a misspecification of the other two \gls{dgp} variants.
The following subsection introduces the study design in more details.
Subsequently, the simulation results will be presented.

\subsection{Study design} \label{sec:design}
%
% General description of DGPs
%
In our simulation study, we use a structural equation model similar to the one used in \citet{Aguirre-Urreta2024}.
Figure~\ref{fig:sim_Aguirre_ex} illustrates this model.
It contains four constructs, three of which are specified as latent variables ($\eta_1, \eta_2$ and~$\eta_3$).
Each latent variable is measured by four indicators with factor loadings of $0.8$.
All measurement errors~$\varepsilon_{i,j}$ are uncorrelated and have a variance of~$0.36$.
Consequently, all indicators have a variance of one.
Choosing these values for variances and factor loadings within and across constructs simplifies the comparisons of the estimation results.
\par
\begin{figure}[ht]
    \centering
    \includegraphics[width=0.95\textwidth]{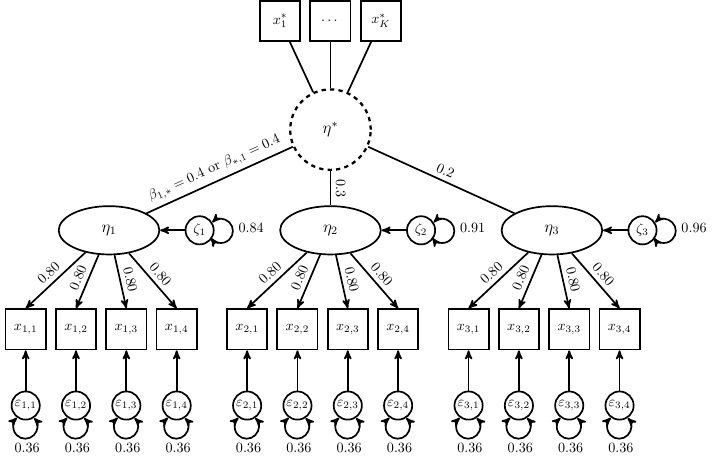}
    %\begin{minipage}{0.95\linewidth}
    %    \input{graphics/sim_sem_horizontal.tikz}
    %\end{minipage}
    \caption{Visualization of the \glspl{dgp} in the simulation study. $\eta^*$ is associated with varying position (e.g., exogenous), \gls{icm} type (e.g., composite), number of indicators and indicator correlations. All standardized structural coefficients and the \gls{icm} of every other latent variable are fixed to avoid confounding factors.}
\label{fig:sim_Aguirre_ex}
\end{figure}
%
% design conditions
%
To answer our research question, we specifically utilize the fourth construct $\eta^*$ and vary its  \textit{type} and \textit{position}.
The construct type is either 'latent variable', 'causal-formative construct' or 'composite'.
The position of the construct $\eta^*$ can be either 'exogenous' (in the following referred to as the exogenous case) or 'endogenous' (referred to as the endogenous case).
Previous research indicates that the position might affect the estimation bias \citep[see, e.g.,][]{MacKenzie2005TheSolutions, Aguirre-Urreta2024, Cadogan2013ImproperVariables}.
When $\eta^*$ is exogenous, the structural equations are given by
\begin{align}
    \eta_1 &= 0.4\cdot \eta^* + \zeta_1 \ , \ \quad
    \eta_2 = 0.3\cdot \eta^* + \zeta_2 \ , \ \quad
    \eta_3 = 0.2\cdot \eta^* + \zeta_3. \tag{exogenous case}
\end{align}
When $\eta^*$ is endogenous instead, the structural equation is as follows:
\begin{align}
    \eta^* &= 0.4\cdot \eta_1 + 0.3\cdot \eta_2 + 0.2\cdot \eta_3 +\zeta^*. \tag{endogenous case}
\end{align}
However, it should be noted that in the endogenous case our causal-formative mode would not be identified, since it does not have '2+ emitted paths'. 
For this reason, we exclude this \gls{dgp} type and the \gls{icm} assumption from further analyses.
Figure~\ref{fig:sim_Aguirre_ex} illustrates the different construct types for $\eta^*$ by its dashed circle and undirected paths connecting it to the indicators $(x^*_{1},\dots, x^*_{K})$.
The different positions are highlighted by undirected paths between $\eta^*$ and other constructs. 
For the sake of clarity, we omit all parameters in Figure~\ref{fig:sim_Aguirre_ex} that are only apparent for certain \glspl{icm} such as the disturbance term~$\zeta^*$ or the  weights~$\bm w^*$.
\par
%
% Parameters/conditions that do not change
To ensure comparability among the different combinations of construct type and position, we fix certain properties of $\eta^*$.
First, all weights associated with $\eta^*$ are homogeneous, i.e.\ $w_1^*=\dots=w_K^*$. 
Moreover, the covariances between $\eta^*$ and its indicators are the same across all construct types. 
The latter is achieved by setting composite weights exactly equal to their causal-formative counterparts (i.e.\ $\bm w^*= \bm \gamma^*$), and factor loadings to the regression coefficients from regressing the indicators on the composite.
If $\eta^*$ is a causal-formative construct, the variance of the disturbance terms is fixed at $0.25$.
Forcing the relationship between constructs and indicators to be equal leads to unequal variances of the constructs.
More specifically, the causal-formative constructs have a variance of $1.25$, whereas the composites and latent variables each have a variance of one.
To ensure that the standardized path coefficients of all \glspl{dgp} are equal, the (unstandardized) population values of the causal-formative models are specified as $0.4\cdot \sqrt{1.25} \approx0.3578$, $0.3\cdot \sqrt{1.25} \approx0.2683$ and $0.2\cdot \sqrt{1.25} \approx0.1789$. 
More details on the derivation and properties of the \glspl{dgp} can be found in the supplementary material.
\par
In addition to the position of $\eta^*$, we also analyze the effect of additional design conditions. 
We vary the \textit{sample size} from $100$, $300$, to $500$ observations.
The \textit{number of indicators $K$} associated with $\eta^*$ varies between $3$, $5$ and $7$.
The \textit{base correlations} $\sigma$ between these indicators is set to $0.1$, $0.3$ or $0.5$.
Lastly, the design condition \textit{correlation homogeneity} describes whether the correlations between indicators are equal (i.e.\ homogeneous) or not. 
While for homogeneous correlations $\text{Cor}(x^*_{i}, x^*_{j})=\sigma$ for all pairs $i,j$, $i\neq j$ of indicators, heterogeneous correlations are defined as an equidistant sequence from $\text{Cor}(x^*_{1}, x^*_{2})=\sigma-0.1$ up to $\text{Cor}(x^*_{K-1}, x^*_{K})=\sigma+0.1$.
In total, this leads to the $108$ design conditions that are indicated in Figure~\ref{fig:sim_flow}.
\par
%
% Estimation
%
For each operationalization of  $\eta^*$, we estimate the parameters of the corresponding structural equation model.
That is, we estimate the parameters of a latent variable model, a composite model, and a causal-formative model, two of which are misspecified and the third represents the true \gls{dgp} for each particular design condition.
The parameter estimates are based on $1000$ samples.
%An estimate is used only if all three models yield admissible results
The simulation study is carried out using the statistical software \texttt{R} version \texttt{4.4.1} \citep{RCoreTeam2024}.
For \gls{ml} the package \texttt{lavaan} version  \texttt{0.6-19} is used.
The Monte Carlo iterations are carried out using the package \texttt{simsem} version \texttt{0.5.16}. % from 24/09/24
The latent variable model and the causal-formative model can be estimated straightforwardly with \texttt{lavaan}. % because both models are identified. 
For the composite model, we derived the refined \gls{hospec} \citep{Yu2023} prior to the \gls{ml} optimization. 

% Estimation with PLS
To ensure that our findings are valid beyond the choice of \gls{ml} for parameter estimation, the simulation study is repeated a second time using \gls{pls}-based estimators. 
In this context, we refer to \gls{plspm} when $\eta^*$ is a composite in the assumed \gls{icm}, and of \gls{plsc} when it is a latent variable.
The estimations are again carried out in \texttt{R} with the package \texttt{cSEM} version \texttt{0.5.0.9000} \citep{Rademaker2020}
However, \gls{pls}-based estimation is inconsistent for causal-formative models.
Hence, we exclude this \gls{icm} assumption from subsequent analyses (i.e.\ no line in Figure~\ref{fig:sim_flow} between the causal-formative assumed \gls{icm} and \gls{plspm}).
Differences between the \gls{pls}-based estimates and that of \gls{ml} estimation are presented at the end of each section.
\par
% DGP specifications
% Axel identification causal-formative!
% My ideas so far (27th November 2024):
% Structurally identified according to the X rule (every structural variable has at most one connection to any other structural variable) -> in fact two paths are restricted!
% ICM identified according to the 2+ emitted path rule (at least two paths from the causal-formative construct on other variables). This is necessary, but not sufficient!
% In the "smallest" data condition, I have 15 obserations -> max. 120 parameters
% I count 36 (free) parameters in the model (that must be estimated) and lavaan states 84 degrees of freedom accordingly. 
%
% Evaluation Criteria
% 1. Bias
To assess the estimation performance, we examine the two components of the \gls{mse}, namely the variance and the squared bias of the $1000$ admissible estimates.
% 2. GoF
Further, we evaluate the model fit of each model using six criteria, which turned out influential in previous studies: (1) the $\chi^2$ test \citep[e.g.,][]{Bollen1989}, (2) the \gls{srmr} \citep[e.g.,][]{Bentler2006EQSManual}, (3) the \gls{cfi} \citep[e.g.,][]{Bentler1990ComparativeModels.}, (4) the \gls{rmsea} \citep[e.g.,][]{Steiger1990StructuralApproach}, (5) the \gls{cr}, and (6) the \gls{ave} \citep[e.g.,][]{Fornell1981EvaluatingError}. 
While (1) is a classical statistical test of the overall model fit, (2)-(4) are descriptive measures of the model-data correspondence \citep{Schermelleh-Engel2003EvaluatingMeasures}.
The criteria (5) and (6) are calculated on the level of a particular construct and only available for latent variables.
More information about fit measures, including common thresholds for an acceptable model fit, can be found in \citet{Schermelleh-Engel2003EvaluatingMeasures}. 
\subsection{Fisher consistency} \label{sec:fisher}
% Intro Fisher consistency
We start our investigation by evaluating whether the considered estimators are in principle able to reconstruct the parameters of the \gls{dgp} in the case of idealized data. 
In other words, does the estimator give the right answer for the population we asked about? 
Only then can an estimation be meaningful for real-world data. 
We carry out this investigation using Fisher consistency.

An estimator is called \textit{Fisher consistent} if it yields the exact same value for each parameter of the \gls{dgp} when calculated on the entire population \citep{Fisher1922OnStatistics, Cox1979TheoreticalStatistics}.
%Such an estimator gives exactly the right answer when the sample represents exactly the population we asked about \citep{Cox1979TheoreticalStatistics}. 
Such a population is synthetically imitated by drawing a finite sample from the (known) \gls{dgp} and  normalizing this sample so that its relevant statistics (such as empirical mean and empirical variance) equal the values of the theoretical counterparts.
We consider these populations for all three construct types in the \gls{dgp} and all design conditions of our study (except the sample size) and then estimate the latent variable model, the causal-formative model and the composite model using \gls{ml} and \gls{pls}-based estimation.
In the following, we focus on the Fisher consistency of the standardized path coefficients.
\par
% Fisher consistency ML
%
% Results of Fisher consistency ML exogenous case
As expected, the \gls{ml} estimator turns out to be Fisher consistent for all three correctly specified \glspl{icm} in the exogenous case (see Figure in supplementary material).
Moreover, the \gls{ml} estimator is Fisher consistent when the causal-formative model is assumed although the \gls{dgp} is of a different construct type.
In contrast, the \gls{ml} estimator is not Fisher consistent when composite models or latent variable models are incorrectly assumed for the exogenous construct, i.e.\ in case of a misspecification as latent variable or composite, respectively.
%
% Description Fisher consistency ML endogenous case
In the endogenous case, the \gls{ml} estimator is Fisher consistent if and only if the \glspl{icm} are correctly specified (see Figure in the supplementary material).
\par
% Description Fisher consistency PLS
When the estimation is carried out with \gls{pls}-based estimators, the assumed construct type is reflected by the choice for \gls{plspm} or \gls{plsc}, respectively.
While the \gls{plspm} estimator builds composites, \gls{plsc} corrects the procedure for attenuation such that consistent estimates can be obtained for latent variables.
Accordingly,  the \gls{plspm} estimator is Fisher consistent if the exogenous construct is a composite, whereas \gls{plsc} is Fisher consistent for exogenous latent variables.
If either of the assumed construct types is a misspecification, none of the approaches remains Fisher consistent.
Especially, both PLS variants are not Fisher consistent for exogenous causal-formative constructs.
In the endogenous case, \gls{plspm} shows exactly the same Fisher consistencies as \gls{ml}.
Visualizations of the Fisher consistencies regarding \gls{plspm} can also be found in the supplementary material.
\par
%
% 
% Interpretation Fisher consistencies
For correctly specified \glspl{icm}, both estimators are equally capable to reproduce the standardized path coefficients in all data conditions.
Once an \gls{icm} is misspecified, both estimators are no longer Fisher consistent;
the only noteworthy exceptions are the estimates of causal-formative models. 
Estimating these models with \gls{ml} yields Fisher consistent standardized path coefficients regardless of the true construct types, presumably because the larger number of parameters also increases modeling flexibility.
In particular, the \gls{ml} estimates when assuming causal-formative \glspl{icm} produce model-implied variance-covariance matrices that are exactly equal to those of the different populations with the same path coefficients as in the \glspl{dgp}.
From the \gls{pls} variants, neither \gls{plspm} nor \gls{plsc} are suited to estimate causal-formative models.
If an exogenous construct is causal-formative, both PLS variants become not Fisher consistent. This may be due to the fact that they neglect the disturbance term.
\subsection{Admissibility of results} \label{sec:admissibility}
%
% General introduction of this section:
%
In addition to the previous section, where we demonstrated the theoretical functionality of the estimation methods, we will now focus on the practical admissibility in the case of simulated but unnormalized data sets. 
As with real-world data, their statistics (such as empirical mean and variance) will differ from those of the whole population.
Based on such samples, we estimate model parameters and examine their admissibility, i.e.\ whether the solution is proper \citep[see, e.g.,][]{Chen2001ImproperStrategies}.
In an application, a model that yields inadmissible results would be discarded by the researcher.
We can thus use inadmissible results to indicate challenging design conditions for a particular model, like an \gls{icm} is misspecification \citep{vanDriel1978OnAnalysis}.

% Weiteres Vorgehen inadmissibility
Estimates from data are considered \textit{admissible} if the optimizer converges, standard errors of parameter estimates are numerically non-negative, and variance-covariance matrices of the constructs and the indicators (or measurement errors, respectively) are numerically positive definite.
This is verified in \texttt{R} using the \texttt{lavaan} function \texttt{lavTech(object, what = "post.check")} and the \texttt{cSEM} function \texttt{verify(.object)}. 
%The latter additionally checks whether (congeneric) reliabilities exceed one. % gestrichen, weil wir nur noch verify()[c(3,5)] nutzen, was das äquivalent zu lavTech machen sollte.
%
% Inadmissible results exogenous case
\begin{figure}[ht]
    \centering
    \includegraphics[width=0.85\linewidth]{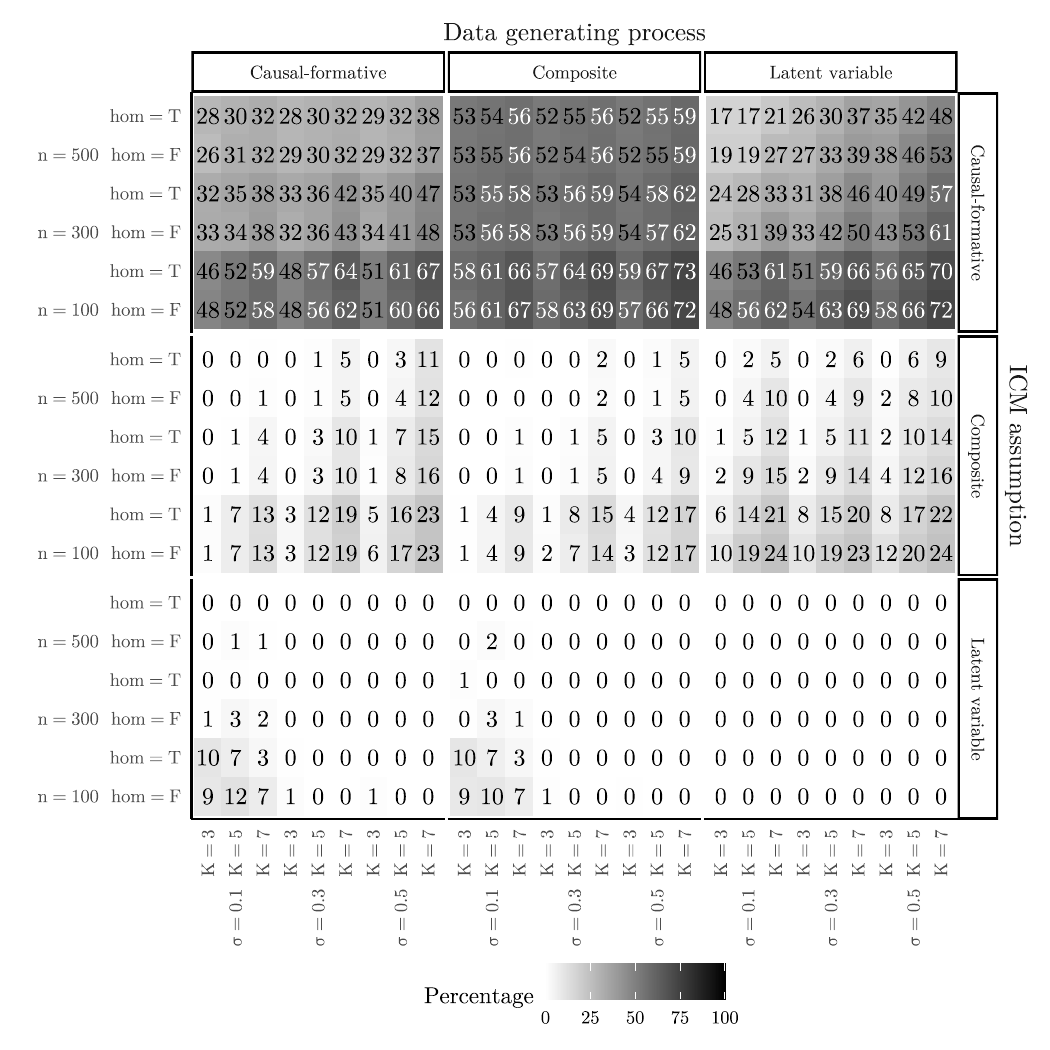}
\caption{Percentage of inadmissible estimations in the total number of estimation attempts based on \gls{ml} in the exogenous case. 
Each tile represents one design condition and is colored according to the percentage of inadmissible results.
%A value of $50\%$ thus means that half of the estimations in a scenario were inadmissible.
The outer $3 \times 3$ grid of tiles represents the combinations of \glspl{dgp} (columns) and \gls{icm} assumptions (rows).
Each cell of this outer grid then consists of an inner $6 \times 9$ grid of the remaining design conditions.
For the inner grid, columns describe combinations of the number of indicators~$K$ and their base correlations~$\sigma$, while the rows represent combinations of sample size~$n$ and correlation homogeneity~\texttt{hom}.}
\label{fig:exog-inadmissible}
\end{figure}

The percentage of inadmissible estimates of \gls{ml} when $\eta^*$ is exogenous is displayed in Figure~\ref{fig:exog-inadmissible}.
The bottom row of the graphic emphasizes that assuming a latent variable model as \gls{icm} leads to the fewest inadmissible results, regardless of the \gls{dgp}.
The only scenarios in which inadmissible estimates occur whatsoever are incorrectly assuming $\eta^*$ to be a latent variable in cases with few indicators and weak correlations between these.
The top row of the Figure displays the assumption of a causal-formative model. 
This modelling choice results in the highest proportion of inadmissible results.
Even if the model is correctly specified, the amount of inadmissible results remains similar.
However, it decreases from $50\%$ to roughly $30\%$ with increasing sample size.
Moreover, a larger number of indicators $K$ increases the percentage of inadmissible results.
The top-center panel in the figure demonstrates that misspecifying a composite as causal-formative construct sharply increases the number of inadmissible results to up to $73\%$. 
Assuming a composite model produces more inadmissible results when many and strongly correlated indicators are associated with $\eta^*$, as highlighted by the center row of Figure \ref{fig:exog-inadmissible}.
Here, the percentage of inadmissible results increases only slightly if the true construct type is causal-formative, but more sharply if it is a latent variable.
Note, however, that under this \gls{icm} assumption almost all inadmissible results are due to non-convergent optimizations.
%

% Inadmissible results endogenous case
A graphical representation of admissibility in the endogenous case is given in Figure~\ref{fig:ML:endog-inadmissible}.
The tiles belonging to design conditions with either an endogenous causal-formative construct in the \gls{dgp} (first column) or an assumed causal-formative \gls{icm} in that position (first row) are deliberately left blank.
The remaining tiles exhibit a pattern similar to the exogenous case, but with generally more inadmissible results.
\begin{figure}[ht]
    \centering
    \includegraphics[width=0.85\linewidth]{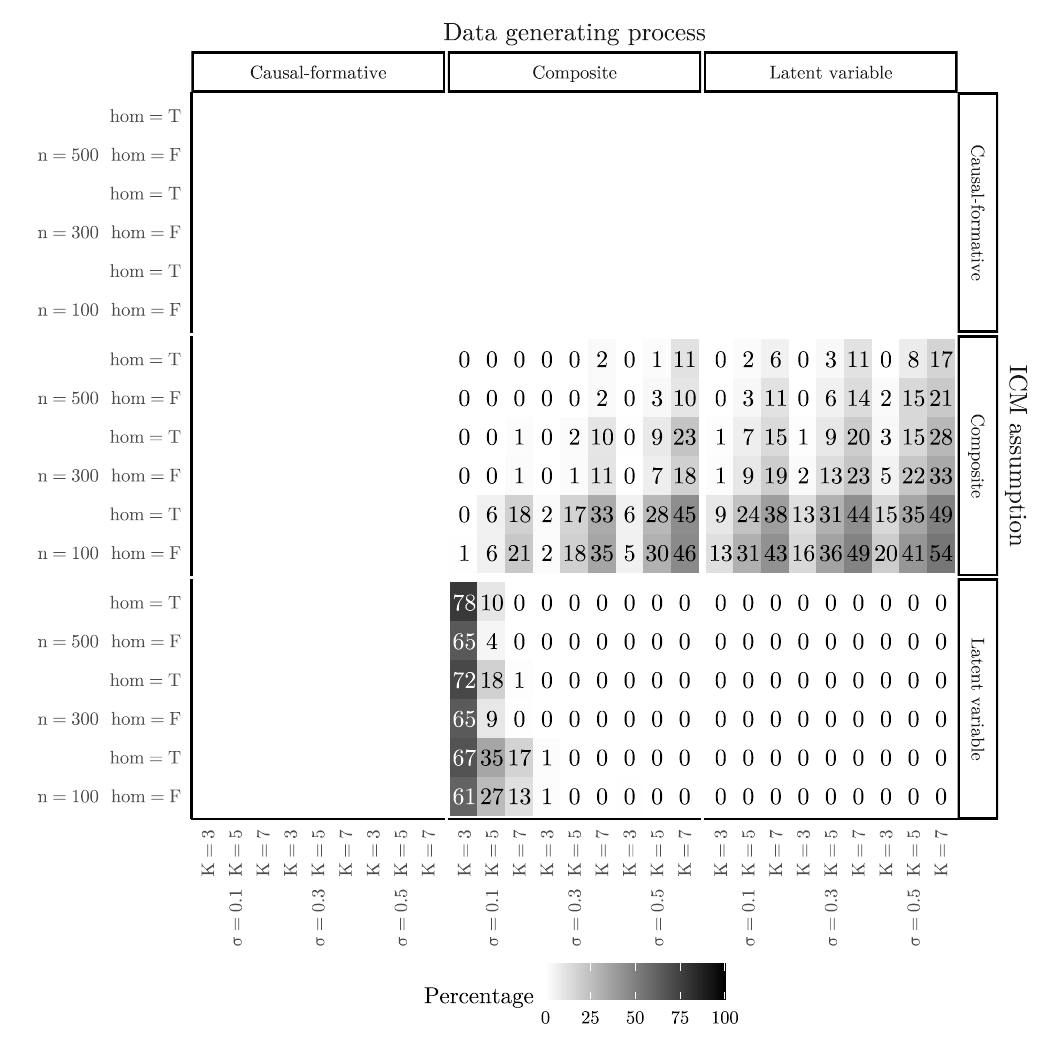}
    \caption{Percentage of inadmissible results from \gls{ml} estimations in the endogenous case. 
Analogously to Figure~\ref{fig:exog-inadmissible}, each tile represents one design condition and is colored according to the percentage of inadmissible estimations until the estimator produced $1000$ admissible results. 
The causal-formative \gls{dgp} and assumed \gls{icm} are deliberately left blank in the outer grid (see Section \ref{sec:design}).}
    \label{fig:ML:endog-inadmissible}
\end{figure}
The latent variable model results in up to $78 \%$ inadmissible results when the true construct type is a composite of few weakly correlated indicators.
In contrast to the exogenous case, a larger sample size does not reduce the amount of inadmissible results but increases it.
\par
%
% Inadmissible results with PLS
The percentage of inadmissible results in the exogenous case using \gls{pls} variants is depicted in Figure~\ref{fig:PLS:exog-inadmissibles} and stresses the similarity between\gls{plspm} (i.e.\ assuming that $\eta^*$ is a composite) and \gls{plsc} (i.e.\ assuming that $\eta^*$ is a latent variable).
\begin{figure}[ht]
    \centering
    \includegraphics[width=0.85\linewidth]{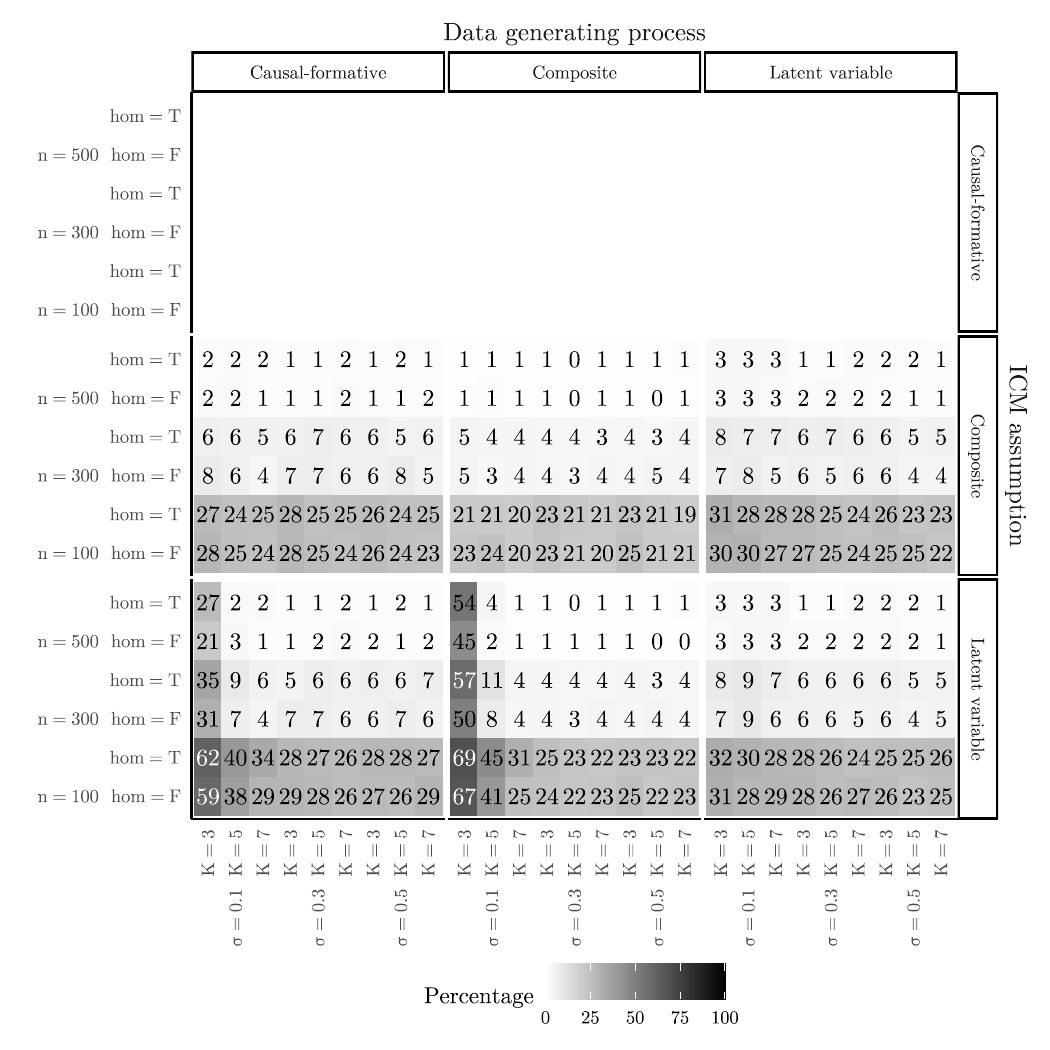}
    \caption{Percentage of inadmissible results from \gls{pls}-based estimations in the exogenous case. 
Analogously to Figure~\ref{fig:exog-inadmissible}, each tile represents one design condition and is colored according to the percentage of inadmissible estimations until the estimator produced $1000$ admissible results. 
The causal-formative model is deliberately left blank in the outer grid because both \gls{pls} variants are not Fisher consistent for this assumed \gls{icm} (see Section \ref{sec:fisher}).}
    \label{fig:PLS:exog-inadmissibles}
\end{figure}
The sample size is the most important design condition for the percentage of inadmissible results.
The smaller the sample, the greater the number of inadmissible results obtained for both \gls{pls} variants. 
As with the \gls{ml} estimator, \gls{plsc} yields many inadmissible results when a latent variable model is incorrectly assumed in a scenario with few, weakly correlated indicators.
Although this is also true for \gls{plspm}, the effect is less pronounced there. 
\par 
%
% Interpretation inadmissibles
%
Presumably, the larger number of free parameters in causal-formative models explains their Fisher consistency as well as their the high percentages of inadmissible results.
This hypothesis would be consistent with the fact that the amount of inadmissible results further increases with the number of indicators $K$.
Especially when the additional parameters are not required, such as a disturbance variance when the construct is actually a composite, the number of inadmissible results becomes higher than for any other \gls{icm} misspecification.
If latent variable models are incorrectly assumed as \gls{icm} and the true construct has small indicator correlations, then the estimated factor loadings are likely to be small.
This, in turn, means that only little information about~$\eta^*$ is reflected by the indicators, which ultimately leads to higher percentages of inadmissible results.
If the assumed \gls{icm} is a composite model, strong indicator collinearities can result in larger estimation error \citep{Hair2021EvaluationModels}.
Thus, the increased percentages of inadmissible results for many (i.e.\ large $K$) correlated (i.e.\ large $\sigma$) indicators, may also be attributed to the issue of multicollinearity.
Since most inadmissible results under this \gls{icm} assumption are from non-convergent estimations, they may also be attributed to the issue of choosing starting values for the parameters in the \gls{hospec}.
The endogenous case appears to be generally more challenging for \gls{ml} and \gls{pls}-based estimators.
Previous studies have often emphasized the superior admissibility rates of \gls{pls}-based estimation compared to \gls{ml} estimation \citep[e.g.,][]{Cho2022ARepresentations, Rigdon2017OnRecommendations}.
While this claim is supported by some of the designs in our study, it seems particularly inaccurate when dealing with small samples and few indicators.
Instead, insufficient sample sizes appear to threaten admissibility regardless of the estimation method.

\clearpage
\subsection{Estimation performance}
% Results of estimation performance 
%
In the previous sections, we have evaluated whether the various combinations of \glspl{dgp} and assumed \glspl{icm} yield Fisher consistent estimates and admissible results for the two applied estimators. 
In this section, we now take a closer at the parameter estimates from these admissible results.
Specifically, we examine the \gls{mse} of $N=1000$ admissible estimates for the standardized path coefficients, which is built upon the bias and the variance as follows:
\begin{align*}
\text{MSE}\left(\hat{\bm\theta}\right) & = \text{Bias}\left(\hat{\bm\theta}, \bm\theta\right)^2 \ \ \ \, + \text{Var}_{\bm\theta}\left(\hat{\bm\theta}\right) \\
& \approx \frac{1}{N}\sum_{t=1}^N\hat{\bm\theta}^{(t)}-\bm\theta \ + \frac{1}{N-1}\sum_{t=1}^N{\left(\hat{\bm\theta}^{(t)}-\bm\theta\right)^2}
\end{align*}
Figure~\ref{fig:exog-biases} depicts \gls{mse} of the \gls{ml} estimates of the standardized path coefficient from $\eta^*$ on $\eta_1$.
The dark gray bars correspond to the squared biases of the estimates, whereas the light gray bars represent the variances.
\begin{figure}[ht]
    \centering
    \includegraphics[width=0.7\linewidth]{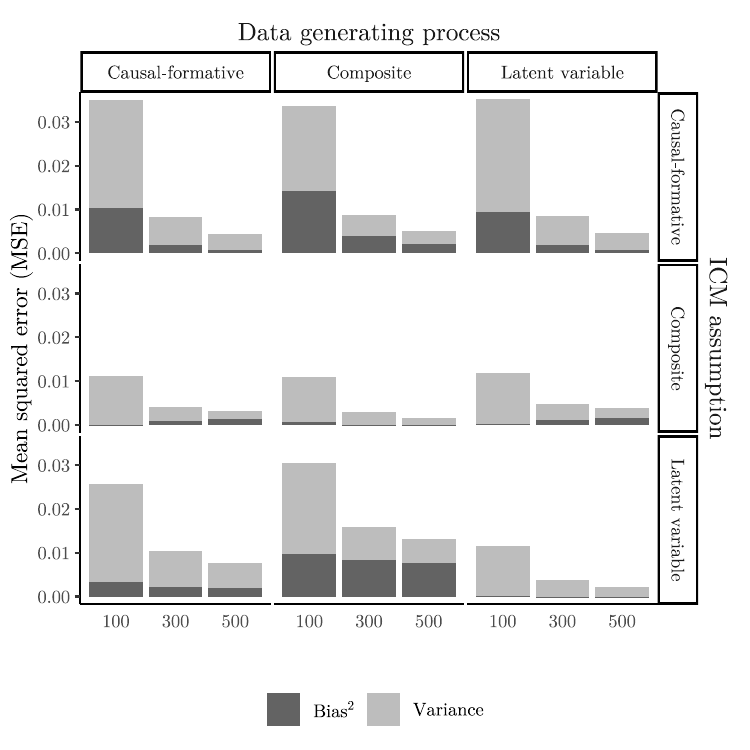}
    \caption{\gls{mse} of the admissible \gls{ml} estimates in the exogenous case of the standardized path coefficient~$\beta_{1,*}$. 
Each bar subsumes all design conditions with the sample size that is displayed on the horizontal axis. 
The outer $3 \times 3$ grid represents the combinations of \glspl{dgp} (columns) and \gls{icm} assumptions (rows).}
    \label{fig:exog-biases}
\end{figure}
The graphic visualizes different sample sizes on the horizontal axis, as well as all nine combinations of \glspl{dgp} and \gls{icm} assumptions in an outer $3 \times 3$ grid.
Consequently, the three graphics on the diagonal from the top left to the bottom right show the scenarios in which the models are correctly specified. 
As expected, the bias of each assumed \gls{icm} (i.e.\ row in Figure~\ref{fig:exog-biases}) is generally smallest if this assumption is correct, i.e.\ the \gls{icm} represents the \gls{dgp} correctly.
The estimates for correctly specified composite models and latent variable models show smaller bias and variance than those of correctly specified causal-formative models.
In any case, both parts of the \gls{mse} diminish with increasing sample sizes.

% causal-formative misspecified
If the causal-formative model is assumed as \gls{icm}, the \gls{mse} is hardly influenced by whether it represents the \gls{dgp} correctly or not.
In particular, the bias and variance in the top row of Figure\ref{fig:exog-biases} are similar when the construct type in the \gls{dgp} is causal-formative or a latent variable.
Only when the construct in the \gls{dgp} is a composite, one obtains larger biases but smaller variances.

The center row of the Figure displays the scenarios in which composite models are assumed.
If this assumption is incorrect, i.e.\ the samples come from one of the other two \glspl{dgp}, the biases are clearly larger than when the model is correctly specified.
Larger sample sizes exacerbate this bias. 
However, the variance of the estimates are generally smaller when assuming a composite model than any other \gls{icm} assumption.
That is, even when the composite model assumption is incorrect, the variance of the estimates are comparable to or even smaller than assuming a different \glspl{icm}.

% LV misspecified
If the latent variable model is assumed, the parameter estimates for both the causal-formative and the composite \glspl{dgp} are clearly biased.
However, the bias for the composite \gls{dgp} is roughly five times larger than for the causal-formative one.
The weaker the correlations between the indicators, the larger these biases.

% Direction of the bias and asymptotics
It is known that incorrectly assuming a latent variable to be a composite leads to path coefficient being estimated closer to zero, while assuming a composite to be a latent variable produces over-estimated coefficients \citep[see, e.g.,][]{Rhemtulla2020}.
For this reason, we investigated the direction of the bias as well.
The corresponding figure in the supplementary material underpins the directions of the biases for our simulation study.
That is, all three path coefficients are underestimated when assuming a composite \gls{icm} on samples from a latent variable \gls{dgp} and hence more likely to be incorrectly considered insignificant in empirical studies.
If, on the other side, a latent variable model is incorrectly assumed, the path coefficients are over-estimated and thus more likely to be considered significant.
\par
\begin{figure}[ht]
    \centering
    \includegraphics[width=0.7\linewidth]{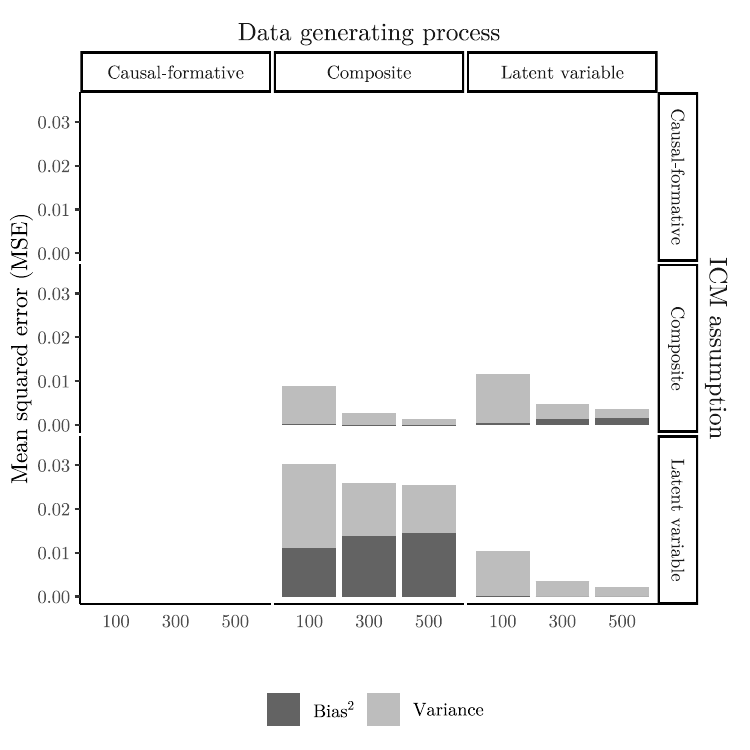}
    \caption{\gls{mse} of the admissible \gls{ml} estimates in the endogenous case of the standardized path coefficient~$\beta_{*,1}$, with axes analogue to Figure~\ref{fig:exog-biases}. The causal-formative \gls{dgp} and assumed \gls{icm} are deliberately left blank (see Section \ref{sec:design}).}
    \label{fig:endog-biases}
\end{figure}
% Estimation bias endogenous case
The \gls{mse} in the endogenous case are similar to the results of the exogenous case, as shown in Figure~\ref{fig:endog-biases}.
The graphic depicts the standardized path coefficient of $\eta_1$ on $\eta^*$.
The correctly specified models again lead to no apparent bias.
However, under both \gls{icm} misspecifications, on the other hand, the biases of the path coefficients is larger.
The estimation performance for the other two standardized path coefficients can be found in the supplementary material.
The smaller the true absolute value of a path coefficient, the smaller its estimation bias.
The estimation variance, on the other hand, does not seem to be affected by the magnitude of path coefficients.
\par
%
% MSE with PLS
%
\begin{figure}[ht]
    \centering
    \includegraphics[width=0.65\linewidth]{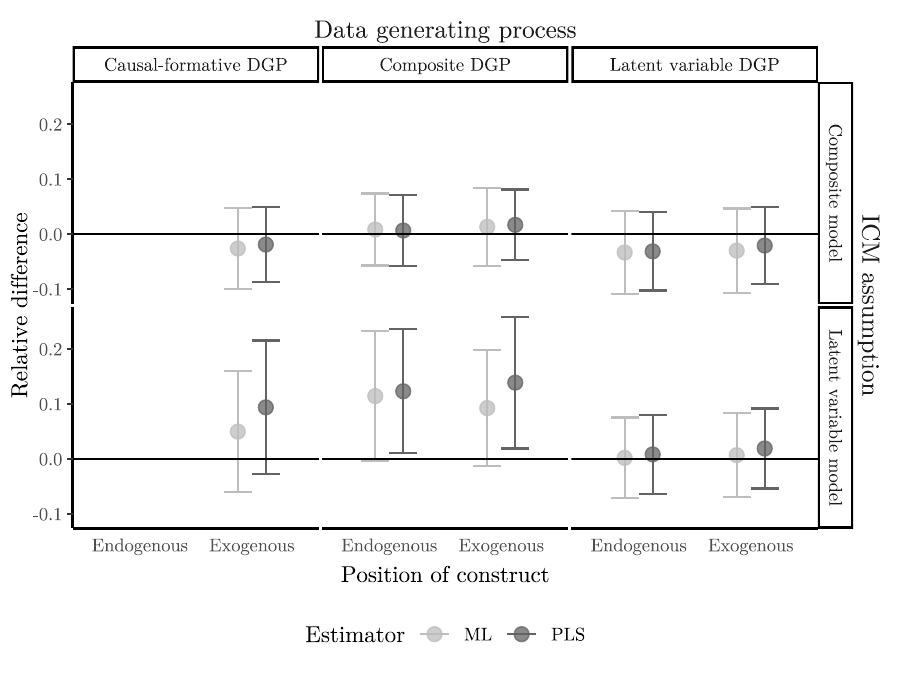}%\includegraphics[width=0.65\linewidth]{graphics/results/exog/PLS/comparison_biasVariance_MLPLS.pdf}
    \caption{Bias (points) with standard deviation (error bars) of admissible estimates of $\beta_{1,*}$ and $\beta_{*,1}$, respectively, using either \gls{ml} or \gls{pls}-based estimators. The estimators perform similarly for all \glspl{dgp} (columns), \gls{icm} assumptions (rows) and construct positions (horizontal axis).}
    \label{fig:MLPLSbiases}
\end{figure}
To evaluate whether the previous results are prone to the use of ML or hold rather independently of the means of estimation, we replicated our study using PLS based estimation.
A comparison of the estimation performances of both estimators is presented in Figure~\ref{fig:MLPLSbiases}.
Overall, both the biases and the variances of the different estimates appear similar.
More precisely, the top row depicts the case of assuming a composite model.
If this assumption is correct, \gls{plspm} yields no apparent bias and a small variance, analogue to the \gls{ml} estimates on the \gls{hospec}.
When the composite \gls{icm} is incorrectly assumed, then both the bias and the variance of \gls{plspm} are slightly smaller than those based on \gls{ml}.
If a latent variable model is correctly assumed, \gls{plsc} performs also unbiased and with variances comparable to those based on \gls{ml}, as indicated by the bottom row of the figure.
If, on the other hand, this assumption is incorrect for an exogenous construct, \gls{plsc} exhibits a larger bias than \gls{ml}.
\par 
%
% Interpretations
%
Overall, our results for the estimation performance can be interpreted as follows:
Since the causal-formative model contains the largest number of parameters, it is not surprising that the \gls{ml} estimator also exhibits the largest variance for small sample sizes.
The particularly large bias when misspecifying an actually composite construct as causal-formative assumed \gls{icm} can again be explained by the estimation of an obsolete disturbance variance.
Only for large sample sizes can this parameter be estimated correctly as zero.
Although the assumption of a latent variable model for \glspl{dgp} with a causal-formative construct yields biased path coefficients even for larger samples, the magnitude of that bias is negligible.
The bias when misspecifying a composite as latent variable is much larger than vice versa (i.e.\ assuming a latent variable to be a composite) and in alignment with previous findings \citep[e.g.,][]{Sarstedt2016}.
An endogenous position of the misspecified construct amplifies the bias even more.
Overall, misspecified \glspl{icm} can bias estimates irrespective of the chosen estimator.
%
%
% Fit measures
%
\subsection{Detection of misspecification}
%
% Intro and general description raster plots
%
The previous section showed that parameter estimates from a misspecified \gls{icm} are likely to be biased and thus harm the conclusions drawn from it.
Hence, it would be of great value to be able to identify such misspecification.
In our study, we considered in total six fit measures.
Out of these six criteria, however, only two are applicable on all \glspl{icm} and exhibit differences between the design conditions.
Consequently, we report in this section (1) the $\chi^2$ test and (2) the \gls{srmr}. 
For the sake of completeness, the results for (3) the \gls{cfi}, (4) the \gls{rmsea}, (5) the \gls{cr}, and (6) the \gls{ave} can be found in the supplementary material.
Since the results for the exogenous and the endogenous case were similar, we only report the results for the exogenous case.
Results concerning the endogenous case can also be found in the supplementary material.
All the criteria mentioned above are designed to evaluate the model fit by comparing an empirical quantity (e.g., a p-value or a residual matrix) to a threshold value (e.g., a significance level or an upper limit for a matrix norm).
Specifically, a model fit is considered poor by the $\chi^2$ test if the p-value is smaller than $0.05$, by the \gls{srmr} if it is larger than $0.08$, by the \gls{cfi} if it is smaller than $0.95$, by the \gls{rmsea} if it is larger than $0.05$, by the \gls{cr} of a construct if it is smaller than $0.7$, and by the \gls{ave} if it is smaller than $0.5$ \citep[see, e.g.,][]{Schermelleh-Engel2003EvaluatingMeasures}.
We interpret a poor model fit based on a criteria for a given sample and threshold as an indication that the assumed \gls{icm} may be misspecified (i.e.\ a flag for potential misspecification).
While the \gls{icm} is the only misspecification present in our Monte Carlo study, real-world application studies can have a poor model fit due to many reasons.
\par
%
% Chi squared test exogenous case
%
\begin{figure}[ht]
    \centering
\includegraphics[width=0.9\linewidth]{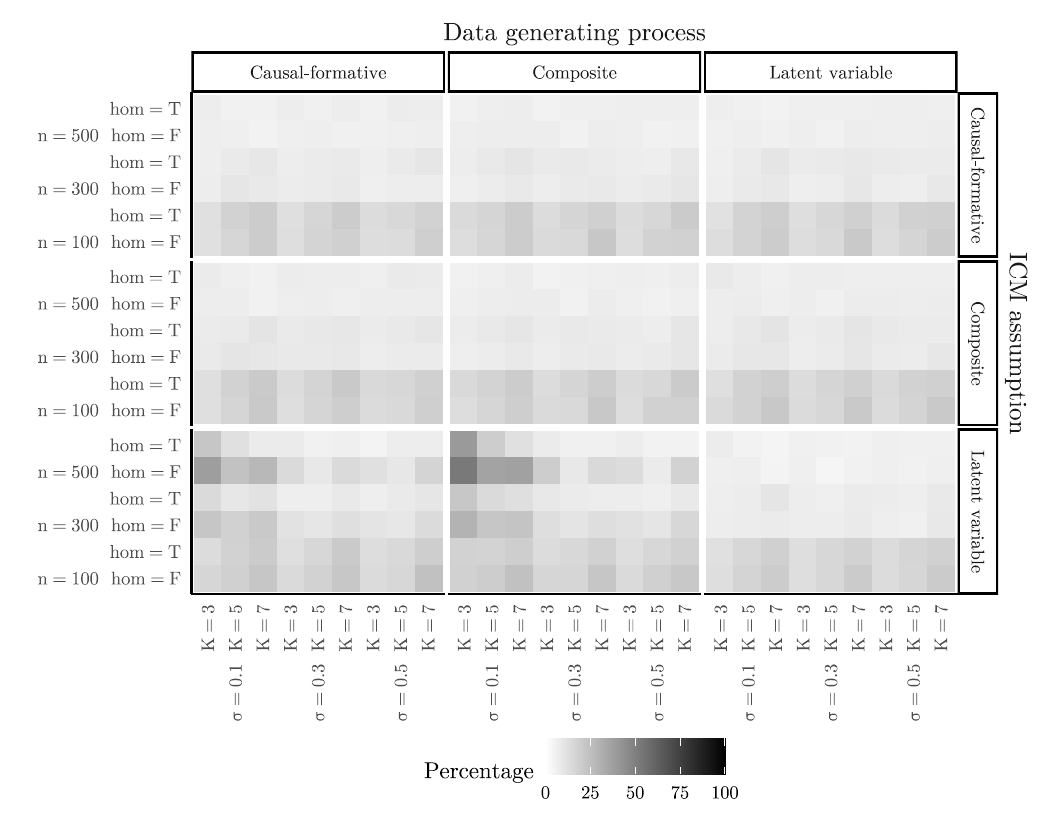}
    \caption{Raster plot of model evaluations based on $\chi^2$ tests in the exogenous case. Each tile represents the percentage of Monte Carlo iterations where the null hypothesis was rejected at $\alpha =0.05$ for a given \gls{dgp} (outer columns), given \gls{icm} assumption (outer rows), given combination of number of indicators and base correlations (inner columns), and given combination of sample size and correlation homogeneity (inner rows). A rejected null hypothesis indicates that an assumed model may be misspecified.}
    \label{fig:exog-chi2}
\end{figure}
In \gls{sem}, the $\chi^2$ test is used to investigate the null hypothesis that the model represents the population variance-covariance matrix of the indicators correctly.
Given that a model is misspecified, the rejection of the null hypothesis means correctly flagging the model as a poor fit for the data.
On the contrary, rejecting the null hypothesis for a correctly specified model means flagging incorrectly.
Figure~\ref{fig:exog-chi2} displays the percentage of flagged \gls{ml} estimations based on $\chi^2$ test decisions in the exogenous case as a raster plot.
This criterion rarely flags when \glspl{icm} are correctly specified, as indicated by the diagonal cells of the figure. 
The larger the sample size, the less estimates are incorrectly flagged.
However, when assuming a causal-formative \gls{icm}, the percentage of estimations which are flagged as potentially misspecified are almost equal for the three \glspl{dgp}.
A similar pattern appears if the composite model is the assumed \gls{icm}.
Only if the assumed \gls{icm} is a latent variable model, the $\chi^2$ test shows differences between the different \glspl{dgp}.
Here, for instance, causal-formative and composite \glspl{dgp} are flagged more often.
The highest flagging frequency occurs when a composite \gls{dgp} with few indicators and small and homogeneous correlations are assumed to be a latent variable model.
The fact that high indicator correlations interfere with the fit of latent variable models is in accordance with existing literature \citep[see, e.g.,][]{Rhemtulla2020}.
\par
%
% Chi squared test endogenous case
%
%
% The detection of \gls{icm} misspecification of an endogenous construct using the $\chi^2$ test performs similarly to the exogenous case.
% In particular, the criterion has overall low flagging frequencies and makes little differences between correctly and misspecified constructs (see supplementary material).
% \par
%
% SRMR exogenous case
%
The second criterion to detect \gls{icm} misspecification is the \gls{srmr}.
The \gls{srmr} is another way of measuring the difference between the sample variance-covariance matrix and the model-implied counterpart.
In particular, this criterion normalizes differences by taking the scale of each indicator into account \citep{Schermelleh-Engel2003EvaluatingMeasures}.
The larger the value, the worse the fit of an assumed structural equation model to the sample.
\begin{figure}[ht]
    \centering
    \includegraphics[width=0.9\linewidth]{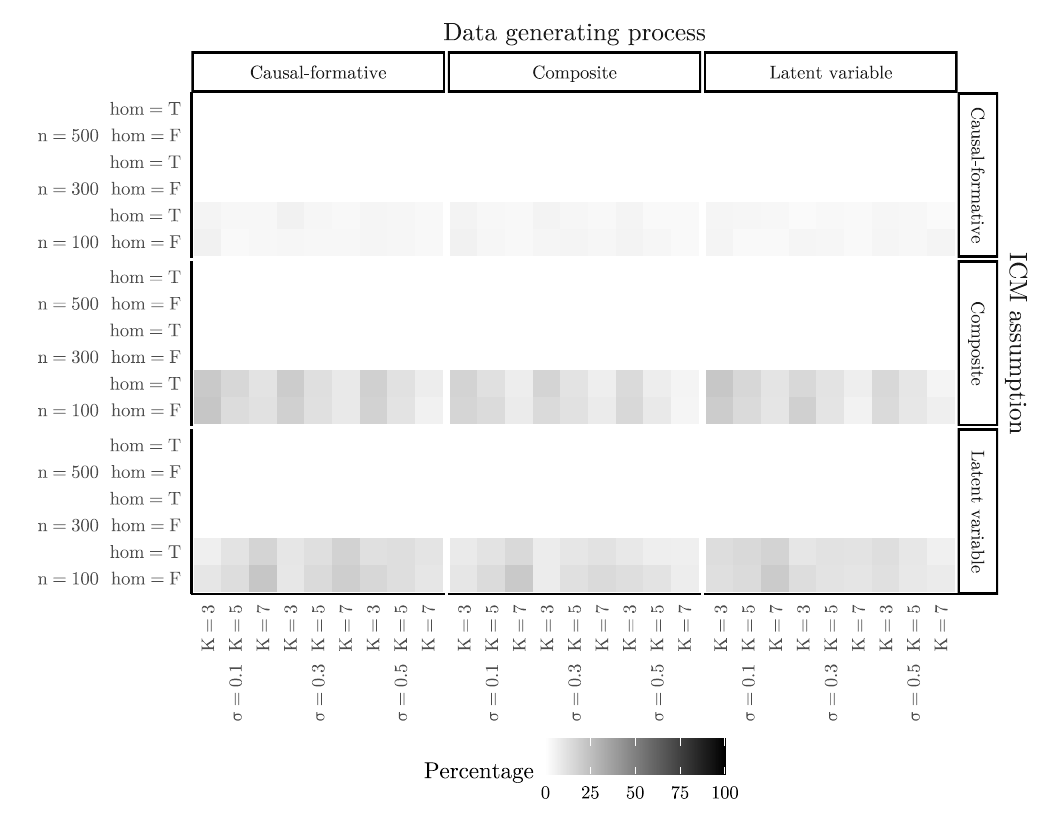}
    \caption{Raster plot of model evaluations based on the \gls{srmr} in the exogenous case, with axes analogue to Figure~\ref{fig:exog-chi2}. Each tile represents the percentage of Monte Carlo iterations where this criterion is larger than $0.08$, which indicates that an assumed model may be misspecified.}
    \label{fig:exog-srmr}
\end{figure}
Figure~\ref{fig:exog-srmr} depicts the percentage of flagged \gls{ml} estimations based on this criterion.
In general, this criterion flags only estimates based on a sample size of $100$.
Although the \gls{srmr} flags very rarely when \glspl{icm} are correctly specified (see diagonal cells in the figure), the figure shows no apparent difference between correctly specified and misspecified (off-diagonal) \glspl{icm}.
Hence, \gls{srmr} is incapable to detect \gls{icm} misspecification accurately.
Assuming a composite \gls{icm} (center row in the figure), the number of indicators~$K$ influences the percentage of flagged estimates.
The smaller $K$, the more estimates are flagged.
If one assumes a latent variable model instead, the number of indicators has an opposite effect on the flagging frequency.
The larger $K$, the more often does \gls{srmr} flag estimates.
\par
%
% SRMR endogenous case
%
% When the \gls{srmr} is used to detect misspecification of an endogenous construct, the flagging rates are overall higher, as indicated by the darker tones in Figure~\ref{app:fig:ML:endog-srmr}.
% Nonetheless, there is no apparent difference between correctly and misspecified models in this criterion.
% \par
%
% Estimation with PLS
%
Both fit measures can also be computed when using \gls{pls}-based estimation.
However, it has been noted that they should be considered with caution, since the estimators optimize different objectives, which can affect the criteria \citep{Hair2019WhenPLS-SEM}.
In line with that, the $\chi^2$ test flags every single Monte Carlo sample in our simulation study as misspecified, regardless of whether that is true or not.
The \gls{srmr}, on the other hand, shows a similar pattern as for \gls{ml}-based estimation. 
However, this criterion flags more \gls{pls}-based estimations than those of \gls{ml} and especially it flags larger samples with week indicator correlations.

% The identification of \gls{icm} misspecification based on significance tests with \gls{plspm} is not in the scope of this article for two reasons.
% First, the results for \gls{ml}-based estimation foreshadow the inadequacy of this criterion due to multiple testing.
% We see no reason to believe otherwise for \gls{plspm}.
% Second, when using \gls{plspm} significance tests would be based on computationally much more expensive bootstrapping.
% \par
%
% Interpretations
%
% chi^2:
Overall, the fit measures considered in this article make almost no difference between whether an assumed \gls{icm} correctly reflects the actual \gls{dgp}.
Although certain design conditions are flagged more often than others, the implications of \gls{icm} misspecification on the $\chi^2$ test and the \gls{srmr} seem negligible.
The remaining fit measures of this article do not overcome this shortcoming. 
While \gls{ave} and the \gls{cr} flag nearly all models irrespective of their correct or misspecification, the opposite is the case using the \gls{rmsea} and the \gls{cfi}.
In general, fit measures are designed to evaluate the fit of a model. 
Their adoption for the detection of \gls{icm} misspecification is based on the assumption that misspecified models yield poorer model fit in the sense of a larger discrepancy between the variance-covariance matrices of the population and the model-implied one.
Our results show that this connection does not necessarily hold.
Especially causal-formative models are richer parameterized and can thus better approximate population matrices based on the other \glspl{icm}.

\section{Discussion and Outlook} \label{sec:discuss}
% Recap of our motivation and the problem
In most statistical analyses the true \gls{dgp} is unknown but hoped to be well reflected by the choice for a particular class of models.
Given a (probably random) sample, one then estimates the parameters of that model.
In such real-world scenarios, many factors can affect a statistical analysis such as the randomness of the data, the correctness of the assumed model and the adequacy of the chosen estimator.
Due to this, Monte Carlo studies are commonly used to investigate isolated aspects of a statistical analysis while controlling the influence of other factors \citep[e.g.,][]{Paxton2001MonteImplementation}.
In this article, for instance, we investigated the consequences of correct and incorrect operationalizations of constructs in \gls{sem}, while accounting for sample effects and the choice of the estimator.
More precisely, we accounted for the sample effect by drawing many random samples; we generated different \glspl{dgp} purposefully to be exactly reflected by one of the three construct operationalizations but incorrectly reflected by the others; we estimated the parameters of the correct and the two misspecified \glspl{icm} once using \gls{ml} and a second time using \gls{pls}-based techniques.
The novelty of this set-up is that it allowed us to separate the issue of \gls{icm} misspecification from the issue of parameter estimation.
In contrast, previous studies often compared different classes of estimators for different \gls{icm} assumptions simultaneously.
Hence, either an incorrect construct type, an inconsistent estimator, or a combination of both could have been responsible for their findings.
\par 
In this article, we presented a comprehensive study of nine combinations between true construct type and assumed \gls{icm}, using two classes of estimators and $108$ Monte Carlo design conditions with respect to Fisher consistency, admissibility, estimation performance and detection of \gls{icm} misspecification using fit measures. 
In doing so, we provided important insights on the consequences of \gls{icm} misspecification which are independent of the choice of the estimator by three control measures.
First, we specified three types of \glspl{dgp} that differed solely with respect to the type of one particular construct but shared other characteristics such as covariances between constructs and indicators.
Hence, the processes and the data they produced differed as little as possible beyond the construct types.
Second, the admissible estimates of the three \glspl{icm} are based on the same samples to reduce the sample effects on our comparison of the correctly specified model with the two misspecifications.
Third, we conducted our simulation study once using \gls{ml} estimation and a second time with \gls{pls}-based estimation to ensure that our findings can be generalized beyond one particular estimator.
% Our findings summarized
Our results highlight two key concerns for researchers utilizing \gls{sem}.
First, correctly specifying a construct leads to the smallest parameter estimation bias, regardless of the construct type.
However, the magnitude of this bias depends on the particular type of misspecification and the sample size.
The largest biases occurred when a composite was misspecified as a latent variable.
And although causal-formative models are Fisher consistent for all construct types, they required generally larger sample sizes for the bias to be negligible.
Since the differences between estimation with \gls{ml} and \gls{pls} variants were negligible in our study, the findings are likely the consequences of \gls{icm} misspecification, and not of the estimator.
Second, researchers cannot rely on fit measures to justify an assumed \gls{icm}.
None of the six commonly used fit measures in our study were able to distinguish correct from incorrect specified models.
%While some criteria flagged too often, others did not flag any model at all.
%
\par
% Context of our findings
In general, model misspecification is an active yet diverse field of research. 
For example, it has been studied in the context of  the distribution of stock returns \citep{Uppal2003ModelUnderdiversification} or exposure effects in observational studies \citep{Vansteelandt2012OnInference}.
Even within the field of  \gls{sem}, models can be misspecified by omitting indicators \citep[e.g.,][]{MacKenzie2005TheSolutions}, omitting structural paths \citep[e.g.,][]{Vilares2013LikelihoodEffects,Hu1998FitMisspecification} or assuming an incorrect \gls{icm} \citep[e.g.,][]{Aguirre-Urreta2024}.
With this article, we contribute to the latter by exploring different forms of misspecification of the \gls{icm}.
The direction of the parameter biases in our study were in line with previous simulations and theoretical considerations \citep{Rhemtulla2020} and are likely to be due to the \gls{icm} misspecification.
While we agree with \citet{Sarstedt2016} that misspecifying a composite as latent variable leads to more severely biased estimates than vice versa, this result appears to be model-specific rather than a flaw of \gls{plsc}.
With respect to the detectability of \gls{icm} misspecification, our findings extend prior research indicating the limitations of selected fit measures \citep[e.g.,][]{Diamantopoulos2008AdvancingModels, MacKenzie2005TheSolutions}. 
More precisely, we demonstrated the lack of efficacy for additional measures and all combinations of construct (mis-)specification.
\par
%
% Limitations
Our findings are based on a simulation study which is naturally limited by its design.
In particular, we used a statistical model and other design conditions from \citet{Aguirre-Urreta2024}, who in turn leaned on
\citet{Jarvis2003AResearch}, \citet{Petter2007SpecifyingResearch}
and others.
Consequently, our results are limited to this particular model.
Moreover, we designed the different construct types in the \glspl{dgp} to share certain properties, but differ in others.
It is yet unknown whether our findings would also hold for \glspl{dgp} that share different or none of the properties.
With respect to parameter estimation, our results are limited to \gls{ml} and \gls{pls}-based estimation.

\par
%
% Implications and future research
Our findings have significant implications for applied \gls{sem} research. The results underline the need for careful operationalization of constructs during the model development phase, as the choice of construct type has a direct impact on parameter estimates. 
Future research should verify the consequences of \gls{icm} misspecification in more complex models, e.g., with higher-order constructs, multi-class models, or mixtures of structural equation models.
These models should be examined with additional design conditions, such as the number of misspecified constructs and the skewness of indicator distributions.
Given our results regarding the usage of fit measures to detect misspecifications, researchers cannot rely on model fit indices for this purpose. 
%Instead, we suggest a rigorous theory-driven justification of \glspl{icm}.
Instead, further investigations regarding the detection of misspecification are required. 
For instance, future research could explore other fit measures, model selection criteria, or likelihood-based statistical tests. 
To further evaluate the influence of the estimation technique on the results, further estimators such as \gls{gsca} \citep{Hwang2004GeneralizedAnalysis} could be analyzed.
\par
% Closing
Overall, \gls{icm} misspecification leads to biased path coefficients, which can remain undetected by many common fit measures. 
This issue is stressed by many application studies with ambiguous construct operationalizations \citep[e.g.,][]{Romdhani2015, Grotzinger2019,Wright2012OperationalizingResearch,Fleuren2018HandlingEmployability, Gurbuz2023SustainableStudy}.
For this reason, further research is required to strengthen the understanding of \gls{icm} misspecification, develop better diagnostic procedures for its detection and ultimately enhance the trust in \gls{sem}-based inference.

% \section*{Acknowledgements}
% We would like to thank Miguel I.\ Aguirre-Urreta, Mikko Rönkkö and George M.\ Marakas for sharing the source code of their simulation study with us.
% %
%
\section*{Data availability statement}
Supplementary material including additional visualizations, the \texttt{R} source files, and the results in \texttt{.RData} format can be found on \href{https://osf.io/syrqk/?view_only=1aee0b1096834b8d947013ae30f5f878}{https://osf.io/syrqk/?view\_only=1aee0b1096834b8d947013ae30f5f878}%
. 
\singlespacing
\ifarxiv
  % Im arXiv-Mode keine ORCID-Anzeige
\else
  \ifanonymize
    % Im anonymisierten Modus keine ORCID-Anzeige
  \else
    \section*{ORCID}

    \begin{tabular}{@{}l l l@{}}
      Jonas Bauer         & \orcidlink{0000-0003-1759-7889} & \href{https://orcid.org/0000-0003-1759-7889}{https://orcid.org/0000-0003-1759-7889} \\
      Axel Mayer          & \orcidlink{0000-0001-9716-878X} & \href{https://orcid.org/0000-0001-9716-878X}{https://orcid.org/0000-0001-9716-878X} \\
      Christiane Fuchs    & \orcidlink{0000-0003-3565-8315} & \href{https://orcid.org/0000-0003-3565-8315}{https://orcid.org/0000-0003-3565-8315} \\
      Tamara Schamberger  & \orcidlink{0000-0002-7845-784X} & \href{https://orcid.org/0000-0002-7845-784X}{https://orcid.org/0000-0002-7845-784X} \\
    \end{tabular}
  \fi
\fi

\renewcommand\bibsection{\section*{\refname}}
\bibliography{references_man}
\newpage
%\appendix
\glsaddall
\printglossaries
\end{document}

%% file: preamble.tex
\iftikz
\documentclass{standalone}
\fi

\ifarxiv
\documentclass{article}
\usepackage{arxiv}

\else
\documentclass[APA,Times1COL]{WileyNJDv5} %STIX1COL,STIX2COL,STIXSMALL
\articletype{Focus on research methods}%
\received{Date Month Year}
\revised{Date Month Year}
\accepted{Date Month Year}
\journal{Information Systems Journal}
\volume{00}
\copyyear{2026}
\startpage{1}
\fi

\usepackage{setspace}

\ifarxiv
  \ifanonymize
  \else
    \author{
      Jonas Bauer\\
      Faculty of Business Administration and Economics\\
      Bielefeld University\\
      Universit\"atsstr.\ 25\\
      33615 Bielefeld\\
      Germany
      \And
      Axel Mayer\\
      Faculty of Psychology and Sports Sciences\\
      Bielefeld University\\
      Universit\"atsstr.\ 25\\
      33615 Bielefeld\\
      Germany
      \And
      Christiane Fuchs\\
      Faculty of Business Administration and Economics\\
      Bielefeld University\\
      Universit\"atsstr.\ 25\\
      33615 Bielefeld\\
      Germany
      \And
      Tamara Schamberger\\
      Faculty of Business Administration and Economics\\
      Bielefeld University\\
      Universit\"atsstr.\ 25\\
      33615 Bielefeld\\
      Germany
    }
  \fi
\else
  \author[1]{Jonas Bauer}
  \author[2]{Axel Mayer}
  \author[1,3]{Christiane Fuchs}
  \author[1]{Tamara Schamberger}

  \authormark{BAUER \textsc{et al.}}

  \address[1]{
    \orgdiv{Faculty of Business Administration and Economics}, 
    \orgname{Bielefeld University}, 
    \orgaddress{\state{Universit\"atsstr.\ 25, 33615 Bielefeld}, \country{Germany}}
  }

  \address[2]{
    \orgdiv{Faculty of Psychology and Sports Sciences}, 
    \orgname{Bielefeld University}, 
    \orgaddress{\state{Universit\"atsstr.\ 25, 33615 Bielefeld}, \country{Germany}}
  }

  \address[3]{
    \orgdiv{Institute of Computational Biology}, 
    \orgname{Helmholtz Munich}, 
    \orgaddress{\state{Ingolst\"adter Landstr.\ 1, 85764 Neuherberg}, \country{Germany}}
  }

  \corres{Corresponding author Tamara Schamberger, Bielefeld University. 
    \email{\href{mailto:tamara.schamberger@uni-bielefeld.de}{tamara.schamberger@uni-bielefeld.de}}}
\fi

\usepackage{amsmath}
\usepackage{amssymb}
\usepackage{mathrsfs}
\usepackage{mathtools}
\usepackage{graphicx}
\usepackage{hyperref}
\usepackage{tikz, pgfplots}	
\pgfplotsset{compat=newest} % otherwise warning appears
\usepackage{amsthm}
\usepackage{multicol}
\usepackage{booktabs} % To thicken table lines
\usepackage{tabularx}
\usepackage{bm} % bold face in math equations
\usepackage[section]{placeins} % avoid that figures leave their sections
\usepackage{multirow}
\usepackage{enumerate}
\usepackage{parskip}
\usepackage{orcidlink} % for the orcid link and icon
\usetikzlibrary{
shapes,
arrows, shapes.geometric,
shapes.arrows,
positioning,
decorations.pathreplacing,
angles,quotes,
fit,
calc, 
shapes.symbols} 

\usepackage{natbib}
\usepackage[abbreviations]{glossaries-extra}
\makeglossaries

\newabbreviation[shortplural={DGPs}]{dgp}{DGP}{data generating process}
\newabbreviation{sem}{SEM}{Structural equation modeling}
\newabbreviation[shortplural={ICMs}]{icm}{ICM}{indicator-construct model}
\newabbreviation{plspm}{PLS-PM}{partial least squares path modeling}
\newabbreviation{pls}{PLS}{partial least squares}
\newabbreviation{ml}{ML}{maximum likelihood}
\newabbreviation{plsc}{PLSc}{consistent partial least squares}
\newabbreviation{gsca}{GSCA}{generalized structured component analysis}
\newabbreviation{cfi}{CFI}{comparative fit index}
\newabbreviation{rmsea}{RMSEA}{root mean square error of approximation}
\newabbreviation{srmr}{SRMR}{standardized root mean square residual}
\newabbreviation{ave}{AVE}{average variance extracted}
\newabbreviation{cr}{CR}{composite reliability}
\newabbreviation{mse}{MSE}{mean squared error}
\newabbreviation{hospec}{H--O specification}{Henseler--Ogasawara specification}

\usepackage[labelformat=simple]{subcaption}
\ifarxiv

\else

\fi
\NewDocumentCommand\Cycle{O{} m m m O{} m m}{%
  \draw[#1,<->, very thick, >=stealth'](#2.{#3+asin(#6/(#4*1.41))}) arc (180+#3-45:180+#3-45-#7:#6/2) #5;
}

%% file: main.bbl
\begin{thebibliography}{}

\bibitem [\protect \citeauthoryear {%
Abate%
\ \protect \BOthers {.}}{%
Abate%
\ \protect \BOthers {.}}{%
{\protect \APACyear {2020}}%
}]{%
Abate2020ValuationValuation}
\APACinsertmetastar {%
Abate2020ValuationValuation}%
\begin{APACrefauthors}%
Abate, T\BPBI G.%
, B{\"{o}}rger, T.%
, Aanesen, M.%
, Falk-Andersson, J.%
, Wyles, K\BPBI J.%
\BCBL {}\ \BBA {} Beaumont, N.%
\end{APACrefauthors}%
\unskip\
\newblock
\APACrefYearMonthDay{2020}{}{}.
\newblock
{\BBOQ}\APACrefatitle {{Valuation of marine plastic pollution in the European
  Arctic: Applying an integrated choice and latent variable model to contingent
  valuation}} {{Valuation of marine plastic pollution in the European Arctic:
  Applying an integrated choice and latent variable model to contingent
  valuation}}.{\BBCQ}
\newblock
\APACjournalVolNumPages{Ecological Economics}{169}{}{106521}.
\newblock
\begin{APACrefDOI} 10.1016/j.ecolecon.2019.106521 \end{APACrefDOI}
\PrintBackRefs{\CurrentBib}

\bibitem [\protect \citeauthoryear {%
Aguirre-Urreta%
\ \BBA {} Marakas%
}{%
Aguirre-Urreta%
\ \BBA {} Marakas%
}{%
{\protect \APACyear {2012}}%
}]{%
Aguirre-Urreta2012RevisitingForm}
\APACinsertmetastar {%
Aguirre-Urreta2012RevisitingForm}%
\begin{APACrefauthors}%
Aguirre-Urreta, M\BPBI I.%
\BCBT {}\ \BBA {} Marakas, G\BPBI M.%
\end{APACrefauthors}%
\unskip\
\newblock
\APACrefYearMonthDay{2012}{}{}.
\newblock
{\BBOQ}\APACrefatitle {{Revisiting bias due to construct misspecification:
  Different results from considering coefficients in standardized form}}
  {{Revisiting bias due to construct misspecification: Different results from
  considering coefficients in standardized form}}.{\BBCQ}
\newblock
\APACjournalVolNumPages{MIS Quarterly: Management Information
  Systems}{36}{}{123--138}.
\newblock
\begin{APACrefDOI} 10.2307/41410409 \end{APACrefDOI}
\PrintBackRefs{\CurrentBib}

\bibitem [\protect \citeauthoryear {%
Aguirre-Urreta%
\ \BBA {} Marakas%
}{%
Aguirre-Urreta%
\ \BBA {} Marakas%
}{%
{\protect \APACyear {2014}}%
}]{%
Aguirre-Urreta2014}
\APACinsertmetastar {%
Aguirre-Urreta2014}%
\begin{APACrefauthors}%
Aguirre-Urreta, M\BPBI I.%
\BCBT {}\ \BBA {} Marakas, G\BPBI M.%
\end{APACrefauthors}%
\unskip\
\newblock
\APACrefYearMonthDay{2014}{}{}.
\newblock
{\BBOQ}\APACrefatitle {{Research note—Partial least squares and models with
  formatively specified endogenous constructs: A cautionary note}} {{Research
  note—Partial least squares and models with formatively specified endogenous
  constructs: A cautionary note}}.{\BBCQ}
\newblock
\APACjournalVolNumPages{Information Systems Research}{25}{}{761--778}.
\newblock
\begin{APACrefDOI} 10.1287/isre.2013.0493 \end{APACrefDOI}
\PrintBackRefs{\CurrentBib}

\bibitem [\protect \citeauthoryear {%
Aguirre-Urreta%
, R{\"{o}}nkk{\"{o}}%
\BCBL {}\ \BBA {} Marakas%
}{%
Aguirre-Urreta%
\ \protect \BOthers {.}}{%
{\protect \APACyear {2024}}%
}]{%
Aguirre-Urreta2024}
\APACinsertmetastar {%
Aguirre-Urreta2024}%
\begin{APACrefauthors}%
Aguirre-Urreta, M\BPBI I.%
, R{\"{o}}nkk{\"{o}}, M.%
\BCBL {}\ \BBA {} Marakas, G\BPBI M.%
\end{APACrefauthors}%
\unskip\
\newblock
\APACrefYearMonthDay{2024}{}{}.
\newblock
{\BBOQ}\APACrefatitle {{Reconsidering the implications of formative versus
  reflective measurement model misspecification}} {{Reconsidering the
  implications of formative versus reflective measurement model
  misspecification}}.{\BBCQ}
\newblock
\APACjournalVolNumPages{Information Systems Journal}{34}{}{533--584}.
\newblock
\begin{APACrefDOI} 10.1111/isj.12487 \end{APACrefDOI}
\PrintBackRefs{\CurrentBib}

\bibitem [\protect \citeauthoryear {%
Arbuckle%
}{%
Arbuckle%
}{%
{\protect \APACyear {2011}}%
}]{%
Arbuckle2011IBMGuide}
\APACinsertmetastar {%
Arbuckle2011IBMGuide}%
\begin{APACrefauthors}%
Arbuckle, J\BPBI L.%
\end{APACrefauthors}%
\unskip\
\newblock
\APACrefYearMonthDay{2011}{}{}.
\newblock
\APACrefbtitle {{IBM SPSS Amos 20 User’s Guide}.} {{IBM SPSS Amos 20 User’s
  Guide}.}
\newblock
\APACaddressPublisher{}{Chicago, IL, USA}.
\PrintBackRefs{\CurrentBib}

\bibitem [\protect \citeauthoryear {%
Arslan%
\ \BBA {} Co{\c{s}}kun%
}{%
Arslan%
\ \BBA {} Co{\c{s}}kun%
}{%
{\protect \APACyear {2020}}%
}]{%
Arslan2020StudentAnalysis}
\APACinsertmetastar {%
Arslan2020StudentAnalysis}%
\begin{APACrefauthors}%
Arslan, G.%
\BCBT {}\ \BBA {} Co{\c{s}}kun, M.%
\end{APACrefauthors}%
\unskip\
\newblock
\APACrefYearMonthDay{2020}{}{}.
\newblock
{\BBOQ}\APACrefatitle {{Student subjective wellbeing, school functioning, and
  psychological adjustment in high school adolescents: A latent variable
  analysis}} {{Student subjective wellbeing, school functioning, and
  psychological adjustment in high school adolescents: A latent variable
  analysis}}.{\BBCQ}
\newblock
\APACjournalVolNumPages{Journal of Positive School Psychology}{4}{}{153--164}.
\newblock
\begin{APACrefDOI} 10.47602/jpsp.v4i2.231 \end{APACrefDOI}
\PrintBackRefs{\CurrentBib}

\bibitem [\protect \citeauthoryear {%
Asif%
\ \protect \BOthers {.}}{%
Asif%
\ \protect \BOthers {.}}{%
{\protect \APACyear {2022}}%
}]{%
Asif2023}
\APACinsertmetastar {%
Asif2023}%
\begin{APACrefauthors}%
Asif, M\BPBI H.%
, Zhongfu, T.%
, Ahmad, B.%
, Irfan, M.%
, Razzaq, A.%
\BCBL {}\ \BBA {} Ameer, W.%
\end{APACrefauthors}%
\unskip\
\newblock
\APACrefYearMonthDay{2022}{}{}.
\newblock
{\BBOQ}\APACrefatitle {{Influencing factors of consumers’ buying intention of
  solar energy: A structural equation modeling approach}} {{Influencing factors
  of consumers’ buying intention of solar energy: A structural equation
  modeling approach}}.{\BBCQ}
\newblock
\APACjournalVolNumPages{Environmental Science and Pollution
  Research}{30}{}{30017--30032}.
\newblock
\begin{APACrefDOI} 10.1007/s11356-022-24286-w \end{APACrefDOI}
\PrintBackRefs{\CurrentBib}

\bibitem [\protect \citeauthoryear {%
Bachmann%
, Rose%
, Maul%
\BCBL {}\ \BBA {} H{\"{o}}lzle%
}{%
Bachmann%
\ \protect \BOthers {.}}{%
{\protect \APACyear {2024}}%
}]{%
Bachmann2024WhatIntention}
\APACinsertmetastar {%
Bachmann2024WhatIntention}%
\begin{APACrefauthors}%
Bachmann, N.%
, Rose, R.%
, Maul, V.%
\BCBL {}\ \BBA {} H{\"{o}}lzle, K.%
\end{APACrefauthors}%
\unskip\
\newblock
\APACrefYearMonthDay{2024}{}{}.
\newblock
{\BBOQ}\APACrefatitle {{What makes for future entrepreneurs? The role of
  digital competencies for entrepreneurial intention}} {{What makes for future
  entrepreneurs? The role of digital competencies for entrepreneurial
  intention}}.{\BBCQ}
\newblock
\APACjournalVolNumPages{Journal of Business Research}{174}{}{114481}.
\newblock
\begin{APACrefDOI} 10.1016/j.jbusres.2023.114481 \end{APACrefDOI}
\PrintBackRefs{\CurrentBib}

\bibitem [\protect \citeauthoryear {%
Barillari%
\ \protect \BOthers {.}}{%
Barillari%
\ \protect \BOthers {.}}{%
{\protect \APACyear {2021}}%
}]{%
Barillari2021AStudy}
\APACinsertmetastar {%
Barillari2021AStudy}%
\begin{APACrefauthors}%
Barillari, M\BPBI R.%
, Bastiani, L.%
, Lechien, J\BPBI R.%
, Mannelli, G.%
, Molteni, G.%
, Cantarella, G.%
\BDBL {}Cammaroto, G.%
\end{APACrefauthors}%
\unskip\
\newblock
\APACrefYearMonthDay{2021}{}{}.
\newblock
{\BBOQ}\APACrefatitle {{A structural equation model to examine the clinical
  features of mild-to-moderate COVID-19: A multicenter Italian study}} {{A
  structural equation model to examine the clinical features of
  mild-to-moderate COVID-19: A multicenter Italian study}}.{\BBCQ}
\newblock
\APACjournalVolNumPages{Journal of Medical Virology}{93}{}{983--994}.
\newblock
\begin{APACrefDOI} 10.1002/jmv.26354 \end{APACrefDOI}
\PrintBackRefs{\CurrentBib}

\bibitem [\protect \citeauthoryear {%
Battistoni%
, Gitto%
, Murgia%
\BCBL {}\ \BBA {} Campisi%
}{%
Battistoni%
\ \protect \BOthers {.}}{%
{\protect \APACyear {2023}}%
}]{%
Battistoni2023AdoptionSME}
\APACinsertmetastar {%
Battistoni2023AdoptionSME}%
\begin{APACrefauthors}%
Battistoni, E.%
, Gitto, S.%
, Murgia, G.%
\BCBL {}\ \BBA {} Campisi, D.%
\end{APACrefauthors}%
\unskip\
\newblock
\APACrefYearMonthDay{2023}{}{}.
\newblock
{\BBOQ}\APACrefatitle {{Adoption paths of digital transformation in
  manufacturing SME}} {{Adoption paths of digital transformation in
  manufacturing SME}}.{\BBCQ}
\newblock
\APACjournalVolNumPages{International Journal of Production
  Economics}{255}{}{108675}.
\newblock
\begin{APACrefDOI} 10.1016/j.ijpe.2022.108675 \end{APACrefDOI}
\PrintBackRefs{\CurrentBib}

\bibitem [\protect \citeauthoryear {%
Becker%
, Rai%
\BCBL {}\ \BBA {} Rigdon%
}{%
Becker%
\ \protect \BOthers {.}}{%
{\protect \APACyear {2013}}%
}]{%
Becker2013PredictiveRelevance}
\APACinsertmetastar {%
Becker2013PredictiveRelevance}%
\begin{APACrefauthors}%
Becker, J\BHBI M.%
, Rai, A.%
\BCBL {}\ \BBA {} Rigdon, E.%
\end{APACrefauthors}%
\unskip\
\newblock
\APACrefYearMonthDay{2013}{}{}.
\newblock
{\BBOQ}\APACrefatitle {{Predictive validity and formative measurement in
  structural equation modeling: Embracing practical relevance}} {{Predictive
  validity and formative measurement in structural equation modeling: Embracing
  practical relevance}}.{\BBCQ}
\newblock
\BIn{} \APACrefbtitle {Proceedings of the International Conference on
  Information Systems: 17. Research Methods and Philosophy.} {Proceedings of
  the international conference on information systems: 17. research methods and
  philosophy.}
\PrintBackRefs{\CurrentBib}

\bibitem [\protect \citeauthoryear {%
Bentler%
}{%
Bentler%
}{%
{\protect \APACyear {1990}}%
}]{%
Bentler1990ComparativeModels.}
\APACinsertmetastar {%
Bentler1990ComparativeModels.}%
\begin{APACrefauthors}%
Bentler, P\BPBI M.%
\end{APACrefauthors}%
\unskip\
\newblock
\APACrefYearMonthDay{1990}{}{}.
\newblock
{\BBOQ}\APACrefatitle {{Comparative fit indexes in structural models.}}
  {{Comparative fit indexes in structural models.}}{\BBCQ}
\newblock
\APACjournalVolNumPages{Psychological Bulletin}{107}{}{238--246}.
\newblock
\begin{APACrefDOI} 10.1037/0033-2909.107.2.238 \end{APACrefDOI}
\PrintBackRefs{\CurrentBib}

\bibitem [\protect \citeauthoryear {%
Bentler%
}{%
Bentler%
}{%
{\protect \APACyear {2006}}%
}]{%
Bentler2006EQSManual}
\APACinsertmetastar {%
Bentler2006EQSManual}%
\begin{APACrefauthors}%
Bentler, P\BPBI M.%
\end{APACrefauthors}%
\unskip\
\newblock
\APACrefYearMonthDay{2006}{}{}.
\newblock
\APACrefbtitle {{EQS 6 structural equations program manual}.} {{EQS 6
  structural equations program manual}.}
\PrintBackRefs{\CurrentBib}

\bibitem [\protect \citeauthoryear {%
Bentler%
\ \BBA {} Speckart%
}{%
Bentler%
\ \BBA {} Speckart%
}{%
{\protect \APACyear {1981}}%
}]{%
Bentler1981AttitudesAnalysis.}
\APACinsertmetastar {%
Bentler1981AttitudesAnalysis.}%
\begin{APACrefauthors}%
Bentler, P\BPBI M.%
\BCBT {}\ \BBA {} Speckart, G.%
\end{APACrefauthors}%
\unskip\
\newblock
\APACrefYearMonthDay{1981}{}{}.
\newblock
{\BBOQ}\APACrefatitle {{Attitudes "cause" behaviors: A structural equation
  analysis.}} {{Attitudes "cause" behaviors: A structural equation
  analysis.}}{\BBCQ}
\newblock
\APACjournalVolNumPages{Journal of Personality and Social
  Psychology}{40}{}{226--238}.
\newblock
\begin{APACrefDOI} 10.1037/0022-3514.40.2.226 \end{APACrefDOI}
\PrintBackRefs{\CurrentBib}

\bibitem [\protect \citeauthoryear {%
Berbekova%
, Kock%
, Assaf%
\BCBL {}\ \BBA {} Josiassen%
}{%
Berbekova%
\ \protect \BOthers {.}}{%
{\protect \APACyear {2025}}%
}]{%
Berbekova2025UnderstandingMisspecification}
\APACinsertmetastar {%
Berbekova2025UnderstandingMisspecification}%
\begin{APACrefauthors}%
Berbekova, A.%
, Kock, F.%
, Assaf, A\BPBI G.%
\BCBL {}\ \BBA {} Josiassen, A.%
\end{APACrefauthors}%
\unskip\
\newblock
\APACrefYearMonthDay{2025}{}{}.
\newblock
{\BBOQ}\APACrefatitle {{Understanding and employing formative constructs:
  Conceptualization, operationalization, and threats of misspecification}}
  {{Understanding and employing formative constructs: Conceptualization,
  operationalization, and threats of misspecification}}.{\BBCQ}
\newblock
\APACjournalVolNumPages{Journal of Hospitality {\&} Tourism
  Research}{49}{}{946--960}.
\newblock
\begin{APACrefDOI} 10.1177/10963480241258510 \end{APACrefDOI}
\PrintBackRefs{\CurrentBib}

\bibitem [\protect \citeauthoryear {%
Bitri{\'{a}}n%
, Buil%
, Catal{\'{a}}n%
\BCBL {}\ \BBA {} Merli%
}{%
Bitri{\'{a}}n%
\ \protect \BOthers {.}}{%
{\protect \APACyear {2024}}%
}]{%
Bitrian2024}
\APACinsertmetastar {%
Bitrian2024}%
\begin{APACrefauthors}%
Bitri{\'{a}}n, P.%
, Buil, I.%
, Catal{\'{a}}n, S.%
\BCBL {}\ \BBA {} Merli, D.%
\end{APACrefauthors}%
\unskip\
\newblock
\APACrefYearMonthDay{2024}{}{}.
\newblock
{\BBOQ}\APACrefatitle {{Gamification in workforce training: Improving
  employees’ self-efficacy and information security and data protection
  behaviours}} {{Gamification in workforce training: Improving employees’
  self-efficacy and information security and data protection
  behaviours}}.{\BBCQ}
\newblock
\APACjournalVolNumPages{Journal of Business Research}{179}{}{114685}.
\newblock
\begin{APACrefDOI} 10.1016/j.jbusres.2024.114685 \end{APACrefDOI}
\PrintBackRefs{\CurrentBib}

\bibitem [\protect \citeauthoryear {%
Bollen%
}{%
Bollen%
}{%
{\protect \APACyear {1989}}%
}]{%
Bollen1989}
\APACinsertmetastar {%
Bollen1989}%
\begin{APACrefauthors}%
Bollen, K\BPBI A.%
\end{APACrefauthors}%
\unskip\
\newblock
\APACrefYear{1989}.
\newblock
\APACrefbtitle {{Structural equations with latent variables}} {{Structural
  equations with latent variables}}.
\newblock
\APACaddressPublisher{}{John Wiley {\&} Sons, Inc.}
\newblock
\begin{APACrefDOI} 10.1002/9781118619179 \end{APACrefDOI}
\PrintBackRefs{\CurrentBib}

\bibitem [\protect \citeauthoryear {%
Bollen%
}{%
Bollen%
}{%
{\protect \APACyear {2002}}%
}]{%
Bollen2002LatentSciences}
\APACinsertmetastar {%
Bollen2002LatentSciences}%
\begin{APACrefauthors}%
Bollen, K\BPBI A.%
\end{APACrefauthors}%
\unskip\
\newblock
\APACrefYearMonthDay{2002}{}{}.
\newblock
{\BBOQ}\APACrefatitle {{Latent Variables in Psychology and the Social
  Sciences}} {{Latent Variables in Psychology and the Social Sciences}}.{\BBCQ}
\newblock
\APACjournalVolNumPages{Annual Review of Psychology}{53}{}{605--634}.
\newblock
\begin{APACrefDOI} 10.1146/annurev.psych.53.100901.135239 \end{APACrefDOI}
\PrintBackRefs{\CurrentBib}

\bibitem [\protect \citeauthoryear {%
Bollen%
\ \BBA {} Bauldry%
}{%
Bollen%
\ \BBA {} Bauldry%
}{%
{\protect \APACyear {2011}}%
}]{%
Bollen2011}
\APACinsertmetastar {%
Bollen2011}%
\begin{APACrefauthors}%
Bollen, K\BPBI A.%
\BCBT {}\ \BBA {} Bauldry, S.%
\end{APACrefauthors}%
\unskip\
\newblock
\APACrefYearMonthDay{2011}{}{}.
\newblock
{\BBOQ}\APACrefatitle {{Three Cs in measurement models: Causal indicators,
  composite indicators, and covariates.}} {{Three Cs in measurement models:
  Causal indicators, composite indicators, and covariates.}}{\BBCQ}
\newblock
\APACjournalVolNumPages{Psychological Methods}{16}{}{265--284}.
\newblock
\begin{APACrefDOI} 10.1037/a0024448 \end{APACrefDOI}
\PrintBackRefs{\CurrentBib}

\bibitem [\protect \citeauthoryear {%
Bollen%
\ \BBA {} Curran%
}{%
Bollen%
\ \BBA {} Curran%
}{%
{\protect \APACyear {2005}}%
}]{%
Bollen2005LatentPerspective}
\APACinsertmetastar {%
Bollen2005LatentPerspective}%
\begin{APACrefauthors}%
Bollen, K\BPBI A.%
\BCBT {}\ \BBA {} Curran, P\BPBI J.%
\end{APACrefauthors}%
\unskip\
\newblock
\APACrefYear{2005}.
\newblock
\APACrefbtitle {{Latent curve models: A structural equation perspective}}
  {{Latent curve models: A structural equation perspective}}.
\newblock
\APACaddressPublisher{}{John Wiley {\&} Sons, Inc.}
\newblock
\begin{APACrefDOI} 10.1002/0471746096 \end{APACrefDOI}
\PrintBackRefs{\CurrentBib}

\bibitem [\protect \citeauthoryear {%
Bollen%
\ \BBA {} Davis%
}{%
Bollen%
\ \BBA {} Davis%
}{%
{\protect \APACyear {2009}}%
}]{%
Bollen2009CausalTesting}
\APACinsertmetastar {%
Bollen2009CausalTesting}%
\begin{APACrefauthors}%
Bollen, K\BPBI A.%
\BCBT {}\ \BBA {} Davis, W\BPBI R.%
\end{APACrefauthors}%
\unskip\
\newblock
\APACrefYearMonthDay{2009}{}{}.
\newblock
{\BBOQ}\APACrefatitle {{Causal indicator models: Identification, estimation,
  and testing}} {{Causal indicator models: Identification, estimation, and
  testing}}.{\BBCQ}
\newblock
\APACjournalVolNumPages{Structural Equation Modeling: A Multidisciplinary
  Journal}{16}{}{498--522}.
\newblock
\begin{APACrefDOI} 10.1080/10705510903008253 \end{APACrefDOI}
\PrintBackRefs{\CurrentBib}

\bibitem [\protect \citeauthoryear {%
Bollen%
\ \BBA {} Diamantopoulos%
}{%
Bollen%
\ \BBA {} Diamantopoulos%
}{%
{\protect \APACyear {2017}}%
}]{%
Bollen2017}
\APACinsertmetastar {%
Bollen2017}%
\begin{APACrefauthors}%
Bollen, K\BPBI A.%
\BCBT {}\ \BBA {} Diamantopoulos, A.%
\end{APACrefauthors}%
\unskip\
\newblock
\APACrefYearMonthDay{2017}{}{}.
\newblock
{\BBOQ}\APACrefatitle {{In defense of causal-formative indicators: A minority
  report.}} {{In defense of causal-formative indicators: A minority
  report.}}{\BBCQ}
\newblock
\APACjournalVolNumPages{Psychological Methods}{22}{}{581--596}.
\newblock
\begin{APACrefDOI} 10.1037/met0000056 \end{APACrefDOI}
\PrintBackRefs{\CurrentBib}

\bibitem [\protect \citeauthoryear {%
Bollen%
\ \BBA {} Lennox%
}{%
Bollen%
\ \BBA {} Lennox%
}{%
{\protect \APACyear {1991}}%
}]{%
Bollen1991ConventionalPerspective}
\APACinsertmetastar {%
Bollen1991ConventionalPerspective}%
\begin{APACrefauthors}%
Bollen, K\BPBI A.%
\BCBT {}\ \BBA {} Lennox, R.%
\end{APACrefauthors}%
\unskip\
\newblock
\APACrefYearMonthDay{1991}{}{}.
\newblock
{\BBOQ}\APACrefatitle {{Conventional wisdom on measurement: A structural
  equation perspective.}} {{Conventional wisdom on measurement: A structural
  equation perspective.}}{\BBCQ}
\newblock
\APACjournalVolNumPages{Psychological Bulletin}{110}{}{305--314}.
\newblock
\begin{APACrefDOI} 10.1037/0033-2909.110.2.305 \end{APACrefDOI}
\PrintBackRefs{\CurrentBib}

\bibitem [\protect \citeauthoryear {%
Bollen%
\ \BBA {} Noble%
}{%
Bollen%
\ \BBA {} Noble%
}{%
{\protect \APACyear {2011}}%
}]{%
Bollen2011StructuralBehavior}
\APACinsertmetastar {%
Bollen2011StructuralBehavior}%
\begin{APACrefauthors}%
Bollen, K\BPBI A.%
\BCBT {}\ \BBA {} Noble, M\BPBI D.%
\end{APACrefauthors}%
\unskip\
\newblock
\APACrefYearMonthDay{2011}{}{}.
\newblock
{\BBOQ}\APACrefatitle {{Structural equation models and the quantification of
  behavior}} {{Structural equation models and the quantification of
  behavior}}.{\BBCQ}
\newblock
\BIn{} \APACrefbtitle {Proceedings of the National Academy of Sciences}
  {Proceedings of the national academy of sciences}\ (\BPGS\ 15639--15646).
\newblock
\begin{APACrefDOI} 10.1073/pnas.1010661108 \end{APACrefDOI}
\PrintBackRefs{\CurrentBib}

\bibitem [\protect \citeauthoryear {%
Bollen%
\ \BBA {} Ting%
}{%
Bollen%
\ \BBA {} Ting%
}{%
{\protect \APACyear {2000}}%
}]{%
Bollen2000AIndicators.}
\APACinsertmetastar {%
Bollen2000AIndicators.}%
\begin{APACrefauthors}%
Bollen, K\BPBI A.%
\BCBT {}\ \BBA {} Ting, K\BHBI f.%
\end{APACrefauthors}%
\unskip\
\newblock
\APACrefYearMonthDay{2000}{}{}.
\newblock
{\BBOQ}\APACrefatitle {{A tetrad test for causal indicators.}} {{A tetrad test
  for causal indicators.}}{\BBCQ}
\newblock
\APACjournalVolNumPages{Psychological Methods}{5}{}{3--22}.
\newblock
\begin{APACrefDOI} 10.1037/1082-989X.5.1.3 \end{APACrefDOI}
\PrintBackRefs{\CurrentBib}

\bibitem [\protect \citeauthoryear {%
Cadogan%
\ \BBA {} Lee%
}{%
Cadogan%
\ \BBA {} Lee%
}{%
{\protect \APACyear {2013}}%
}]{%
Cadogan2013ImproperVariables}
\APACinsertmetastar {%
Cadogan2013ImproperVariables}%
\begin{APACrefauthors}%
Cadogan, J\BPBI W.%
\BCBT {}\ \BBA {} Lee, N.%
\end{APACrefauthors}%
\unskip\
\newblock
\APACrefYearMonthDay{2013}{}{}.
\newblock
{\BBOQ}\APACrefatitle {{Improper use of endogenous formative variables}}
  {{Improper use of endogenous formative variables}}.{\BBCQ}
\newblock
\APACjournalVolNumPages{Journal of Business Research}{66}{}{233--241}.
\newblock
\begin{APACrefDOI} 10.1016/j.jbusres.2012.08.006 \end{APACrefDOI}
\PrintBackRefs{\CurrentBib}

\bibitem [\protect \citeauthoryear {%
Carvalho%
\ \BBA {} Chima%
}{%
Carvalho%
\ \BBA {} Chima%
}{%
{\protect \APACyear {2014}}%
}]{%
Carvalho2014ApplicationsResearch}
\APACinsertmetastar {%
Carvalho2014ApplicationsResearch}%
\begin{APACrefauthors}%
Carvalho, d\BPBI J.%
\BCBT {}\ \BBA {} Chima, O\BPBI F.%
\end{APACrefauthors}%
\unskip\
\newblock
\APACrefYearMonthDay{2014}{}{}.
\newblock
{\BBOQ}\APACrefatitle {{Applications of structural equation modeling in social
  sciences research}} {{Applications of structural equation modeling in social
  sciences research}}.{\BBCQ}
\newblock
\APACjournalVolNumPages{American International Journal of Contemporary
  Research}{4}{}{6--11}.
\PrintBackRefs{\CurrentBib}

\bibitem [\protect \citeauthoryear {%
Cenfetelli%
\ \BBA {} Bassellier%
}{%
Cenfetelli%
\ \BBA {} Bassellier%
}{%
{\protect \APACyear {2009}}%
}]{%
Cenfetelli2009InterpretationResearch}
\APACinsertmetastar {%
Cenfetelli2009InterpretationResearch}%
\begin{APACrefauthors}%
Cenfetelli, R\BPBI T.%
\BCBT {}\ \BBA {} Bassellier, G.%
\end{APACrefauthors}%
\unskip\
\newblock
\APACrefYearMonthDay{2009}{}{}.
\newblock
{\BBOQ}\APACrefatitle {{Interpretation of formative measurement in information
  systems research}} {{Interpretation of formative measurement in information
  systems research}}.{\BBCQ}
\newblock
\APACjournalVolNumPages{MIS Quarterly}{33}{}{689--707}.
\PrintBackRefs{\CurrentBib}

\bibitem [\protect \citeauthoryear {%
Chand%
, Kumar%
, Thakkar%
\BCBL {}\ \BBA {} Ghosh%
}{%
Chand%
\ \protect \BOthers {.}}{%
{\protect \APACyear {2022}}%
}]{%
Chand2022DirectTheory}
\APACinsertmetastar {%
Chand2022DirectTheory}%
\begin{APACrefauthors}%
Chand, P.%
, Kumar, A.%
, Thakkar, J.%
\BCBL {}\ \BBA {} Ghosh, K\BPBI K.%
\end{APACrefauthors}%
\unskip\
\newblock
\APACrefYearMonthDay{2022}{}{}.
\newblock
{\BBOQ}\APACrefatitle {{Direct and mediation effect of supply chain complexity
  drivers on supply chain performance: An empirical evidence of
  organizational complexity theory}} {{Direct and mediation effect of supply
  chain complexity drivers on supply chain performance: An empirical evidence
  of organizational complexity theory}}.{\BBCQ}
\newblock
\APACjournalVolNumPages{International Journal of Operations {\&} Production
  Management}{42}{}{797--825}.
\newblock
\begin{APACrefDOI} 10.1108/IJOPM-11-2021-0681 \end{APACrefDOI}
\PrintBackRefs{\CurrentBib}

\bibitem [\protect \citeauthoryear {%
Cheah%
, Thurasamy%
, Memon%
, Chuah%
\BCBL {}\ \BBA {} Ting%
}{%
Cheah%
\ \protect \BOthers {.}}{%
{\protect \APACyear {2020}}%
}]{%
Cheah2020}
\APACinsertmetastar {%
Cheah2020}%
\begin{APACrefauthors}%
Cheah, J\BHBI H.%
, Thurasamy, R.%
, Memon, M\BPBI A.%
, Chuah, F.%
\BCBL {}\ \BBA {} Ting, H.%
\end{APACrefauthors}%
\unskip\
\newblock
\APACrefYearMonthDay{2020}{}{}.
\newblock
{\BBOQ}\APACrefatitle {{Multigroup analysis using SmartPLS: Step-by-step
  guidelines for business research}} {{Multigroup analysis using SmartPLS:
  Step-by-step guidelines for business research}}.{\BBCQ}
\newblock
\APACjournalVolNumPages{Asian Journal of Business Research}{10(3)}{}{1--19}.
\newblock
\begin{APACrefDOI} 10.14707/ajbr.200087 \end{APACrefDOI}
\PrintBackRefs{\CurrentBib}

\bibitem [\protect \citeauthoryear {%
Chen%
, Bollen%
, Paxton%
, Curran%
\BCBL {}\ \BBA {} Kirby%
}{%
Chen%
\ \protect \BOthers {.}}{%
{\protect \APACyear {2001}}%
}]{%
Chen2001ImproperStrategies}
\APACinsertmetastar {%
Chen2001ImproperStrategies}%
\begin{APACrefauthors}%
Chen, F.%
, Bollen, K\BPBI A.%
, Paxton, P.%
, Curran, P\BPBI J.%
\BCBL {}\ \BBA {} Kirby, J\BPBI B.%
\end{APACrefauthors}%
\unskip\
\newblock
\APACrefYearMonthDay{2001}{}{}.
\newblock
{\BBOQ}\APACrefatitle {{Improper solutions in structural equation models:
  Causes, consequences, and strategies}} {{Improper solutions in structural
  equation models: Causes, consequences, and strategies}}.{\BBCQ}
\newblock
\APACjournalVolNumPages{Sociological Methods and Research}{29}{}{468--508}.
\newblock
\begin{APACrefDOI} 10.1177/0049124101029004003 \end{APACrefDOI}
\PrintBackRefs{\CurrentBib}

\bibitem [\protect \citeauthoryear {%
Chin%
}{%
Chin%
}{%
{\protect \APACyear {1998}}%
}]{%
Chin1998TheModeling}
\APACinsertmetastar {%
Chin1998TheModeling}%
\begin{APACrefauthors}%
Chin, W\BPBI W.%
\end{APACrefauthors}%
\unskip\
\newblock
\APACrefYearMonthDay{1998}{}{}.
\newblock
{\BBOQ}\APACrefatitle {{The partial least squares approach to structural
  equation modeling}} {{The partial least squares approach to structural
  equation modeling}}.{\BBCQ}
\newblock
\BIn{} G\BPBI A.~Marcoulides\ (\BED), \APACrefbtitle {Modern Methods for
  Business Research} {Modern methods for business research}\ (\BPGS\ 295--336).
\newblock
\APACaddressPublisher{}{Psychology Press}.
\newblock
\begin{APACrefDOI} 10.4324/9781410604385-10 \end{APACrefDOI}
\PrintBackRefs{\CurrentBib}

\bibitem [\protect \citeauthoryear {%
Cho%
\ \BBA {} Choi%
}{%
Cho%
\ \BBA {} Choi%
}{%
{\protect \APACyear {2020}}%
}]{%
Cho2020AnModels}
\APACinsertmetastar {%
Cho2020AnModels}%
\begin{APACrefauthors}%
Cho, G.%
\BCBT {}\ \BBA {} Choi, J\BPBI Y.%
\end{APACrefauthors}%
\unskip\
\newblock
\APACrefYearMonthDay{2020}{}{}.
\newblock
{\BBOQ}\APACrefatitle {{An empirical comparison of generalized structured
  component analysis and partial least squares path modeling under
  variance-based structural equation models}} {{An empirical comparison of
  generalized structured component analysis and partial least squares path
  modeling under variance-based structural equation models}}.{\BBCQ}
\newblock
\APACjournalVolNumPages{Behaviormetrika}{47}{}{243--272}.
\newblock
\begin{APACrefDOI} 10.1007/s41237-019-00098-0 \end{APACrefDOI}
\PrintBackRefs{\CurrentBib}

\bibitem [\protect \citeauthoryear {%
Cho%
, Sarstedt%
\BCBL {}\ \BBA {} Hwang%
}{%
Cho%
\ \protect \BOthers {.}}{%
{\protect \APACyear {2022}}%
}]{%
Cho2022ARepresentations}
\APACinsertmetastar {%
Cho2022ARepresentations}%
\begin{APACrefauthors}%
Cho, G.%
, Sarstedt, M.%
\BCBL {}\ \BBA {} Hwang, H.%
\end{APACrefauthors}%
\unskip\
\newblock
\APACrefYearMonthDay{2022}{}{}.
\newblock
{\BBOQ}\APACrefatitle {{A comparative evaluation of factor‐ and
  component‐based structural equation modelling approaches under (in)correct
  construct representations}} {{A comparative evaluation of factor‐ and
  component‐based structural equation modelling approaches under (in)correct
  construct representations}}.{\BBCQ}
\newblock
\APACjournalVolNumPages{British Journal of Mathematical and Statistical
  Psychology}{75}{}{220--251}.
\newblock
\begin{APACrefDOI} 10.1111/bmsp.12255 \end{APACrefDOI}
\PrintBackRefs{\CurrentBib}

\bibitem [\protect \citeauthoryear {%
Cox%
\ \BBA {} Hinkley%
}{%
Cox%
\ \BBA {} Hinkley%
}{%
{\protect \APACyear {1979}}%
}]{%
Cox1979TheoreticalStatistics}
\APACinsertmetastar {%
Cox1979TheoreticalStatistics}%
\begin{APACrefauthors}%
Cox, D.%
\BCBT {}\ \BBA {} Hinkley, D.%
\end{APACrefauthors}%
\unskip\
\newblock
\APACrefYear{1979}.
\newblock
\APACrefbtitle {{Theoretical Statistics}} {{Theoretical Statistics}}.
\newblock
\begin{APACrefDOI} 10.1201/b14832 \end{APACrefDOI}
\PrintBackRefs{\CurrentBib}

\bibitem [\protect \citeauthoryear {%
{Diamantopoulos}%
}{%
{Diamantopoulos}%
}{%
{\protect \APACyear {2011}}%
}]{%
Diamantopoulos2011IncorporatingModels}
\APACinsertmetastar {%
Diamantopoulos2011IncorporatingModels}%
\begin{APACrefauthors}%
{Diamantopoulos}.%
\end{APACrefauthors}%
\unskip\
\newblock
\APACrefYearMonthDay{2011}{}{}.
\newblock
{\BBOQ}\APACrefatitle {{Incorporating formative measures into covariance-based
  structural equation models}} {{Incorporating formative measures into
  covariance-based structural equation models}}.{\BBCQ}
\newblock
\APACjournalVolNumPages{MIS Quarterly}{35}{}{335--358}.
\newblock
\begin{APACrefDOI} 10.2307/23044046 \end{APACrefDOI}
\PrintBackRefs{\CurrentBib}

\bibitem [\protect \citeauthoryear {%
Diamantopoulos%
, Riefler%
\BCBL {}\ \BBA {} Roth%
}{%
Diamantopoulos%
\ \protect \BOthers {.}}{%
{\protect \APACyear {2008}}%
}]{%
Diamantopoulos2008AdvancingModels}
\APACinsertmetastar {%
Diamantopoulos2008AdvancingModels}%
\begin{APACrefauthors}%
Diamantopoulos, A.%
, Riefler, P.%
\BCBL {}\ \BBA {} Roth, K\BPBI P.%
\end{APACrefauthors}%
\unskip\
\newblock
\APACrefYearMonthDay{2008}{}{}.
\newblock
{\BBOQ}\APACrefatitle {{Advancing formative measurement models}} {{Advancing
  formative measurement models}}.{\BBCQ}
\newblock
\APACjournalVolNumPages{Journal of Business Research}{61}{}{1203--1218}.
\newblock
\begin{APACrefDOI} 10.1016/j.jbusres.2008.01.009 \end{APACrefDOI}
\PrintBackRefs{\CurrentBib}

\bibitem [\protect \citeauthoryear {%
Dijkstra%
}{%
Dijkstra%
}{%
{\protect \APACyear {2017}}%
}]{%
Dijkstra2017}
\APACinsertmetastar {%
Dijkstra2017}%
\begin{APACrefauthors}%
Dijkstra, T\BPBI K.%
\end{APACrefauthors}%
\unskip\
\newblock
\APACrefYearMonthDay{2017}{}{}.
\newblock
{\BBOQ}\APACrefatitle {{A perfect match between a model and a mode}} {{A
  perfect match between a model and a mode}}.{\BBCQ}
\newblock
\BIn{} \APACrefbtitle {Partial Least Squares Path Modeling} {Partial least
  squares path modeling}\ (\BPGS\ 55--80).
\newblock
\APACaddressPublisher{Cham}{Springer International Publishing}.
\newblock
\begin{APACrefDOI} 10.1007/978-3-319-64069-3{\_}4 \end{APACrefDOI}
\PrintBackRefs{\CurrentBib}

\bibitem [\protect \citeauthoryear {%
Dijkstra%
\ \BBA {} Henseler%
}{%
Dijkstra%
\ \BBA {} Henseler%
}{%
{\protect \APACyear {2015}}%
}]{%
Dijkstra2015ConsistentModeling}
\APACinsertmetastar {%
Dijkstra2015ConsistentModeling}%
\begin{APACrefauthors}%
Dijkstra, T\BPBI K.%
\BCBT {}\ \BBA {} Henseler, J.%
\end{APACrefauthors}%
\unskip\
\newblock
\APACrefYearMonthDay{2015}{}{}.
\newblock
{\BBOQ}\APACrefatitle {{Consistent partial least squares path modeling}}
  {{Consistent partial least squares path modeling}}.{\BBCQ}
\newblock
\APACjournalVolNumPages{JSTOR}{39}{}{297--316}.
\PrintBackRefs{\CurrentBib}

\bibitem [\protect \citeauthoryear {%
Dijkstra%
\ \BBA {} Schermelleh-Engel%
}{%
Dijkstra%
\ \BBA {} Schermelleh-Engel%
}{%
{\protect \APACyear {2014}}%
}]{%
Dijkstra2014ConsistentModels}
\APACinsertmetastar {%
Dijkstra2014ConsistentModels}%
\begin{APACrefauthors}%
Dijkstra, T\BPBI K.%
\BCBT {}\ \BBA {} Schermelleh-Engel, K.%
\end{APACrefauthors}%
\unskip\
\newblock
\APACrefYearMonthDay{2014}{}{}.
\newblock
{\BBOQ}\APACrefatitle {{Consistent partial least squares for nonlinear
  structural equation models}} {{Consistent partial least squares for nonlinear
  structural equation models}}.{\BBCQ}
\newblock
\APACjournalVolNumPages{Psychometrika}{79}{}{585--604}.
\newblock
\begin{APACrefDOI} 10.1007/s11336-013-9370-0 \end{APACrefDOI}
\PrintBackRefs{\CurrentBib}

\bibitem [\protect \citeauthoryear {%
Edwards%
}{%
Edwards%
}{%
{\protect \APACyear {2001}}%
}]{%
Edwards2001MultidimensionalFramework}
\APACinsertmetastar {%
Edwards2001MultidimensionalFramework}%
\begin{APACrefauthors}%
Edwards, J\BPBI R.%
\end{APACrefauthors}%
\unskip\
\newblock
\APACrefYearMonthDay{2001}{}{}.
\newblock
{\BBOQ}\APACrefatitle {{Multidimensional constructs in organizational behavior
  research: An integrative analytical framework}} {{Multidimensional constructs
  in organizational behavior research: An integrative analytical
  framework}}.{\BBCQ}
\newblock
\APACjournalVolNumPages{Organizational Research Methods}{4}{}{144--192}.
\newblock
\begin{APACrefDOI} 10.1177/109442810142004 \end{APACrefDOI}
\PrintBackRefs{\CurrentBib}

\bibitem [\protect \citeauthoryear {%
Fisher%
}{%
Fisher%
}{%
{\protect \APACyear {1922}}%
}]{%
Fisher1922OnStatistics}
\APACinsertmetastar {%
Fisher1922OnStatistics}%
\begin{APACrefauthors}%
Fisher, R\BPBI A.%
\end{APACrefauthors}%
\unskip\
\newblock
\APACrefYearMonthDay{1922}{}{}.
\newblock
{\BBOQ}\APACrefatitle {{On the mathematical foundations of theoretical
  statistics}} {{On the mathematical foundations of theoretical
  statistics}}.{\BBCQ}
\newblock
\APACjournalVolNumPages{Philosophical Transactions of the Royal Society of
  London. Series A, Containing Papers of a Mathematical or Physical
  Character}{222}{}{309--368}.
\newblock
\begin{APACrefDOI} 10.1098/rsta.1922.0009 \end{APACrefDOI}
\PrintBackRefs{\CurrentBib}

\bibitem [\protect \citeauthoryear {%
Fleuren%
, van Amelsvoort%
, Zijlstra%
, de Grip%
\BCBL {}\ \BBA {} Kant%
}{%
Fleuren%
\ \protect \BOthers {.}}{%
{\protect \APACyear {2018}}%
}]{%
Fleuren2018HandlingEmployability}
\APACinsertmetastar {%
Fleuren2018HandlingEmployability}%
\begin{APACrefauthors}%
Fleuren, B\BPBI P.%
, van Amelsvoort, L\BPBI G.%
, Zijlstra, F\BPBI R.%
, de Grip, A.%
\BCBL {}\ \BBA {} Kant, I.%
\end{APACrefauthors}%
\unskip\
\newblock
\APACrefYearMonthDay{2018}{}{}.
\newblock
{\BBOQ}\APACrefatitle {{Handling the reflective-formative measurement
  conundrum: A practical illustration based on sustainable employability}}
  {{Handling the reflective-formative measurement conundrum: A practical
  illustration based on sustainable employability}}.{\BBCQ}
\newblock
\APACjournalVolNumPages{Journal of Clinical Epidemiology}{103}{}{71--81}.
\newblock
\begin{APACrefDOI} 10.1016/j.jclinepi.2018.07.007 \end{APACrefDOI}
\PrintBackRefs{\CurrentBib}

\bibitem [\protect \citeauthoryear {%
Fornell%
\ \BBA {} Larcker%
}{%
Fornell%
\ \BBA {} Larcker%
}{%
{\protect \APACyear {1981}}%
}]{%
Fornell1981EvaluatingError}
\APACinsertmetastar {%
Fornell1981EvaluatingError}%
\begin{APACrefauthors}%
Fornell, C.%
\BCBT {}\ \BBA {} Larcker, D\BPBI F.%
\end{APACrefauthors}%
\unskip\
\newblock
\APACrefYearMonthDay{1981}{}{}.
\newblock
{\BBOQ}\APACrefatitle {{Evaluating structural equation models with unobservable
  variables and measurement error}} {{Evaluating structural equation models
  with unobservable variables and measurement error}}.{\BBCQ}
\newblock
\APACjournalVolNumPages{Journal of Marketing Research}{18}{}{39--50}.
\newblock
\begin{APACrefDOI} 10.1177/002224378101800104 \end{APACrefDOI}
\PrintBackRefs{\CurrentBib}

\bibitem [\protect \citeauthoryear {%
Gao%
, Zhang%
, Guan%
, Feng%
\BCBL {}\ \BBA {} Mardani%
}{%
Gao%
\ \protect \BOthers {.}}{%
{\protect \APACyear {2023}}%
}]{%
Gao2023TheTransformation}
\APACinsertmetastar {%
Gao2023TheTransformation}%
\begin{APACrefauthors}%
Gao, J.%
, Zhang, W.%
, Guan, T.%
, Feng, Q.%
\BCBL {}\ \BBA {} Mardani, A.%
\end{APACrefauthors}%
\unskip\
\newblock
\APACrefYearMonthDay{2023}{}{}.
\newblock
{\BBOQ}\APACrefatitle {{The effect of manufacturing agent heterogeneity on
  enterprise innovation performance and competitive advantage in the era of
  digital transformation}} {{The effect of manufacturing agent heterogeneity on
  enterprise innovation performance and competitive advantage in the era of
  digital transformation}}.{\BBCQ}
\newblock
\APACjournalVolNumPages{Journal of Business Research}{155}{}{113387}.
\newblock
\begin{APACrefDOI} 10.1016/j.jbusres.2022.113387 \end{APACrefDOI}
\PrintBackRefs{\CurrentBib}

\bibitem [\protect \citeauthoryear {%
Grace%
\ \BBA {} Bollen%
}{%
Grace%
\ \BBA {} Bollen%
}{%
{\protect \APACyear {2008}}%
}]{%
Grace2008RepresentingVariables}
\APACinsertmetastar {%
Grace2008RepresentingVariables}%
\begin{APACrefauthors}%
Grace, J\BPBI B.%
\BCBT {}\ \BBA {} Bollen, K\BPBI A.%
\end{APACrefauthors}%
\unskip\
\newblock
\APACrefYearMonthDay{2008}{}{}.
\newblock
{\BBOQ}\APACrefatitle {{Representing general theoretical concepts in structural
  equation models: The role of composite variables}} {{Representing general
  theoretical concepts in structural equation models: The role of composite
  variables}}.{\BBCQ}
\newblock
\APACjournalVolNumPages{Environmental and Ecological
  Statistics}{15}{}{191--213}.
\newblock
\begin{APACrefDOI} 10.1007/s10651-007-0047-7 \end{APACrefDOI}
\PrintBackRefs{\CurrentBib}

\bibitem [\protect \citeauthoryear {%
Grotzinger%
\ \protect \BOthers {.}}{%
Grotzinger%
\ \protect \BOthers {.}}{%
{\protect \APACyear {2019}}%
}]{%
Grotzinger2019}
\APACinsertmetastar {%
Grotzinger2019}%
\begin{APACrefauthors}%
Grotzinger, A\BPBI D.%
, Rhemtulla, M.%
, de Vlaming, R.%
, Ritchie, S\BPBI J.%
, Mallard, T\BPBI T.%
, Hill, W\BPBI D.%
\BDBL {}Tucker-Drob, E\BPBI M.%
\end{APACrefauthors}%
\unskip\
\newblock
\APACrefYearMonthDay{2019}{}{}.
\newblock
{\BBOQ}\APACrefatitle {{Genomic structural equation modelling provides insights
  into the multivariate genetic architecture of complex traits}} {{Genomic
  structural equation modelling provides insights into the multivariate genetic
  architecture of complex traits}}.{\BBCQ}
\newblock
\APACjournalVolNumPages{Nature Human Behaviour}{3}{}{513--525}.
\newblock
\begin{APACrefDOI} 10.1038/s41562-019-0566-x \end{APACrefDOI}
\PrintBackRefs{\CurrentBib}

\bibitem [\protect \citeauthoryear {%
G{\"{u}}rb{\"{u}}z%
, Bakker%
, Demerouti%
\BCBL {}\ \BBA {} Brouwers%
}{%
G{\"{u}}rb{\"{u}}z%
\ \protect \BOthers {.}}{%
{\protect \APACyear {2023}}%
}]{%
Gurbuz2023SustainableStudy}
\APACinsertmetastar {%
Gurbuz2023SustainableStudy}%
\begin{APACrefauthors}%
G{\"{u}}rb{\"{u}}z, S.%
, Bakker, A\BPBI B.%
, Demerouti, E.%
\BCBL {}\ \BBA {} Brouwers, E\BPBI P\BPBI M.%
\end{APACrefauthors}%
\unskip\
\newblock
\APACrefYearMonthDay{2023}{}{}.
\newblock
{\BBOQ}\APACrefatitle {{Sustainable employability and work engagement: A
  three-wave study}} {{Sustainable employability and work engagement: A
  three-wave study}}.{\BBCQ}
\newblock
\APACjournalVolNumPages{Frontiers in Psychology}{14}{}{1188728}.
\newblock
\begin{APACrefDOI} 10.3389/fpsyg.2023.1188728 \end{APACrefDOI}
\PrintBackRefs{\CurrentBib}

\bibitem [\protect \citeauthoryear {%
Hair%
\ \protect \BOthers {.}}{%
Hair%
\ \protect \BOthers {.}}{%
{\protect \APACyear {2021}}%
}]{%
Hair2021EvaluationModels}
\APACinsertmetastar {%
Hair2021EvaluationModels}%
\begin{APACrefauthors}%
Hair, J\BPBI F.%
, Hult, G\BPBI T\BPBI M.%
, Ringle, C\BPBI M.%
, Sarstedt, M.%
, Danks, N\BPBI P.%
\BCBL {}\ \BBA {} Ray, S.%
\end{APACrefauthors}%
\unskip\
\newblock
\APACrefYearMonthDay{2021}{}{}.
\newblock
{\BBOQ}\APACrefatitle {{Evaluation of formative measurement models}}
  {{Evaluation of formative measurement models}}.{\BBCQ}
\newblock
\BIn{} \APACrefbtitle {Partial Least Squares Structural Equation Modeling
  (PLS-SEM) Using R} {Partial least squares structural equation modeling
  (pls-sem) using r}\ (\BPGS\ 91--113).
\newblock
\APACaddressPublisher{}{Springer, Cham}.
\newblock
\begin{APACrefDOI} 10.1007/978-3-030-80519-7{\_}5 \end{APACrefDOI}
\PrintBackRefs{\CurrentBib}

\bibitem [\protect \citeauthoryear {%
Hair%
, Hult%
, Ringle%
, Sarstedt%
\BCBL {}\ \BBA {} Thiele%
}{%
Hair%
\ \protect \BOthers {.}}{%
{\protect \APACyear {2017}}%
}]{%
Hair2017}
\APACinsertmetastar {%
Hair2017}%
\begin{APACrefauthors}%
Hair, J\BPBI F.%
, Hult, G\BPBI T\BPBI M.%
, Ringle, C\BPBI M.%
, Sarstedt, M.%
\BCBL {}\ \BBA {} Thiele, K\BPBI O.%
\end{APACrefauthors}%
\unskip\
\newblock
\APACrefYearMonthDay{2017}{}{}.
\newblock
{\BBOQ}\APACrefatitle {{Mirror, mirror on the wall: A comparative evaluation of
  composite-based structural equation modeling methods}} {{Mirror, mirror on
  the wall: A comparative evaluation of composite-based structural equation
  modeling methods}}.{\BBCQ}
\newblock
\APACjournalVolNumPages{Journal of the Academy of Marketing
  Science}{45}{}{616--632}.
\newblock
\begin{APACrefDOI} 10.1007/s11747-017-0517-x \end{APACrefDOI}
\PrintBackRefs{\CurrentBib}

\bibitem [\protect \citeauthoryear {%
Hair%
, Risher%
, Sarstedt%
\BCBL {}\ \BBA {} Ringle%
}{%
Hair%
\ \protect \BOthers {.}}{%
{\protect \APACyear {2019}}%
}]{%
Hair2019WhenPLS-SEM}
\APACinsertmetastar {%
Hair2019WhenPLS-SEM}%
\begin{APACrefauthors}%
Hair, J\BPBI F.%
, Risher, J\BPBI J.%
, Sarstedt, M.%
\BCBL {}\ \BBA {} Ringle, C\BPBI M.%
\end{APACrefauthors}%
\unskip\
\newblock
\APACrefYearMonthDay{2019}{}{}.
\newblock
{\BBOQ}\APACrefatitle {{When to use and how to report the results of PLS-SEM}}
  {{When to use and how to report the results of PLS-SEM}}.{\BBCQ}
\newblock
\APACjournalVolNumPages{European Business Review}{31}{}{2--24}.
\newblock
\begin{APACrefDOI} 10.1108/EBR-11-2018-0203 \end{APACrefDOI}
\PrintBackRefs{\CurrentBib}

\bibitem [\protect \citeauthoryear {%
Hair%
, Sarstedt%
, Ringle%
\BCBL {}\ \BBA {} Mena%
}{%
Hair%
\ \protect \BOthers {.}}{%
{\protect \APACyear {2012}}%
}]{%
Hair2012AnResearch}
\APACinsertmetastar {%
Hair2012AnResearch}%
\begin{APACrefauthors}%
Hair, J\BPBI F.%
, Sarstedt, M.%
, Ringle, C\BPBI M.%
\BCBL {}\ \BBA {} Mena, J\BPBI A.%
\end{APACrefauthors}%
\unskip\
\newblock
\APACrefYearMonthDay{2012}{}{}.
\newblock
{\BBOQ}\APACrefatitle {{An assessment of the use of partial least squares
  structural equation modeling in marketing research}} {{An assessment of the
  use of partial least squares structural equation modeling in marketing
  research}}.{\BBCQ}
\newblock
\APACjournalVolNumPages{Journal of the Academy of Marketing
  Science}{40}{}{414--433}.
\newblock
\begin{APACrefDOI} 10.1007/s11747-011-0261-6 \end{APACrefDOI}
\PrintBackRefs{\CurrentBib}

\bibitem [\protect \citeauthoryear {%
Henseler%
}{%
Henseler%
}{%
{\protect \APACyear {2020}}%
}]{%
Henseler2021}
\APACinsertmetastar {%
Henseler2021}%
\begin{APACrefauthors}%
Henseler, J.%
\end{APACrefauthors}%
\unskip\
\newblock
\APACrefYear{2020}.
\newblock
\APACrefbtitle {{Composite-based structural equation modeling: Analyzing latent
  and emergent variables}} {{Composite-based structural equation modeling:
  Analyzing latent and emergent variables}}.
\newblock
\APACaddressPublisher{}{Guilford Publications}.
\PrintBackRefs{\CurrentBib}

\bibitem [\protect \citeauthoryear {%
Hitchcock%
\ \BBA {} R{\'{e}}dei%
}{%
Hitchcock%
\ \BBA {} R{\'{e}}dei%
}{%
{\protect \APACyear {2021}}%
}]{%
Hitchcock2021ReichenbachsPrinciple}
\APACinsertmetastar {%
Hitchcock2021ReichenbachsPrinciple}%
\begin{APACrefauthors}%
Hitchcock, C.%
\BCBT {}\ \BBA {} R{\'{e}}dei, M.%
\end{APACrefauthors}%
\unskip\
\newblock
\APACrefYearMonthDay{2021}{}{}.
\newblock
{\BBOQ}\APACrefatitle {{Reichenbach’s common cause principle}}
  {{Reichenbach’s common cause principle}}.{\BBCQ}
\newblock
\BIn{} E\BPBI N.~Zalta\ (\BED), \APACrefbtitle {The Stanford Encyclopedia of
  Philosophy} {The stanford encyclopedia of philosophy}\ (\BPGS\ 158--159).
\newblock
\APACaddressPublisher{}{Metaphysics Research Lab, Stanford University}.
\PrintBackRefs{\CurrentBib}

\bibitem [\protect \citeauthoryear {%
Hu%
\ \BBA {} Bentler%
}{%
Hu%
\ \BBA {} Bentler%
}{%
{\protect \APACyear {1998}}%
}]{%
Hu1998FitMisspecification}
\APACinsertmetastar {%
Hu1998FitMisspecification}%
\begin{APACrefauthors}%
Hu, L\BPBI T.%
\BCBT {}\ \BBA {} Bentler, P\BPBI M.%
\end{APACrefauthors}%
\unskip\
\newblock
\APACrefYearMonthDay{1998}{}{}.
\newblock
{\BBOQ}\APACrefatitle {{Fit indices in covariance structure modeling:
  Sensitivity to underparameterized model misspecification}} {{Fit indices in
  covariance structure modeling: Sensitivity to underparameterized model
  misspecification}}.{\BBCQ}
\newblock
\APACjournalVolNumPages{Psychological Methods}{3}{}{424}.
\newblock
\begin{APACrefDOI} 10.1037/1082-989X.3.4.424 \end{APACrefDOI}
\PrintBackRefs{\CurrentBib}

\bibitem [\protect \citeauthoryear {%
Hwang%
\ \protect \BOthers {.}}{%
Hwang%
\ \protect \BOthers {.}}{%
{\protect \APACyear {2021}}%
}]{%
Hwang2021AnAnalysis}
\APACinsertmetastar {%
Hwang2021AnAnalysis}%
\begin{APACrefauthors}%
Hwang, H.%
, Cho, G.%
, Jung, K.%
, Falk, C\BPBI F.%
, Flake, J\BPBI K.%
, Jin, M\BPBI J.%
\BCBL {}\ \BBA {} Lee, S\BPBI H.%
\end{APACrefauthors}%
\unskip\
\newblock
\APACrefYearMonthDay{2021}{}{}.
\newblock
{\BBOQ}\APACrefatitle {{An approach to structural equation modeling with both
  factors and components: Integrated generalized structured component
  analysis.}} {{An approach to structural equation modeling with both factors
  and components: Integrated generalized structured component
  analysis.}}{\BBCQ}
\newblock
\APACjournalVolNumPages{Psychological Methods}{26}{}{273--294}.
\newblock
\begin{APACrefDOI} 10.1037/met0000336 \end{APACrefDOI}
\PrintBackRefs{\CurrentBib}

\bibitem [\protect \citeauthoryear {%
Hwang%
, Malhotra%
, Kim%
, Tomiuk%
\BCBL {}\ \BBA {} Hong%
}{%
Hwang%
\ \protect \BOthers {.}}{%
{\protect \APACyear {2010}}%
}]{%
Hwang2010AModeling}
\APACinsertmetastar {%
Hwang2010AModeling}%
\begin{APACrefauthors}%
Hwang, H.%
, Malhotra, N\BPBI K.%
, Kim, Y.%
, Tomiuk, M\BPBI A.%
\BCBL {}\ \BBA {} Hong, S.%
\end{APACrefauthors}%
\unskip\
\newblock
\APACrefYearMonthDay{2010}{}{}.
\newblock
{\BBOQ}\APACrefatitle {{A comparative study on parameter recovery of three
  approaches to structural equation modeling}} {{A comparative study on
  parameter recovery of three approaches to structural equation
  modeling}}.{\BBCQ}
\newblock
\APACjournalVolNumPages{Journal of Marketing Research}{47}{}{699--712}.
\newblock
\begin{APACrefDOI} 10.1509/jmkr.47.4.699 \end{APACrefDOI}
\PrintBackRefs{\CurrentBib}

\bibitem [\protect \citeauthoryear {%
Hwang%
\ \BBA {} Takane%
}{%
Hwang%
\ \BBA {} Takane%
}{%
{\protect \APACyear {2004}}%
}]{%
Hwang2004GeneralizedAnalysis}
\APACinsertmetastar {%
Hwang2004GeneralizedAnalysis}%
\begin{APACrefauthors}%
Hwang, H.%
\BCBT {}\ \BBA {} Takane, Y.%
\end{APACrefauthors}%
\unskip\
\newblock
\APACrefYearMonthDay{2004}{}{}.
\newblock
{\BBOQ}\APACrefatitle {{Generalized structured component analysis}}
  {{Generalized structured component analysis}}.{\BBCQ}
\newblock
\APACjournalVolNumPages{Psychometrika}{69}{}{81--99}.
\newblock
\begin{APACrefDOI} 10.1007/BF02295841 \end{APACrefDOI}
\PrintBackRefs{\CurrentBib}

\bibitem [\protect \citeauthoryear {%
Jarvis%
, MacKenzie%
\BCBL {}\ \BBA {} Podsakoff%
}{%
Jarvis%
\ \protect \BOthers {.}}{%
{\protect \APACyear {2003}}%
}]{%
Jarvis2003AResearch}
\APACinsertmetastar {%
Jarvis2003AResearch}%
\begin{APACrefauthors}%
Jarvis, C.%
, MacKenzie, S.%
\BCBL {}\ \BBA {} Podsakoff, P.%
\end{APACrefauthors}%
\unskip\
\newblock
\APACrefYearMonthDay{2003}{}{}.
\newblock
{\BBOQ}\APACrefatitle {{A critical review of construct indicators and
  measurement model misspecification in marketing and consumer research}} {{A
  critical review of construct indicators and measurement model
  misspecification in marketing and consumer research}}.{\BBCQ}
\newblock
\APACjournalVolNumPages{Journal of Consumer Research}{30}{}{199--218}.
\newblock
\begin{APACrefDOI} 10.1086/376806 \end{APACrefDOI}
\PrintBackRefs{\CurrentBib}

\bibitem [\protect \citeauthoryear {%
J{\"{o}}reskog%
}{%
J{\"{o}}reskog%
}{%
{\protect \APACyear {1969}}%
}]{%
Joreskog1969AAnalysis}
\APACinsertmetastar {%
Joreskog1969AAnalysis}%
\begin{APACrefauthors}%
J{\"{o}}reskog, K\BPBI G.%
\end{APACrefauthors}%
\unskip\
\newblock
\APACrefYearMonthDay{1969}{}{}.
\newblock
{\BBOQ}\APACrefatitle {{A general approach to confirmatory maximum likelihood
  factor analysis}} {{A general approach to confirmatory maximum likelihood
  factor analysis}}.{\BBCQ}
\newblock
\APACjournalVolNumPages{Psychometrika}{34}{}{183--202}.
\newblock
\begin{APACrefDOI} 10.1007/BF02289343 \end{APACrefDOI}
\PrintBackRefs{\CurrentBib}

\bibitem [\protect \citeauthoryear {%
J{\"{o}}reskog%
}{%
J{\"{o}}reskog%
}{%
{\protect \APACyear {1970}}%
}]{%
Joreskog1970a}
\APACinsertmetastar {%
Joreskog1970a}%
\begin{APACrefauthors}%
J{\"{o}}reskog, K\BPBI G.%
\end{APACrefauthors}%
\unskip\
\newblock
\APACrefYearMonthDay{1970}{}{}.
\newblock
{\BBOQ}\APACrefatitle {{A general method for analysis of covariance
  structures}} {{A general method for analysis of covariance
  structures}}.{\BBCQ}
\newblock
\APACjournalVolNumPages{Biometrika}{57}{}{239--251}.
\newblock
\begin{APACrefDOI} 10.1093/biomet/57.2.239 \end{APACrefDOI}
\PrintBackRefs{\CurrentBib}

\bibitem [\protect \citeauthoryear {%
J{\"{o}}reskog%
\ \BBA {} Goldberger%
}{%
J{\"{o}}reskog%
\ \BBA {} Goldberger%
}{%
{\protect \APACyear {1975}}%
}]{%
Joreskog1975EstimationVariable}
\APACinsertmetastar {%
Joreskog1975EstimationVariable}%
\begin{APACrefauthors}%
J{\"{o}}reskog, K\BPBI G.%
\BCBT {}\ \BBA {} Goldberger, A\BPBI S.%
\end{APACrefauthors}%
\unskip\
\newblock
\APACrefYearMonthDay{1975}{}{}.
\newblock
{\BBOQ}\APACrefatitle {{Estimation of a model with multiple indicators and
  multiple causes of a single latent variable}} {{Estimation of a model with
  multiple indicators and multiple causes of a single latent variable}}.{\BBCQ}
\newblock
\APACjournalVolNumPages{Journal of the American Statistical
  Association}{70}{}{631--639}.
\newblock
\begin{APACrefDOI} 10.1080/01621459.1975.10482485 \end{APACrefDOI}
\PrintBackRefs{\CurrentBib}

\bibitem [\protect \citeauthoryear {%
J{\"{o}}reskog%
\ \BBA {} S{\"{o}}rbom%
}{%
J{\"{o}}reskog%
\ \BBA {} S{\"{o}}rbom%
}{%
{\protect \APACyear {2018}}%
}]{%
Joreskog2018LISRELWindows}
\APACinsertmetastar {%
Joreskog2018LISRELWindows}%
\begin{APACrefauthors}%
J{\"{o}}reskog, K\BPBI G.%
\BCBT {}\ \BBA {} S{\"{o}}rbom, D.%
\end{APACrefauthors}%
\unskip\
\newblock
\APACrefYearMonthDay{2018}{}{}.
\newblock
\APACrefbtitle {{LISREL 10 for windows}.} {{LISREL 10 for windows}.}
\newblock
\APACaddressPublisher{}{Scientific software international}.
\PrintBackRefs{\CurrentBib}

\bibitem [\protect \citeauthoryear {%
J{\"{o}}reskog%
\ \BBA {} Yang%
}{%
J{\"{o}}reskog%
\ \BBA {} Yang%
}{%
{\protect \APACyear {1996}}%
}]{%
Joreskog1996NonlinearEffects}
\APACinsertmetastar {%
Joreskog1996NonlinearEffects}%
\begin{APACrefauthors}%
J{\"{o}}reskog, K\BPBI G.%
\BCBT {}\ \BBA {} Yang, F.%
\end{APACrefauthors}%
\unskip\
\newblock
\APACrefYearMonthDay{1996}{}{}.
\newblock
{\BBOQ}\APACrefatitle {{Nonlinear structural equation models: The Kenny-Judd
  model with interaction effects}} {{Nonlinear structural equation models: The
  Kenny-Judd model with interaction effects}}.{\BBCQ}
\newblock
\BIn{} \APACrefbtitle {Advanced structural equation modeling.} {Advanced
  structural equation modeling.}\ (\BPGS\ 57--58).
\newblock
\APACaddressPublisher{}{Psychology Press}.
\PrintBackRefs{\CurrentBib}

\bibitem [\protect \citeauthoryear {%
Kaplan%
\ \BBA {} Depaoli%
}{%
Kaplan%
\ \BBA {} Depaoli%
}{%
{\protect \APACyear {2013}}%
}]{%
Kaplan2013BayesianMethods}
\APACinsertmetastar {%
Kaplan2013BayesianMethods}%
\begin{APACrefauthors}%
Kaplan, D.%
\BCBT {}\ \BBA {} Depaoli, S.%
\end{APACrefauthors}%
\unskip\
\newblock
\APACrefYearMonthDay{2013}{}{}.
\newblock
{\BBOQ}\APACrefatitle {{Bayesian statistical methods}} {{Bayesian statistical
  methods}}.{\BBCQ}
\newblock
\BIn{} \APACrefbtitle {The Oxford Handbook of Quantitative Methods} {The oxford
  handbook of quantitative methods}\ (\BVOL\ 1: Foundations).
\newblock
\begin{APACrefDOI} 10.1093/oxfordhb/9780199934874.013.0020 \end{APACrefDOI}
\PrintBackRefs{\CurrentBib}

\bibitem [\protect \citeauthoryear {%
Khatri%
\ \BBA {} Gupta%
}{%
Khatri%
\ \BBA {} Gupta%
}{%
{\protect \APACyear {2019}}%
}]{%
Khatri2019DevelopmentModel}
\APACinsertmetastar {%
Khatri2019DevelopmentModel}%
\begin{APACrefauthors}%
Khatri, P.%
\BCBT {}\ \BBA {} Gupta, P.%
\end{APACrefauthors}%
\unskip\
\newblock
\APACrefYearMonthDay{2019}{}{}.
\newblock
{\BBOQ}\APACrefatitle {{Development and validation of employee wellbeing scale
  – a formative measurement model}} {{Development and validation of employee
  wellbeing scale – a formative measurement model}}.{\BBCQ}
\newblock
\APACjournalVolNumPages{International Journal of Workplace Health
  Management}{12}{}{352--368}.
\newblock
\begin{APACrefDOI} 10.1108/IJWHM-12-2018-0161 \end{APACrefDOI}
\PrintBackRefs{\CurrentBib}

\bibitem [\protect \citeauthoryear {%
Kleck%
, Tark%
\BCBL {}\ \BBA {} Bellows%
}{%
Kleck%
\ \protect \BOthers {.}}{%
{\protect \APACyear {2006}}%
}]{%
Kleck2006WhatJustice}
\APACinsertmetastar {%
Kleck2006WhatJustice}%
\begin{APACrefauthors}%
Kleck, G.%
, Tark, J.%
\BCBL {}\ \BBA {} Bellows, J\BPBI J.%
\end{APACrefauthors}%
\unskip\
\newblock
\APACrefYearMonthDay{2006}{}{}.
\newblock
{\BBOQ}\APACrefatitle {{What methods are most frequently used in research in
  criminology and criminal justice?}} {{What methods are most frequently used
  in research in criminology and criminal justice?}}{\BBCQ}
\newblock
\APACjournalVolNumPages{Journal of Criminal Justice}{34}{}{147--152}.
\newblock
\begin{APACrefDOI} 10.1016/j.jcrimjus.2006.01.007 \end{APACrefDOI}
\PrintBackRefs{\CurrentBib}

\bibitem [\protect \citeauthoryear {%
Kline%
}{%
Kline%
}{%
{\protect \APACyear {2023}}%
}]{%
Kline2023}
\APACinsertmetastar {%
Kline2023}%
\begin{APACrefauthors}%
Kline, R\BPBI B.%
\end{APACrefauthors}%
\unskip\
\newblock
\APACrefYear{2023}.
\newblock
\APACrefbtitle {{Principles and practice of structural equation modeling (5th
  edition)}} {{Principles and practice of structural equation modeling (5th
  edition)}}.
\newblock
\APACaddressPublisher{}{Guilford publications}.
\PrintBackRefs{\CurrentBib}

\bibitem [\protect \citeauthoryear {%
Kroonenberg%
\ \BBA {} {Pieter M}%
}{%
Kroonenberg%
\ \BBA {} {Pieter M}%
}{%
{\protect \APACyear {1990}}%
}]{%
Kroonenberg1990LatentSquares}
\APACinsertmetastar {%
Kroonenberg1990LatentSquares}%
\begin{APACrefauthors}%
Kroonenberg, P\BPBI M.%
\BCBT {}\ \BBA {} {Pieter M}.%
\end{APACrefauthors}%
\unskip\
\newblock
\APACrefYearMonthDay{1990}{}{}.
\newblock
{\BBOQ}\APACrefatitle {{Latent variable path modeling with partial least
  squares}} {{Latent variable path modeling with partial least
  squares}}.{\BBCQ}
\newblock
\APACjournalVolNumPages{Journal of the American Statistical
  Association}{85}{}{909--911}.
\newblock
\begin{APACrefDOI} 10.2307/2290049 \end{APACrefDOI}
\PrintBackRefs{\CurrentBib}

\bibitem [\protect \citeauthoryear {%
Liu%
, Schuberth%
, Liu%
\BCBL {}\ \BBA {} Henseler%
}{%
Liu%
\ \protect \BOthers {.}}{%
{\protect \APACyear {2022}}%
}]{%
Liu2022ModelingAnalysis}
\APACinsertmetastar {%
Liu2022ModelingAnalysis}%
\begin{APACrefauthors}%
Liu, Y.%
, Schuberth, F.%
, Liu, Y.%
\BCBL {}\ \BBA {} Henseler, J.%
\end{APACrefauthors}%
\unskip\
\newblock
\APACrefYearMonthDay{2022}{}{}.
\newblock
{\BBOQ}\APACrefatitle {{Modeling and assessing forged concepts in tourism and
  hospitality using confirmatory composite analysis}} {{Modeling and assessing
  forged concepts in tourism and hospitality using confirmatory composite
  analysis}}.{\BBCQ}
\newblock
\APACjournalVolNumPages{Journal of Business Research}{152}{}{221--230}.
\newblock
\begin{APACrefDOI} 10.1016/j.jbusres.2022.07.040 \end{APACrefDOI}
\PrintBackRefs{\CurrentBib}

\bibitem [\protect \citeauthoryear {%
MacKenzie%
, Podsakoff%
\BCBL {}\ \BBA {} Jarvis%
}{%
MacKenzie%
\ \protect \BOthers {.}}{%
{\protect \APACyear {2005}}%
}]{%
MacKenzie2005TheSolutions}
\APACinsertmetastar {%
MacKenzie2005TheSolutions}%
\begin{APACrefauthors}%
MacKenzie, S\BPBI B.%
, Podsakoff, P\BPBI M.%
\BCBL {}\ \BBA {} Jarvis, C\BPBI B.%
\end{APACrefauthors}%
\unskip\
\newblock
\APACrefYearMonthDay{2005}{}{}.
\newblock
{\BBOQ}\APACrefatitle {{The problem of measurement model misspecification in
  behavioral and organizational research and some recommended solutions}} {{The
  problem of measurement model misspecification in behavioral and
  organizational research and some recommended solutions}}.{\BBCQ}
\newblock
\APACjournalVolNumPages{Journal of Applied Psychology}{90}{}{710--730}.
\newblock
\begin{APACrefDOI} 10.1037/0021-9010.90.4.710 \end{APACrefDOI}
\PrintBackRefs{\CurrentBib}

\bibitem [\protect \citeauthoryear {%
Martens%
}{%
Martens%
}{%
{\protect \APACyear {2005}}%
}]{%
Martens2005TheResearch}
\APACinsertmetastar {%
Martens2005TheResearch}%
\begin{APACrefauthors}%
Martens, M\BPBI P.%
\end{APACrefauthors}%
\unskip\
\newblock
\APACrefYearMonthDay{2005}{}{}.
\newblock
{\BBOQ}\APACrefatitle {{The use of structural equation modeling in counseling
  psychology research}} {{The use of structural equation modeling in counseling
  psychology research}}.{\BBCQ}
\newblock
\APACjournalVolNumPages{The Counseling Psychologist}{33}{}{269--298}.
\newblock
\begin{APACrefDOI} 10.1177/0011000004272260 \end{APACrefDOI}
\PrintBackRefs{\CurrentBib}

\bibitem [\protect \citeauthoryear {%
Mayer%
, Steyer%
\BCBL {}\ \BBA {} Mueller%
}{%
Mayer%
\ \protect \BOthers {.}}{%
{\protect \APACyear {2012}}%
}]{%
Mayer2012AComponents}
\APACinsertmetastar {%
Mayer2012AComponents}%
\begin{APACrefauthors}%
Mayer, A.%
, Steyer, R.%
\BCBL {}\ \BBA {} Mueller, H.%
\end{APACrefauthors}%
\unskip\
\newblock
\APACrefYearMonthDay{2012}{}{}.
\newblock
{\BBOQ}\APACrefatitle {{A general approach to defining latent growth
  components}} {{A general approach to defining latent growth
  components}}.{\BBCQ}
\newblock
\APACjournalVolNumPages{Structural Equation Modeling: A Multidisciplinary
  Journal}{19}{}{513--533}.
\newblock
\begin{APACrefDOI} 10.1080/10705511.2012.713242 \end{APACrefDOI}
\PrintBackRefs{\CurrentBib}

\bibitem [\protect \citeauthoryear {%
McDonald%
}{%
McDonald%
}{%
{\protect \APACyear {1996}}%
}]{%
McDonald1996PathVariables}
\APACinsertmetastar {%
McDonald1996PathVariables}%
\begin{APACrefauthors}%
McDonald, R\BPBI P.%
\end{APACrefauthors}%
\unskip\
\newblock
\APACrefYearMonthDay{1996}{}{}.
\newblock
{\BBOQ}\APACrefatitle {{Path analysis with composite variables}} {{Path
  analysis with composite variables}}.{\BBCQ}
\newblock
\APACjournalVolNumPages{Multivariate Behavioral Research}{31}{}{239--270}.
\newblock
\begin{APACrefDOI} 10.1207/s15327906mbr3102{\_}5 \end{APACrefDOI}
\PrintBackRefs{\CurrentBib}

\bibitem [\protect \citeauthoryear {%
Millsap%
\ \BBA {} Yun-Tein%
}{%
Millsap%
\ \BBA {} Yun-Tein%
}{%
{\protect \APACyear {2004}}%
}]{%
Millsap2004AssessingMeasures}
\APACinsertmetastar {%
Millsap2004AssessingMeasures}%
\begin{APACrefauthors}%
Millsap, R\BPBI E.%
\BCBT {}\ \BBA {} Yun-Tein, J.%
\end{APACrefauthors}%
\unskip\
\newblock
\APACrefYearMonthDay{2004}{}{}.
\newblock
{\BBOQ}\APACrefatitle {{Assessing factorial invariance in ordered-categorical
  measures}} {{Assessing factorial invariance in ordered-categorical
  measures}}.{\BBCQ}
\newblock
\APACjournalVolNumPages{Multivariate Behavioral Research}{39}{}{479--515}.
\newblock
\begin{APACrefDOI} 10.1207/S15327906MBR3903{\_}4 \end{APACrefDOI}
\PrintBackRefs{\CurrentBib}

\bibitem [\protect \citeauthoryear {%
Mustillo%
, Li%
\BCBL {}\ \BBA {} Ferraro%
}{%
Mustillo%
\ \protect \BOthers {.}}{%
{\protect \APACyear {2021}}%
}]{%
Mustillo2021EvaluatingApproach}
\APACinsertmetastar {%
Mustillo2021EvaluatingApproach}%
\begin{APACrefauthors}%
Mustillo, S.%
, Li, M.%
\BCBL {}\ \BBA {} Ferraro, K\BPBI F.%
\end{APACrefauthors}%
\unskip\
\newblock
\APACrefYearMonthDay{2021}{}{}.
\newblock
{\BBOQ}\APACrefatitle {{Evaluating the cumulative impact of childhood
  misfortune: A structural equation modeling approach}} {{Evaluating the
  cumulative impact of childhood misfortune: A structural equation modeling
  approach}}.{\BBCQ}
\newblock
\APACjournalVolNumPages{Sociological Methods and Research}{50}{}{1073--1109}.
\newblock
\begin{APACrefDOI} 10.1177/0049124119875957 \end{APACrefDOI}
\PrintBackRefs{\CurrentBib}

\bibitem [\protect \citeauthoryear {%
Nascimento%
\ \BBA {} Maria Correia~Loureiro%
}{%
Nascimento%
\ \BBA {} Maria Correia~Loureiro%
}{%
{\protect \APACyear {2024}}%
}]{%
Nascimento2024}
\APACinsertmetastar {%
Nascimento2024}%
\begin{APACrefauthors}%
Nascimento, J.%
\BCBT {}\ \BBA {} Maria Correia~Loureiro, S.%
\end{APACrefauthors}%
\unskip\
\newblock
\APACrefYearMonthDay{2024}{}{}.
\newblock
{\BBOQ}\APACrefatitle {{Understanding the desire for green consumption: Norms,
  emotions, and attitudes}} {{Understanding the desire for green consumption:
  Norms, emotions, and attitudes}}.{\BBCQ}
\newblock
\APACjournalVolNumPages{Journal of Business Research}{178}{}{114675}.
\newblock
\begin{APACrefDOI} 10.1016/j.jbusres.2024.114675 \end{APACrefDOI}
\PrintBackRefs{\CurrentBib}

\bibitem [\protect \citeauthoryear {%
Nunkoo%
, Ramkissoon%
\BCBL {}\ \BBA {} Gursoy%
}{%
Nunkoo%
\ \protect \BOthers {.}}{%
{\protect \APACyear {2013}}%
}]{%
Nunkoo2013UseFuture}
\APACinsertmetastar {%
Nunkoo2013UseFuture}%
\begin{APACrefauthors}%
Nunkoo, R.%
, Ramkissoon, H.%
\BCBL {}\ \BBA {} Gursoy, D.%
\end{APACrefauthors}%
\unskip\
\newblock
\APACrefYearMonthDay{2013}{}{}.
\newblock
{\BBOQ}\APACrefatitle {{Use of structural equation modeling in tourism
  research: Past, present, and future}} {{Use of structural equation modeling
  in tourism research: Past, present, and future}}.{\BBCQ}
\newblock
\APACjournalVolNumPages{Journal of Travel Research}{52}{}{759--771}.
\newblock
\begin{APACrefDOI} 10.1177/0047287513478503 \end{APACrefDOI}
\PrintBackRefs{\CurrentBib}

\bibitem [\protect \citeauthoryear {%
Paxton%
, Curran%
, Bollen%
, Kirby%
\BCBL {}\ \BBA {} Chen%
}{%
Paxton%
\ \protect \BOthers {.}}{%
{\protect \APACyear {2001}}%
}]{%
Paxton2001MonteImplementation}
\APACinsertmetastar {%
Paxton2001MonteImplementation}%
\begin{APACrefauthors}%
Paxton, P.%
, Curran, P\BPBI J.%
, Bollen, K\BPBI A.%
, Kirby, J.%
\BCBL {}\ \BBA {} Chen, F.%
\end{APACrefauthors}%
\unskip\
\newblock
\APACrefYearMonthDay{2001}{}{}.
\newblock
{\BBOQ}\APACrefatitle {{Monte Carlo experiments: Design and implementation}}
  {{Monte Carlo experiments: Design and implementation}}.{\BBCQ}
\newblock
\APACjournalVolNumPages{Structural Equation Modeling: A Multidisciplinary
  Journal}{8}{}{287--312}.
\newblock
\begin{APACrefDOI} 10.1207/S15328007SEM0802{\_}7 \end{APACrefDOI}
\PrintBackRefs{\CurrentBib}

\bibitem [\protect \citeauthoryear {%
Petter%
, Straub%
\BCBL {}\ \BBA {} Rai%
}{%
Petter%
\ \protect \BOthers {.}}{%
{\protect \APACyear {2007}}%
}]{%
Petter2007SpecifyingResearch}
\APACinsertmetastar {%
Petter2007SpecifyingResearch}%
\begin{APACrefauthors}%
Petter, S.%
, Straub, D.%
\BCBL {}\ \BBA {} Rai, A.%
\end{APACrefauthors}%
\unskip\
\newblock
\APACrefYearMonthDay{2007}{}{}.
\newblock
{\BBOQ}\APACrefatitle {{Specifying formative constructs in information systems
  research}} {{Specifying formative constructs in information systems
  research}}.{\BBCQ}
\newblock
\APACjournalVolNumPages{MIS Quarterly}{31}{}{623--656}.
\newblock
\begin{APACrefDOI} 10.2307/25148814 \end{APACrefDOI}
\PrintBackRefs{\CurrentBib}

\bibitem [\protect \citeauthoryear {%
{R Core Team}%
}{%
{R Core Team}%
}{%
{\protect \APACyear {2024}}%
}]{%
RCoreTeam2024}
\APACinsertmetastar {%
RCoreTeam2024}%
\begin{APACrefauthors}%
{R Core Team}.%
\end{APACrefauthors}%
\unskip\
\newblock
\APACrefYearMonthDay{2024}{}{}.
\newblock
{\BBOQ}\APACrefatitle {{R: A language and environment for statistical
  computing.}} {{R: A language and environment for statistical
  computing.}}{\BBCQ}\ [\bibcomputersoftwaremanual].
\newblock
\APACaddressPublisher{Vienna, Austria}{}.
\PrintBackRefs{\CurrentBib}

\bibitem [\protect \citeauthoryear {%
Rademaker%
\ \BBA {} Schuberth%
}{%
Rademaker%
\ \BBA {} Schuberth%
}{%
{\protect \APACyear {2020}}%
}]{%
Rademaker2020}
\APACinsertmetastar {%
Rademaker2020}%
\begin{APACrefauthors}%
Rademaker, M\BPBI E.%
\BCBT {}\ \BBA {} Schuberth, F.%
\end{APACrefauthors}%
\unskip\
\newblock
\APACrefYearMonthDay{2020}{}{}.
\newblock
{\BBOQ}\APACrefatitle {{cSEM: Composite-based structural equation modeling}}
  {{cSEM: Composite-based structural equation modeling}}{\BBCQ}\
  [\bibcomputersoftwaremanual].
\PrintBackRefs{\CurrentBib}

\bibitem [\protect \citeauthoryear {%
Rahman%
, Langner%
\BCBL {}\ \BBA {} Temme%
}{%
Rahman%
\ \protect \BOthers {.}}{%
{\protect \APACyear {2021}}%
}]{%
Rahman2021BrandModel}
\APACinsertmetastar {%
Rahman2021BrandModel}%
\begin{APACrefauthors}%
Rahman, R.%
, Langner, T.%
\BCBL {}\ \BBA {} Temme, D.%
\end{APACrefauthors}%
\unskip\
\newblock
\APACrefYearMonthDay{2021}{}{}.
\newblock
{\BBOQ}\APACrefatitle {{Brand love: Conceptual and empirical investigation of a
  holistic causal model}} {{Brand love: Conceptual and empirical investigation
  of a holistic causal model}}.{\BBCQ}
\newblock
\APACjournalVolNumPages{Journal of Brand Management}{28}{}{609--642}.
\newblock
\begin{APACrefDOI} 10.1057/s41262-021-00237-7 \end{APACrefDOI}
\PrintBackRefs{\CurrentBib}

\bibitem [\protect \citeauthoryear {%
Rasheed%
, Hameed%
, Kaur%
\BCBL {}\ \BBA {} Dhir%
}{%
Rasheed%
\ \protect \BOthers {.}}{%
{\protect \APACyear {2024}}%
}]{%
Rasheed2024TooQuality}
\APACinsertmetastar {%
Rasheed2024TooQuality}%
\begin{APACrefauthors}%
Rasheed, M\BPBI I.%
, Hameed, Z.%
, Kaur, P.%
\BCBL {}\ \BBA {} Dhir, A.%
\end{APACrefauthors}%
\unskip\
\newblock
\APACrefYearMonthDay{2024}{}{}.
\newblock
{\BBOQ}\APACrefatitle {{Too sleepy to be innovative? Ethical leadership and
  employee service innovation behavior: A dual-path model moderated by sleep
  quality}} {{Too sleepy to be innovative? Ethical leadership and employee
  service innovation behavior: A dual-path model moderated by sleep
  quality}}.{\BBCQ}
\newblock
\APACjournalVolNumPages{Human Relations}{77}{}{739--767}.
\newblock
\begin{APACrefDOI} 10.1177/00187267231163040 \end{APACrefDOI}
\PrintBackRefs{\CurrentBib}

\bibitem [\protect \citeauthoryear {%
Reichenbach%
}{%
Reichenbach%
}{%
{\protect \APACyear {1956}}%
}]{%
Reichenbach1956TheTime.}
\APACinsertmetastar {%
Reichenbach1956TheTime.}%
\begin{APACrefauthors}%
Reichenbach, H.%
\end{APACrefauthors}%
\unskip\
\newblock
\APACrefYear{1956}.
\newblock
\APACrefbtitle {{The Direction of Time.}} {{The Direction of Time.}}\ (\BVOL~8;
  M.~Reichenbach, \BED{}).
\newblock
\APACaddressPublisher{}{Dover Publications}.
\PrintBackRefs{\CurrentBib}

\bibitem [\protect \citeauthoryear {%
Reinartz%
, Haenlein%
\BCBL {}\ \BBA {} Henseler%
}{%
Reinartz%
\ \protect \BOthers {.}}{%
{\protect \APACyear {2009}}%
}]{%
Reinartz2009AnSEM}
\APACinsertmetastar {%
Reinartz2009AnSEM}%
\begin{APACrefauthors}%
Reinartz, W.%
, Haenlein, M.%
\BCBL {}\ \BBA {} Henseler, J.%
\end{APACrefauthors}%
\unskip\
\newblock
\APACrefYearMonthDay{2009}{}{}.
\newblock
{\BBOQ}\APACrefatitle {{An empirical comparison of the efficacy of
  covariance-based and variance-based SEM}} {{An empirical comparison of the
  efficacy of covariance-based and variance-based SEM}}.{\BBCQ}
\newblock
\APACjournalVolNumPages{International Journal of Research in
  Marketing}{26}{}{332--344}.
\newblock
\begin{APACrefDOI} 10.1016/j.ijresmar.2009.08.001 \end{APACrefDOI}
\PrintBackRefs{\CurrentBib}

\bibitem [\protect \citeauthoryear {%
Rhemtulla%
, van Bork%
\BCBL {}\ \BBA {} Borsboom%
}{%
Rhemtulla%
\ \protect \BOthers {.}}{%
{\protect \APACyear {2020}}%
}]{%
Rhemtulla2020}
\APACinsertmetastar {%
Rhemtulla2020}%
\begin{APACrefauthors}%
Rhemtulla, M.%
, van Bork, R.%
\BCBL {}\ \BBA {} Borsboom, D.%
\end{APACrefauthors}%
\unskip\
\newblock
\APACrefYearMonthDay{2020}{}{}.
\newblock
{\BBOQ}\APACrefatitle {{Worse than measurement error: Consequences of
  inappropriate latent variable measurement models.}} {{Worse than measurement
  error: Consequences of inappropriate latent variable measurement
  models.}}{\BBCQ}
\newblock
\APACjournalVolNumPages{Psychological Methods}{25}{}{30--45}.
\newblock
\begin{APACrefDOI} 10.1037/met0000220 \end{APACrefDOI}
\PrintBackRefs{\CurrentBib}

\bibitem [\protect \citeauthoryear {%
Rigdon%
}{%
Rigdon%
}{%
{\protect \APACyear {2016}}%
}]{%
Rigdon2016}
\APACinsertmetastar {%
Rigdon2016}%
\begin{APACrefauthors}%
Rigdon, E\BPBI E.%
\end{APACrefauthors}%
\unskip\
\newblock
\APACrefYearMonthDay{2016}{}{}.
\newblock
{\BBOQ}\APACrefatitle {{Choosing PLS path modeling as analytical method in
  European management research: A realist perspective}} {{Choosing PLS path
  modeling as analytical method in European management research: A realist
  perspective}}.{\BBCQ}
\newblock
\APACjournalVolNumPages{European Management Journal}{34}{}{598--605}.
\newblock
\begin{APACrefDOI} 10.1016/j.emj.2016.05.006 \end{APACrefDOI}
\PrintBackRefs{\CurrentBib}

\bibitem [\protect \citeauthoryear {%
Rigdon%
, Sarstedt%
\BCBL {}\ \BBA {} Ringle%
}{%
Rigdon%
\ \protect \BOthers {.}}{%
{\protect \APACyear {2017}}%
}]{%
Rigdon2017OnRecommendations}
\APACinsertmetastar {%
Rigdon2017OnRecommendations}%
\begin{APACrefauthors}%
Rigdon, E\BPBI E.%
, Sarstedt, M.%
\BCBL {}\ \BBA {} Ringle, C\BPBI M.%
\end{APACrefauthors}%
\unskip\
\newblock
\APACrefYearMonthDay{2017}{}{}.
\newblock
{\BBOQ}\APACrefatitle {{On comparing results from CB-SEM and PLS-SEM: Five
  perspectives and five recommendations}} {{On comparing results from CB-SEM
  and PLS-SEM: Five perspectives and five recommendations}}.{\BBCQ}
\newblock
\APACjournalVolNumPages{Journal of Research and Management}{39}{}{4--16}.
\newblock
\begin{APACrefDOI} 10.2307/26426850 \end{APACrefDOI}
\PrintBackRefs{\CurrentBib}

\bibitem [\protect \citeauthoryear {%
Romdhani%
, Hwang%
, Paradis%
, Roy‐Gagnon%
\BCBL {}\ \BBA {} Labbe%
}{%
Romdhani%
\ \protect \BOthers {.}}{%
{\protect \APACyear {2015}}%
}]{%
Romdhani2015}
\APACinsertmetastar {%
Romdhani2015}%
\begin{APACrefauthors}%
Romdhani, H.%
, Hwang, H.%
, Paradis, G.%
, Roy‐Gagnon, M.%
\BCBL {}\ \BBA {} Labbe, A.%
\end{APACrefauthors}%
\unskip\
\newblock
\APACrefYearMonthDay{2015}{}{}.
\newblock
{\BBOQ}\APACrefatitle {{Pathway‐based association study of multiple candidate
  genes and multiple traits using structural equation models}}
  {{Pathway‐based association study of multiple candidate genes and multiple
  traits using structural equation models}}.{\BBCQ}
\newblock
\APACjournalVolNumPages{Genetic Epidemiology}{39}{}{101--113}.
\newblock
\begin{APACrefDOI} 10.1002/gepi.21872 \end{APACrefDOI}
\PrintBackRefs{\CurrentBib}

\bibitem [\protect \citeauthoryear {%
R{\"{o}}nkk{\"{o}}%
\ \BBA {} Evermann%
}{%
R{\"{o}}nkk{\"{o}}%
\ \BBA {} Evermann%
}{%
{\protect \APACyear {2013}}%
}]{%
Ronkko2013AModeling}
\APACinsertmetastar {%
Ronkko2013AModeling}%
\begin{APACrefauthors}%
R{\"{o}}nkk{\"{o}}, M.%
\BCBT {}\ \BBA {} Evermann, J.%
\end{APACrefauthors}%
\unskip\
\newblock
\APACrefYearMonthDay{2013}{}{}.
\newblock
{\BBOQ}\APACrefatitle {{A critical examination of common beliefs about partial
  least squares path modeling}} {{A critical examination of common beliefs
  about partial least squares path modeling}}.{\BBCQ}
\newblock
\APACjournalVolNumPages{Organizational Research Methods}{16}{}{425--448}.
\newblock
\begin{APACrefDOI} 10.1177/1094428112474693 \end{APACrefDOI}
\PrintBackRefs{\CurrentBib}

\bibitem [\protect \citeauthoryear {%
Rose%
, Wagner%
, Mayer%
\BCBL {}\ \BBA {} Nagengast%
}{%
Rose%
\ \protect \BOthers {.}}{%
{\protect \APACyear {2019}}%
}]{%
Rose2019Model-BasedModels}
\APACinsertmetastar {%
Rose2019Model-BasedModels}%
\begin{APACrefauthors}%
Rose, N.%
, Wagner, W.%
, Mayer, A.%
\BCBL {}\ \BBA {} Nagengast, B.%
\end{APACrefauthors}%
\unskip\
\newblock
\APACrefYearMonthDay{2019}{}{}.
\newblock
{\BBOQ}\APACrefatitle {{Model-based manifest and latent composite scores in
  structural equation models}} {{Model-based manifest and latent composite
  scores in structural equation models}}.{\BBCQ}
\newblock
\APACjournalVolNumPages{Collabra: Psychology}{5}{}{9}.
\newblock
\begin{APACrefDOI} 10.1525/collabra.143 \end{APACrefDOI}
\PrintBackRefs{\CurrentBib}

\bibitem [\protect \citeauthoryear {%
Rosseel%
}{%
Rosseel%
}{%
{\protect \APACyear {2012}}%
}]{%
Rosseel2012}
\APACinsertmetastar {%
Rosseel2012}%
\begin{APACrefauthors}%
Rosseel, Y.%
\end{APACrefauthors}%
\unskip\
\newblock
\APACrefYearMonthDay{2012}{}{}.
\newblock
{\BBOQ}\APACrefatitle {{lavaan: An R package for structural equation modeling}}
  {{lavaan: An R package for structural equation modeling}}.{\BBCQ}
\newblock
\APACjournalVolNumPages{Journal of Statistical Software}{48(1)}{}{1--36}.
\newblock
\begin{APACrefDOI} 10.18637/jss.v048.i02 \end{APACrefDOI}
\PrintBackRefs{\CurrentBib}

\bibitem [\protect \citeauthoryear {%
Sarstedt%
, Hair%
, Ringle%
, Thiele%
\BCBL {}\ \BBA {} Gudergan%
}{%
Sarstedt%
\ \protect \BOthers {.}}{%
{\protect \APACyear {2016}}%
}]{%
Sarstedt2016}
\APACinsertmetastar {%
Sarstedt2016}%
\begin{APACrefauthors}%
Sarstedt, M.%
, Hair, J\BPBI F.%
, Ringle, C\BPBI M.%
, Thiele, K\BPBI O.%
\BCBL {}\ \BBA {} Gudergan, S\BPBI P.%
\end{APACrefauthors}%
\unskip\
\newblock
\APACrefYearMonthDay{2016}{}{}.
\newblock
{\BBOQ}\APACrefatitle {{Estimation issues with PLS and CBSEM: Where the bias
  lies!}} {{Estimation issues with PLS and CBSEM: Where the bias lies!}}{\BBCQ}
\newblock
\APACjournalVolNumPages{Journal of Business Research}{69}{}{3998--4010}.
\newblock
\begin{APACrefDOI} 10.1016/j.jbusres.2016.06.007 \end{APACrefDOI}
\PrintBackRefs{\CurrentBib}

\bibitem [\protect \citeauthoryear {%
Schamberger%
, Schuberth%
, Henseler%
\BCBL {}\ \BBA {} Dijkstra%
}{%
Schamberger%
\ \protect \BOthers {.}}{%
{\protect \APACyear {2020}}%
}]{%
Schamberger2020}
\APACinsertmetastar {%
Schamberger2020}%
\begin{APACrefauthors}%
Schamberger, T.%
, Schuberth, F.%
, Henseler, J.%
\BCBL {}\ \BBA {} Dijkstra, T\BPBI K.%
\end{APACrefauthors}%
\unskip\
\newblock
\APACrefYearMonthDay{2020}{}{}.
\newblock
{\BBOQ}\APACrefatitle {{Robust partial least squares path modeling}} {{Robust
  partial least squares path modeling}}.{\BBCQ}
\newblock
\APACjournalVolNumPages{Behaviormetrika}{47}{}{307--334}.
\newblock
\begin{APACrefDOI} 10.1007/s41237-019-00088-2 \end{APACrefDOI}
\PrintBackRefs{\CurrentBib}

\bibitem [\protect \citeauthoryear {%
Schermelleh-Engel%
, Moosbrugger%
\BCBL {}\ \BBA {} M{\"{u}}ller%
}{%
Schermelleh-Engel%
\ \protect \BOthers {.}}{%
{\protect \APACyear {2003}}%
}]{%
Schermelleh-Engel2003EvaluatingMeasures}
\APACinsertmetastar {%
Schermelleh-Engel2003EvaluatingMeasures}%
\begin{APACrefauthors}%
Schermelleh-Engel, K.%
, Moosbrugger, H.%
\BCBL {}\ \BBA {} M{\"{u}}ller, H.%
\end{APACrefauthors}%
\unskip\
\newblock
\APACrefYearMonthDay{2003}{}{}.
\newblock
{\BBOQ}\APACrefatitle {{Evaluating the fit of structural equation models: Tests
  of significance and descriptive goodness-of-fit measures}} {{Evaluating the
  fit of structural equation models: Tests of significance and descriptive
  goodness-of-fit measures}}.{\BBCQ}
\newblock
\APACjournalVolNumPages{Methods of Psychological Research}{8}{}{23--74}.
\PrintBackRefs{\CurrentBib}

\bibitem [\protect \citeauthoryear {%
Schleiden%
\ \BBA {} Neiberger%
}{%
Schleiden%
\ \BBA {} Neiberger%
}{%
{\protect \APACyear {2020}}%
}]{%
Schleiden2020DoesBehaviour}
\APACinsertmetastar {%
Schleiden2020DoesBehaviour}%
\begin{APACrefauthors}%
Schleiden, V.%
\BCBT {}\ \BBA {} Neiberger, C.%
\end{APACrefauthors}%
\unskip\
\newblock
\APACrefYearMonthDay{2020}{}{}.
\newblock
{\BBOQ}\APACrefatitle {{Does sustainability matter? A structural equation model
  for cross-border online purchasing behaviour}} {{Does sustainability matter?
  A structural equation model for cross-border online purchasing
  behaviour}}.{\BBCQ}
\newblock
\APACjournalVolNumPages{The International Review of Retail, Distribution and
  Consumer Research}{30}{}{46--67}.
\newblock
\begin{APACrefDOI} 10.1080/09593969.2019.1635907 \end{APACrefDOI}
\PrintBackRefs{\CurrentBib}

\bibitem [\protect \citeauthoryear {%
Schuberth%
}{%
Schuberth%
}{%
{\protect \APACyear {2023}}%
}]{%
Schuberth2021a}
\APACinsertmetastar {%
Schuberth2021a}%
\begin{APACrefauthors}%
Schuberth, F.%
\end{APACrefauthors}%
\unskip\
\newblock
\APACrefYearMonthDay{2023}{}{}.
\newblock
{\BBOQ}\APACrefatitle {{The Henseler-Ogasawara specification of composites in
  structural equation modeling: A tutorial.}} {{The Henseler-Ogasawara
  specification of composites in structural equation modeling: A
  tutorial.}}{\BBCQ}
\newblock
\APACjournalVolNumPages{Psychological Methods}{28}{}{843--859}.
\newblock
\begin{APACrefDOI} 10.1037/met0000432 \end{APACrefDOI}
\PrintBackRefs{\CurrentBib}

\bibitem [\protect \citeauthoryear {%
Schuberth%
, Schamberger%
, Kem{\'{e}}ny%
\BCBL {}\ \BBA {} Henseler%
}{%
Schuberth%
\ \protect \BOthers {.}}{%
{\protect \APACyear {2025}}%
}]{%
Schuberth2025TheModeling}
\APACinsertmetastar {%
Schuberth2025TheModeling}%
\begin{APACrefauthors}%
Schuberth, F.%
, Schamberger, T.%
, Kem{\'{e}}ny, I.%
\BCBL {}\ \BBA {} Henseler, J.%
\end{APACrefauthors}%
\unskip\
\newblock
\APACrefYearMonthDay{2025}{}{}.
\newblock
{\BBOQ}\APACrefatitle {{The sum score model: Specifying and testing equally
  weighted composites using structural equation modeling}} {{The sum score
  model: Specifying and testing equally weighted composites using structural
  equation modeling}}.{\BBCQ}
\newblock
\APACjournalVolNumPages{Psychometrika}{90}{}{358--383}.
\newblock
\begin{APACrefDOI} 10.1017/psy.2024.5 \end{APACrefDOI}
\PrintBackRefs{\CurrentBib}

\bibitem [\protect \citeauthoryear {%
Steiger%
}{%
Steiger%
}{%
{\protect \APACyear {1990}}%
}]{%
Steiger1990StructuralApproach}
\APACinsertmetastar {%
Steiger1990StructuralApproach}%
\begin{APACrefauthors}%
Steiger, J\BPBI H.%
\end{APACrefauthors}%
\unskip\
\newblock
\APACrefYearMonthDay{1990}{}{}.
\newblock
{\BBOQ}\APACrefatitle {{Structural model evaluation and modification: An
  interval estimation approach}} {{Structural model evaluation and
  modification: An interval estimation approach}}.{\BBCQ}
\newblock
\APACjournalVolNumPages{Multivariate Behavioral Research}{25}{}{173--180}.
\newblock
\begin{APACrefDOI} 10.1207/s15327906mbr2502{\_}4 \end{APACrefDOI}
\PrintBackRefs{\CurrentBib}

\bibitem [\protect \citeauthoryear {%
Steyer%
, Mayer%
, Geiser%
\BCBL {}\ \BBA {} Cole%
}{%
Steyer%
\ \protect \BOthers {.}}{%
{\protect \APACyear {2015}}%
}]{%
Steyer2015ARevised}
\APACinsertmetastar {%
Steyer2015ARevised}%
\begin{APACrefauthors}%
Steyer, R.%
, Mayer, A.%
, Geiser, C.%
\BCBL {}\ \BBA {} Cole, D\BPBI A.%
\end{APACrefauthors}%
\unskip\
\newblock
\APACrefYearMonthDay{2015}{}{}.
\newblock
{\BBOQ}\APACrefatitle {{A theory of states and traits — revised}} {{A theory
  of states and traits — revised}}.{\BBCQ}
\newblock
\APACjournalVolNumPages{Annual Review of Clinical Psychology}{11}{}{71--98}.
\newblock
\begin{APACrefDOI} 10.1146/annurev-clinpsy-032813-153719 \end{APACrefDOI}
\PrintBackRefs{\CurrentBib}

\bibitem [\protect \citeauthoryear {%
Stojanovic%
, Milosevic%
, Arsic%
, Urosevic%
\BCBL {}\ \BBA {} Mihajlovic%
}{%
Stojanovic%
\ \protect \BOthers {.}}{%
{\protect \APACyear {2020}}%
}]{%
Stojanovic2020CorporatePerformance}
\APACinsertmetastar {%
Stojanovic2020CorporatePerformance}%
\begin{APACrefauthors}%
Stojanovic, A.%
, Milosevic, I.%
, Arsic, S.%
, Urosevic, S.%
\BCBL {}\ \BBA {} Mihajlovic, I.%
\end{APACrefauthors}%
\unskip\
\newblock
\APACrefYearMonthDay{2020}{}{}.
\newblock
{\BBOQ}\APACrefatitle {{Corporate social responsibility as a determinant of
  employee loyalty and business performance}} {{Corporate social responsibility
  as a determinant of employee loyalty and business performance}}.{\BBCQ}
\newblock
\APACjournalVolNumPages{Journal of Competitiveness}{12}{}{149--166}.
\newblock
\begin{APACrefDOI} 10.7441/joc.2020.02.09 \end{APACrefDOI}
\PrintBackRefs{\CurrentBib}

\bibitem [\protect \citeauthoryear {%
Sutton%
}{%
Sutton%
}{%
{\protect \APACyear {1998}}%
}]{%
Sutton1998}
\APACinsertmetastar {%
Sutton1998}%
\begin{APACrefauthors}%
Sutton, S.%
\end{APACrefauthors}%
\unskip\
\newblock
\APACrefYearMonthDay{1998}{}{}.
\newblock
{\BBOQ}\APACrefatitle {{Predicting and explaining intentions and behavior: How
  well are we doing?}} {{Predicting and explaining intentions and behavior: How
  well are we doing?}}{\BBCQ}
\newblock
\APACjournalVolNumPages{Journal of Applied Social
  Psychology}{28}{}{1317--1338}.
\newblock
\begin{APACrefDOI} 10.1111/j.1559-1816.1998.tb01679.x \end{APACrefDOI}
\PrintBackRefs{\CurrentBib}

\bibitem [\protect \citeauthoryear {%
Tarka%
}{%
Tarka%
}{%
{\protect \APACyear {2018}}%
}]{%
Tarka2018AnSciences}
\APACinsertmetastar {%
Tarka2018AnSciences}%
\begin{APACrefauthors}%
Tarka, P.%
\end{APACrefauthors}%
\unskip\
\newblock
\APACrefYearMonthDay{2018}{}{}.
\newblock
{\BBOQ}\APACrefatitle {{An overview of structural equation modeling: Its
  beginnings, historical development, usefulness and controversies in the
  social sciences}} {{An overview of structural equation modeling: Its
  beginnings, historical development, usefulness and controversies in the
  social sciences}}.{\BBCQ}
\newblock
\APACjournalVolNumPages{Quality {\&} Quantity}{52}{}{313--354}.
\newblock
\begin{APACrefDOI} 10.1007/s11135-017-0469-8 \end{APACrefDOI}
\PrintBackRefs{\CurrentBib}

\bibitem [\protect \citeauthoryear {%
Tenenhaus%
, Vinzi%
, Chatelin%
\BCBL {}\ \BBA {} Lauro%
}{%
Tenenhaus%
\ \protect \BOthers {.}}{%
{\protect \APACyear {2005}}%
}]{%
Tenenhaus2005PLSModeling}
\APACinsertmetastar {%
Tenenhaus2005PLSModeling}%
\begin{APACrefauthors}%
Tenenhaus, M.%
, Vinzi, V\BPBI E.%
, Chatelin, Y\BHBI M.%
\BCBL {}\ \BBA {} Lauro, C.%
\end{APACrefauthors}%
\unskip\
\newblock
\APACrefYearMonthDay{2005}{}{}.
\newblock
{\BBOQ}\APACrefatitle {{PLS path modeling}} {{PLS path modeling}}.{\BBCQ}
\newblock
\APACjournalVolNumPages{Computational Statistics {\&} Data
  Analysis}{48}{}{159--205}.
\newblock
\begin{APACrefDOI} 10.1016/j.csda.2004.03.005 \end{APACrefDOI}
\PrintBackRefs{\CurrentBib}

\bibitem [\protect \citeauthoryear {%
Testa%
\ \protect \BOthers {.}}{%
Testa%
\ \protect \BOthers {.}}{%
{\protect \APACyear {2021}}%
}]{%
Testa2021ResponseIndicators}
\APACinsertmetastar {%
Testa2021ResponseIndicators}%
\begin{APACrefauthors}%
Testa, S.%
, Di~Cuonzo, D.%
, Ritorto, G.%
, Fanchini, L.%
, Bustreo, S.%
, Racca, P.%
\BCBL {}\ \BBA {} Rosato, R.%
\end{APACrefauthors}%
\unskip\
\newblock
\APACrefYearMonthDay{2021}{}{}.
\newblock
{\BBOQ}\APACrefatitle {{Response shift in health-related quality of life
  measures in the presence of formative indicators}} {{Response shift in
  health-related quality of life measures in the presence of formative
  indicators}}.{\BBCQ}
\newblock
\APACjournalVolNumPages{Health and Quality of Life Outcomes}{19}{}{9}.
\newblock
\begin{APACrefDOI} 10.1186/s12955-020-01663-y \end{APACrefDOI}
\PrintBackRefs{\CurrentBib}

\bibitem [\protect \citeauthoryear {%
Tian%
\ \BBA {} Henseler%
}{%
Tian%
\ \BBA {} Henseler%
}{%
{\protect \APACyear {2025}}%
}]{%
Tian2025AImplementation}
\APACinsertmetastar {%
Tian2025AImplementation}%
\begin{APACrefauthors}%
Tian, Y.%
\BCBT {}\ \BBA {} Henseler, J.%
\end{APACrefauthors}%
\unskip\
\newblock
\APACrefYearMonthDay{2025}{}{}.
\newblock
{\BBOQ}\APACrefatitle {{A food well-being index for sustainable eating
  behavior: Construction, validation, and implementation}} {{A food well-being
  index for sustainable eating behavior: Construction, validation, and
  implementation}}.{\BBCQ}
\newblock
\APACjournalVolNumPages{Food Quality and Preference}{122}{}{105295}.
\newblock
\begin{APACrefDOI} 10.1016/J.FOODQUAL.2024.105295 \end{APACrefDOI}
\PrintBackRefs{\CurrentBib}

\bibitem [\protect \citeauthoryear {%
Tuma%
\ \BBA {} Decker%
}{%
Tuma%
\ \BBA {} Decker%
}{%
{\protect \APACyear {2013}}%
}]{%
Tuma2013}
\APACinsertmetastar {%
Tuma2013}%
\begin{APACrefauthors}%
Tuma, M.%
\BCBT {}\ \BBA {} Decker, R.%
\end{APACrefauthors}%
\unskip\
\newblock
\APACrefYearMonthDay{2013}{}{}.
\newblock
{\BBOQ}\APACrefatitle {{Finite mixture models in market segmentation: A review
  and suggestions for best practices}} {{Finite mixture models in market
  segmentation: A review and suggestions for best practices}}.{\BBCQ}
\newblock
\APACjournalVolNumPages{Electronic Journal of Business Research
  Methods}{11}{}{2--15}.
\PrintBackRefs{\CurrentBib}

\bibitem [\protect \citeauthoryear {%
Uppal%
\ \BBA {} Wang%
}{%
Uppal%
\ \BBA {} Wang%
}{%
{\protect \APACyear {2003}}%
}]{%
Uppal2003ModelUnderdiversification}
\APACinsertmetastar {%
Uppal2003ModelUnderdiversification}%
\begin{APACrefauthors}%
Uppal, R.%
\BCBT {}\ \BBA {} Wang, T.%
\end{APACrefauthors}%
\unskip\
\newblock
\APACrefYearMonthDay{2003}{}{}.
\newblock
{\BBOQ}\APACrefatitle {{Model misspecification and underdiversification}}
  {{Model misspecification and underdiversification}}.{\BBCQ}
\newblock
\APACjournalVolNumPages{The Journal of Finance}{58}{}{2465--2486}.
\newblock
\begin{APACrefDOI} 10.1046/j.1540-6261.2003.00612.x \end{APACrefDOI}
\PrintBackRefs{\CurrentBib}

\bibitem [\protect \citeauthoryear {%
van Driel%
}{%
van Driel%
}{%
{\protect \APACyear {1978}}%
}]{%
vanDriel1978OnAnalysis}
\APACinsertmetastar {%
vanDriel1978OnAnalysis}%
\begin{APACrefauthors}%
van Driel, O\BPBI P.%
\end{APACrefauthors}%
\unskip\
\newblock
\APACrefYearMonthDay{1978}{}{}.
\newblock
{\BBOQ}\APACrefatitle {{On various causes of improper solutions in maximum
  likelihood factor analysis}} {{On various causes of improper solutions in
  maximum likelihood factor analysis}}.{\BBCQ}
\newblock
\APACjournalVolNumPages{Psychometrika}{43}{}{225--243}.
\newblock
\begin{APACrefDOI} 10.1007/BF02293865 \end{APACrefDOI}
\PrintBackRefs{\CurrentBib}

\bibitem [\protect \citeauthoryear {%
Vansteelandt%
, Bekaert%
\BCBL {}\ \BBA {} Claeskens%
}{%
Vansteelandt%
\ \protect \BOthers {.}}{%
{\protect \APACyear {2012}}%
}]{%
Vansteelandt2012OnInference}
\APACinsertmetastar {%
Vansteelandt2012OnInference}%
\begin{APACrefauthors}%
Vansteelandt, S.%
, Bekaert, M.%
\BCBL {}\ \BBA {} Claeskens, G.%
\end{APACrefauthors}%
\unskip\
\newblock
\APACrefYearMonthDay{2012}{}{}.
\newblock
{\BBOQ}\APACrefatitle {{On model selection and model misspecification in causal
  inference}} {{On model selection and model misspecification in causal
  inference}}.{\BBCQ}
\newblock
\APACjournalVolNumPages{Statistical Methods in Medical Research}{21}{}{7--30}.
\newblock
\begin{APACrefDOI} 10.1177/0962280210387717 \end{APACrefDOI}
\PrintBackRefs{\CurrentBib}

\bibitem [\protect \citeauthoryear {%
Vilares%
, Almeida%
\BCBL {}\ \BBA {} Coelho%
}{%
Vilares%
\ \protect \BOthers {.}}{%
{\protect \APACyear {2010}}%
}]{%
Vilares2010}
\APACinsertmetastar {%
Vilares2010}%
\begin{APACrefauthors}%
Vilares, M\BPBI J.%
, Almeida, M\BPBI H.%
\BCBL {}\ \BBA {} Coelho, P\BPBI S.%
\end{APACrefauthors}%
\unskip\
\newblock
\APACrefYearMonthDay{2010}{}{}.
\newblock
{\BBOQ}\APACrefatitle {{Comparison of likelihood and PLS estimators for
  structural equation modeling: A simulation with customer satisfaction data}}
  {{Comparison of likelihood and PLS estimators for structural equation
  modeling: A simulation with customer satisfaction data}}.{\BBCQ}
\newblock
\BIn{} \APACrefbtitle {Handbook of Partial Least Squares} {Handbook of partial
  least squares}\ (\BPGS\ 289--305).
\newblock
\APACaddressPublisher{}{Springer Berlin Heidelberg}.
\newblock
\begin{APACrefDOI} 10.1007/978-3-540-32827-8{\_}14 \end{APACrefDOI}
\PrintBackRefs{\CurrentBib}

\bibitem [\protect \citeauthoryear {%
Vilares%
\ \BBA {} Coelho%
}{%
Vilares%
\ \BBA {} Coelho%
}{%
{\protect \APACyear {2013}}%
}]{%
Vilares2013LikelihoodEffects}
\APACinsertmetastar {%
Vilares2013LikelihoodEffects}%
\begin{APACrefauthors}%
Vilares, M\BPBI J.%
\BCBT {}\ \BBA {} Coelho, P\BPBI S.%
\end{APACrefauthors}%
\unskip\
\newblock
\APACrefYearMonthDay{2013}{}{}.
\newblock
{\BBOQ}\APACrefatitle {{Likelihood and PLS estimators for structural equation
  modeling: An assessment of sample size, skewness and model misspecification
  effects}} {{Likelihood and PLS estimators for structural equation modeling:
  An assessment of sample size, skewness and model misspecification
  effects}}.{\BBCQ}
\newblock
\BIn{} \APACrefbtitle {Studies in Theoretical and Applied Statistics, Selected
  Papers of the Statistical Societies} {Studies in theoretical and applied
  statistics, selected papers of the statistical societies}\ (\BPGS\ 11--33).
\newblock
\APACaddressPublisher{}{Springer International Publishing}.
\newblock
\begin{APACrefDOI} 10.1007/978-3-642-34904-1{\_}2 \end{APACrefDOI}
\PrintBackRefs{\CurrentBib}

\bibitem [\protect \citeauthoryear {%
Willoughby%
\ \BBA {} Blair%
}{%
Willoughby%
\ \BBA {} Blair%
}{%
{\protect \APACyear {2016}}%
}]{%
Willoughby2016MeasuringMeasurement.}
\APACinsertmetastar {%
Willoughby2016MeasuringMeasurement.}%
\begin{APACrefauthors}%
Willoughby, M\BPBI T.%
\BCBT {}\ \BBA {} Blair, C\BPBI B.%
\end{APACrefauthors}%
\unskip\
\newblock
\APACrefYearMonthDay{2016}{}{}.
\newblock
{\BBOQ}\APACrefatitle {{Measuring executive function in early childhood: A case
  for formative measurement.}} {{Measuring executive function in early
  childhood: A case for formative measurement.}}{\BBCQ}
\newblock
\APACjournalVolNumPages{Psychological Assessment}{28}{}{319--330}.
\newblock
\begin{APACrefDOI} 10.1037/pas0000152 \end{APACrefDOI}
\PrintBackRefs{\CurrentBib}

\bibitem [\protect \citeauthoryear {%
Wold%
}{%
Wold%
}{%
{\protect \APACyear {1975}}%
}]{%
Wold1975}
\APACinsertmetastar {%
Wold1975}%
\begin{APACrefauthors}%
Wold, H.%
\end{APACrefauthors}%
\unskip\
\newblock
\APACrefYearMonthDay{1975}{}{}.
\newblock
{\BBOQ}\APACrefatitle {{Path models with latent variables: The NIPALS
  approach}} {{Path models with latent variables: The NIPALS approach}}.{\BBCQ}
\newblock
\BIn{} \APACrefbtitle {Quantitative Sociology} {Quantitative sociology}\
  (\BPGS\ 307--357).
\newblock
\APACaddressPublisher{}{Elsevier}.
\newblock
\begin{APACrefDOI} 10.1016/B978-0-12-103950-9.50017-4 \end{APACrefDOI}
\PrintBackRefs{\CurrentBib}

\bibitem [\protect \citeauthoryear {%
Wold%
}{%
Wold%
}{%
{\protect \APACyear {1982}}%
}]{%
Wold1982}
\APACinsertmetastar {%
Wold1982}%
\begin{APACrefauthors}%
Wold, H.%
\end{APACrefauthors}%
\unskip\
\newblock
\APACrefYearMonthDay{1982}{}{}.
\newblock
{\BBOQ}\APACrefatitle {{Soft modelling: The basic design and some extensions}}
  {{Soft modelling: The basic design and some extensions}}.{\BBCQ}
\newblock
\APACjournalVolNumPages{Systems under indirect observation: Part
  II}{}{}{36--37}.
\PrintBackRefs{\CurrentBib}

\bibitem [\protect \citeauthoryear {%
Wright%
, Campbell%
, Thatcher%
\BCBL {}\ \BBA {} Roberts%
}{%
Wright%
\ \protect \BOthers {.}}{%
{\protect \APACyear {2012}}%
}]{%
Wright2012OperationalizingResearch}
\APACinsertmetastar {%
Wright2012OperationalizingResearch}%
\begin{APACrefauthors}%
Wright, R\BPBI T.%
, Campbell, D\BPBI E.%
, Thatcher, J\BPBI B.%
\BCBL {}\ \BBA {} Roberts, N.%
\end{APACrefauthors}%
\unskip\
\newblock
\APACrefYearMonthDay{2012}{}{}.
\newblock
{\BBOQ}\APACrefatitle {{Operationalizing multidimensional constructs in
  structural equation modeling: Recommendations for IS research}}
  {{Operationalizing multidimensional constructs in structural equation
  modeling: Recommendations for IS research}}.{\BBCQ}
\newblock
\APACjournalVolNumPages{Communications of the Association for Information
  Systems}{30}{}{23}.
\newblock
\begin{APACrefDOI} 10.17705/1CAIS.03023 \end{APACrefDOI}
\PrintBackRefs{\CurrentBib}

\bibitem [\protect \citeauthoryear {%
Yoshikuni%
, Dwivedi%
\BCBL {}\ \BBA {} Dwivedi%
}{%
Yoshikuni%
\ \protect \BOthers {.}}{%
{\protect \APACyear {2024}}%
}]{%
Yoshikuni2024StrategicPerformance}
\APACinsertmetastar {%
Yoshikuni2024StrategicPerformance}%
\begin{APACrefauthors}%
Yoshikuni, A\BPBI C.%
, Dwivedi, R.%
\BCBL {}\ \BBA {} Dwivedi, Y\BPBI K.%
\end{APACrefauthors}%
\unskip\
\newblock
\APACrefYearMonthDay{2024}{}{}.
\newblock
{\BBOQ}\APACrefatitle {{Strategic knowledge, IT capabilities and innovation
  ambidexterity: Role of business process performance}} {{Strategic knowledge,
  IT capabilities and innovation ambidexterity: Role of business process
  performance}}.{\BBCQ}
\newblock
\APACjournalVolNumPages{Industrial Management {\&} Data
  Systems}{124}{}{915--948}.
\newblock
\begin{APACrefDOI} 10.1108/IMDS-01-2023-0056 \end{APACrefDOI}
\PrintBackRefs{\CurrentBib}

\bibitem [\protect \citeauthoryear {%
Yu%
, Schuberth%
\BCBL {}\ \BBA {} Henseler%
}{%
Yu%
\ \protect \BOthers {.}}{%
{\protect \APACyear {2023}}%
}]{%
Yu2023}
\APACinsertmetastar {%
Yu2023}%
\begin{APACrefauthors}%
Yu, X.%
, Schuberth, F.%
\BCBL {}\ \BBA {} Henseler, J.%
\end{APACrefauthors}%
\unskip\
\newblock
\APACrefYearMonthDay{2023}{}{}.
\newblock
{\BBOQ}\APACrefatitle {{Specifying composites in structural equation modeling:
  A refinement of the Henseler–Ogasawara specification}} {{Specifying
  composites in structural equation modeling: A refinement of the
  Henseler–Ogasawara specification}}.{\BBCQ}
\newblock
\APACjournalVolNumPages{Statistical Analysis and Data Mining: The ASA Data
  Science Journal}{16}{}{348--357}.
\newblock
\begin{APACrefDOI} 10.1002/sam.11608 \end{APACrefDOI}
\PrintBackRefs{\CurrentBib}

\end{thebibliography}


\begin{thebibliography}{10}
\providecommand \doibase [0]{http://dx.doi.org/}%

\bibitem{Hirt1974}
Hirt CW, Amsden AA, Cook JL. An arbitrary {L}agrangian-{E}ulerian computing
  method for all flow speeds. {\it J {C}omput {P}hys.}
  1974\string;14(3)\string:227--253.

\bibitem{Liska2010}
Liska R, Shashkov M, Vachal P, et al. Optimization-based synchronized
  flux-corrected conservative interpolation (remapping) of mass and momentum
  for arbitrary {L}agrangian-{E}ulerian methods. {\it J {C}omput {P}hys.}
  2010\string;229(5)\string:1467--1497.

\bibitem{Taylor1937}
Taylor GI, Green AE. Mechanism of the production of small eddies from large
  ones. {\it P {R}oy {S}oc {L}ond {A} {M}at.}
  1937\string;158(895)\string:499--521.
\newblock \url{https://doi.org/10.1098/rspa.1937.0036},
  \url{http://rspa.royalsocietypublishing.org/content/158/895/499}.

\bibitem{Knupp1999}
Knupp PM. Winslow smoothing on two-dimensional unstructured meshes. {\it Eng
  {C}omput.} 1999\string;15\string:263--268.

\bibitem{Kamm2000}
Kamm J. Evaluation of the {S}edov-von {N}eumann-{T}aylor blast wave solution.
  Tech. Rep. Technical {R}eport LA-UR-00-6055, Los {A}lamos {N}ational
  {L}aboratory; The address:   2000.

\bibitem{Kucharik2003}
Kucharik M, Shashkov M, Wendroff B. An efficient linearity-and-bound-preserving
  remapping method. {\it J {C}omput {P}hys.}
  2003\string;188(2)\string:462--471.

\bibitem{Blanchard2015}
Blanchard G, Loubere R. High-Order {C}onservative {R}emapping with a posteriori
  {MOOD} stabilization on polygonal meshes. Details on how published;  2015.
\newblock Accessed January 13, 2016.
  \url{https://hal.archives-ouvertes.fr/hal-01207156}, the {HAL} {O}pen
  {A}rchive, hal-01207156.

\bibitem{Burton2013}
Burton DE, Kenamond MA, Morgan NR, Carney TC, Shashkov MJ, Author AB. An
  intersection based {ALE} scheme {(xALE)} for cell centered hydrodynamics
  {(CCH)}. In: Talk at {M}ultimat 2013, {I}nternational {C}onference on
  {N}umerical {M}ethods for {M}ulti-{M}aterial {F}luid {F}lows. The
  Organization.  September 2--6, 2013; San {F}rancisco.
\newblock LA-UR-13-26756.2.

\bibitem{Berndt2011}
Berndt M, Breil J, Galera S, Kucharik M, Maire PH, Shashkov M. Two-step hybrid
  conservative remapping for multimaterial arbitrary {L}agrangian-{E}ulerian
  methods. {\it J {C}omput {P}hys.} 2011\string;230(17)\string:6664--6687.

\bibitem{Kucharik2012}
Kucharik M, Shashkov M. One-step hybrid remapping algorithm for multi-material
  arbitrary {L}agrangian-{E}ulerian methods. {\it J {C}omput {P}hys.}
  2012\string;231(7)\string:2851--2864.

\bibitem{Breil2015}
Breil J, Alcin H, Maire PH. A swept intersection-based remapping method for
  axisymmetric {ReALE} computation. {\it Int {J} {N}umer {M}eth {F}l.}
  2015\string;77(11)\string:694--706.
\newblock Fld.3996.

\bibitem{Barth1997}
Barth TJ. Numerical methods for gasdynamic systems on unstructured meshes. In:
  Kroner D, Rohde C, Ohlberger M. \kern-2pt, eds. {\it An {I}ntroduction to
  {R}ecent {D}evelopments in {T}heory and {N}umerics for {C}onservation {L}aws,
  {P}roceedings of the {I}nternational {S}chool on {T}heory and {N}umerics for
  {C}onservation {L}aws}, 2~ed., Lecture {N}otes in {C}omputational {S}cience
  and {E}ngineering. Springer,  1997.

\bibitem{Lauritzen2011}
Lauritzen P, Erath C, Mittal R. On simplifying `incremental remap'-based
  transport schemes. {\it J {C}omput {P}hys.}
  2011\string;230(22)\string:7957--7963.

\bibitem{Klima2017}
Klima M, Kucharik M, Shashkov M. Local error analysis and comparison of the
  swept- and intersection-based remapping methods. {\it Commun {C}omput
  {P}hys.} 2017\string;21(2)\string:526--558.

\bibitem{Dukowicz2000}
Dukowicz JK, Baumgardner JR. Incremental remapping as a transport/advection
  algorithm. {\it J {C}omput {P}hys.} 2000\string;160(1)\string:318--335.

\bibitem{Kucharik2011}
Kucharik M, Shashkov M. Flux-based approach for conservative remap of
  multi-material quantities in {2D} arbitrary {L}agrangian-{E}ulerian
  simulations. In:  Fo\v{r}t J, F{\"{u}}rst J, Halama J, Herbin R, Hubert F.
  \kern-2pt, eds. {\it Finite {V}olumes for {C}omplex {A}pplications {VI}
  {P}roblems \& {P}erspectives}, 1~ed., Springer {P}roceedings in
  {M}athematics. Springer,  2011\string:623--631.

\bibitem{Kucharik2014}
Kucharik M, Shashkov M. Conservative multi-material remap for staggered
  multi-material arbitrary {L}agrangian-{E}ulerian methods. {\it J {C}omput
  {P}hys.} 2014\string;258\string:268--304.

\bibitem{Loubere2005}
Loubere R, Shashkov M. A subcell remapping method on staggered polygonal grids
  for arbitrary-{L}agrangian-{E}ulerian methods. {\it J {C}omput {P}hys.}
  2005\string;209(1)\string:105--138.

\bibitem{Caramana1998}
Caramana EJ, Shashkov MJ. Elimination of artificial grid distortion and
  hourglass-type motions by means of {L}agrangian subzonal masses and
  pressures. {\it J {C}omput {P}hys.} 1998\string;142(2)\string:521--561.

\bibitem{Hoch2009}
Hoch P. An arbitrary {L}agrangian-{E}ulerian strategy to solve compressible
  fluid flows. Tech. Rep. Technical {R}eport, CEA; The address:   2009.
\newblock Accessed January 13, 2016. HAL: hal-00366858.
  https://hal.archives-ouvertes.fr/docs/00/36/68/58/PDF/ale2d.pdf.

\bibitem{Shashkov1996}
Shashkov M. {\it Conservative {F}inite-{D}ifference {M}ethods on {G}eneral
  {G}rids}.
\newblock CRC {P}ress, 1996.

\bibitem{Benson1992}
Benson DJ. Computational methods in {L}agrangian and {E}ulerian hydrocodes.
  {\it Comput {M}ethod {A}ppl {M}.} 1992\string;99(2--3)\string:235--394.

\bibitem{Margolin2003}
Margolin LG, Shashkov M. Second-order sign-preserving conservative
  interpolation (remapping) on general grids. {\it J {C}omput {P}hys.}
  2003\string;184(1)\string:266--298.

\bibitem{Kenamond2013}
Kenamond MA, Burton DE. Exact intersection remapping of multi-material
  domain-decomposed polygonal meshes. In: Talk at {M}ultimat 2013,
  {I}nternational {C}onference on {N}umerical {M}ethods for {M}ulti-{M}aterial
  {F}luid {F}lows. The Organization.  September 2--6, 2013; San {F}rancisco.
\newblock LA-UR-13-26794.

\bibitem{Dukowicz1984}
Dukowicz J. Conservative rezoning (remapping) for general quadrilateral meshes.
  {\it J {C}omput {P}hys.} 1984\string;54(3)\string:411--424.

\bibitem{Margolin2002}
Margolin LG, Shashkov M. Second-order sign-preserving remapping on general
  grids. Tech. Rep. Technical Report LA-UR-02-525, Los {A}lamos {N}ational
  {L}aboratory; The address:   2002.

\bibitem{Mavriplis2003}
Mavriplis DJ. Revisiting the least-squares procedure for gradient
  reconstruction on unstructured meshes. In: AIAA 2003-3986. 16th {AIAA}
  {C}omputational {F}luid {D}ynamics {C}onference. The organization.  June
  23--26, 2003; Orlando, {F}lorida.

\bibitem{Scovazzi2008}
Scovazzi G, Love E, Shashkov M. Multi-scale {L}agrangian shock hydrodynamics on
  {Q1/P0} finite elements: {T}heoretical framework and two-dimensional
  computations. {\it Comput {M}ethod {A}ppl {M}.}
  2008\string;197(9--12)\string:1056--1079.

\end{thebibliography}
